\documentclass[12pt]{iopart}
\usepackage{iopams}  
\usepackage{graphicx}
\usepackage{bm}
\usepackage{amsbsy}
\usepackage{color}
\usepackage{amssymb}

\begin{document}
\newcommand{\aap}{Astron. \& Astrophys.}
\newcommand{\araa}{Annu. Rev. Aston. \& Astroph.}
\newcommand{\aapr}{Astron. \& Astrophys. Rev.}
\newcommand{\mnras}{Mon. Not. Roy. astron. Soc.}
\newcommand{\apjl}{Astrophys. J. Lett.}
\newcommand{\apj}{Astrophys. J.}
\newcommand{\apjs}{Astrophys. J. Suppl.}
\newcommand{\pre}{Phys. Rev. E}
\newcommand{\nat}{Nature}
\newcommand{\prl}{Phys. Rev. Lett.}
\newcommand{\jfm}{J. Fluid Mech.}
\newcommand{\jcp}{J. Comput. Phys.}

\title[Magnetized ISM turbulence]{The structure and statistics of interstellar turbulence}

\author{A G Kritsuk$^1$, S D Ustyugov$^2$ and M L Norman$^{1,3}$}

\address{$^1$ Center for Astrophysics and Space Sciences, University of California, San Diego,\\ 9500 Gilman Drive, La Jolla, CA 92093-0424, U.S.A.}
\address{$^2$ Keldysh Institute of applied Mathematics, Moscow, 4 Miusskaya Sq., 125047, Moscow, Russia}
\address{$^3$ San Diego Supercomputer Center, University of California, San Diego, 10100 Hopkins Drive, La Jolla, CA 92093-0505, U.S.A.}
\ead{akritsuk@ucsd.edu}
\vspace{10pt}

\begin{abstract}
We explore the structure and statistics of multiphase, magnetized ISM turbulence in the local Milky Way by means of driven periodic box numerical MHD simulations. Using the higher order-accurate piecewise-parabolic method on a local stencil (PPML), we carry out a small parameter survey varying the mean magnetic field strength and density while fixing the rms velocity to observed values. We quantify numerous characteristics of the transient and steady-state turbulence, including its thermodynamics and phase structure, kinetic and magnetic energy power spectra, structure functions, and distribution functions of density, column density, pressure, and magnetic field strength.  The simulations reproduce many observables of the local ISM, including molecular clouds, such as the ratio of turbulent to mean magnetic field at 100 pc scale, the mass and volume fractions of thermally stable H{\sc i}, the lognormal distribution of column densities, the mass-weighted distribution of thermal pressure, and the linewidth-size relationship for molecular clouds. Our models predict the shape of magnetic field probability density functions (PDFs), which are strongly non-Gaussian, and the relative alignment of magnetic field and density structures. Finally, our models show how the observed low rates of star formation per free-fall time are controlled by the multiphase thermodynamics and large-scale turbulence.
 \end{abstract}

%

%
%
%

\section{Introduction} 
Turbulence is ubiquitous in the interstellar medium (ISM) \cite{elmegreen.04}. It organizes the ISM structure to optimize and direct energy fluxes across an enormous range of length scales within the Milky Way disk. Interstellar turbulence, however, differs from the familiar Kolmogorov homogeneous and isotropic incompressible case \cite{kolmogorov41,frisch95} in several important respects: it is neither isotropic nor homogeneous \cite{kaplan.70}; the interstellar gas is highly compressible and magnetized; and its thermal energy is not strictly conserved. Instead, the ISM-specific radiative cooling and volumetric heating functions determine an equilibrium multiphase nature of the neutral ISM \cite{zeldovich.69,goldsmith..69,wolfire...03}. Nonlinear advection couples with nonlinear radiative cooling, further complicating the treatment of the ISM. We hence deal with a classical example of a hierarchical nonequilibrium dissipative system, receiving energy primarily at large scales ($\sim100$~pc) from stellar feedback, extragalactic gas accretion, large-scale shear, and gravitational instabilities within the disk and loosing energy through small-scale dissipation ($10^{-4}$~pc) and local radiative cooling \cite{maclow.04}. 

The concept of a hierarchy of  `eddies' loosely defined as coherent patterns in the velocity field of incompressible turbulence, is replaced in the case of the ISM by an even more indistinct hierarchy of cloud complexes, clouds, clumps and cores \cite{hennebelle.12}. Observational cloud definitions are usually tied to specific tracers and are essentially two-dimensional --due to complex line-of-sight convolutions-- making straightforward physical interpretation of observations extremely difficult  \cite{padoan+14}. Since interstellar turbulence controls the structure, dynamics and chemistry of the ISM, it is a key building block of any future successful theory of star formation in molecular clouds \cite{maclow.04,mckee.07,paredes...07,krumholz14}.

To explore this complexity, we develop self-consistent models based on very idealized numerical experiments, linking together scales relevant to molecular cloud formation and fragmentation. We exploit the concept of self-organization in nonequilibrium nonlinear dissipative systems \cite{nicolis.77} in application to the ISM, which is long overdue \cite{shull87}. We treat interstellar turbulence as an agent that imposes `order' in the form of coherent structures and correlations between flow fields emerging in a simple periodic box simulation when a statistically stationary state fully develops. In this case, the details of initial conditions are no longer important. Instead the steady state provides realistic initial conditions for star formation. Unlike various flavors of the popular converging flow scenario \cite{walder.98,heitsch....06,hennebelle....08,banerjee...09}, this approach 
allows us to reproduce basic ISM observables, including properties of local molecular clouds, with only a few control parameters. The model can be readily extended to include more physics (e.g., radiative transfer and chemistry), to augment the range of resolved scales, and to close the star formation feedback loop.

%

The paper is organized as follows. Section~\ref{num} contains the details of our numerical models, including equations, initial and boundary conditions, and parameters. Section~\ref{res} presents the results of statistical analysis and comparison with observations for a number of ISM diagnostics. \S~\ref{s-glo} describes the transition to turbulence and global properties of the statistically stationary state. \S~\ref{s-mag-glo} summarizes the evolution of magnetic field under the action of random large-scale external forcing. In \S~\ref{s-phase} we discuss thermodynamics and phase structure of the ISM emerging in the simulations. \S~\ref{s-mach} accesses the turbulent Mach number regimes of various thermal phases. \S~\ref{s-sfreg} presents predictions for star formation rate controlled by the turbulence and thermodynamic properties of the ISM. \S\S~\ref{s-dpdf} and \ref{s-pth} discuss density, column density and thermal pressure distributions. \S\S~\ref{s-pre}, \ref{align}, and \ref{s-mpdf} present statistics of the magnetic field. \S\S~\ref{larson}, \ref{s-rvspec}, and \ref{s-espec} discuss scaling properties of multiphase turbulence and energy spectral densities. We conclude with a summary of results in Section~\ref{s-con}.

\section{Numerical models\label{num}}
\subsection{Governing equations, initial and boundary conditions}
The following system of compressible MHD conservation laws for the mass, momentum, magnetic flux, and energy is solved in a cubic domain of linear size $L$ with triply periodic boundary conditions. The equations include external random forcing terms in the r.h.s. of (\ref{mome}) and (\ref{ener}), as well as the ISM-specific generalized volumetric cooling function in the r.h.s. of (\ref{ener}):
\numparts
\begin{eqnarray}
\partial_t \rho&+&{\bm \nabla}\cdot(\rho {\bm u}) =0, \label{mass}\\
\partial_t (\rho \bm u)&+&{\bm \nabla\cdot}\left[\rho \bm u\bm u -
\bm B\bm B+ \left(p+\frac{\bm B^2}{2}\right)\bm I\right]=\bm f, \label{mome}\\
\partial_t {\bm B}&+&\bm \nabla\cdot(\bm u\bm B - \bm B\bm u) = 0, \label{induc}\\
\partial_t {\cal E}&+&{\bm \nabla}\cdot\left[\left({\cal E}+p+\frac{\bm B^2}{2}\right)\bm u - \left(\bm B\cdot\bm u\right)\bm B\right] = \bm u\cdot\bm f-\rho{\cal L}(\rho,T). \label{ener}
\end{eqnarray}
\endnumparts
Here  $\rho$ is the gas density, $\bm u$ -- velocity, pressure is given by the ideal gas equation of state $p\equiv (\gamma-1)e\rho$, where  $\gamma=5/3$ is the specific heats ratio and  $e$ -- the specific internal energy density. 
The total energy density includes kinetic, internal and magnetic components ${\cal E}=\rho u^2/2 +\rho e+ B^2/2$ and $\langle{\cal E}\rangle=K+U+M$ is one of the invariants of an ideal system (with zero r.h.s.).
The generalized heat-loss function $\rho{\cal L}=n^2\Lambda(T)-n\Gamma$, where $n=\rho/m_{\rm H}$ is the H{\sc i} number density, is introduced to mimic uniform photoelectric and cosmic ray heating rates and local radiative cooling of the gas under typical interstellar conditions \cite{wolfire...03}.
The magnetic field strength, $\bm B$, is subject to the usual solenoidal constraint $\bm\nabla\bm\cdot\bm B=0$. The following initial conditions: $\rho_0+\delta\rho$ (where $\delta\rho$ represent 20\% random density perturbations), $p_{\rm 0}$, ${\bm u}_0=0$, and ${\bm B}_0=(B_0,0,0)$ are applied.
Random forcing implementation assumes zero momentum input $\bm f\equiv\rho\bm a-\langle\rho{\bm a}\rangle$ and applies a fixed-in-time large-scale solenoidal non-helical acceleration $\bm a(x)$ with proper normalization, ensuing a constant kinetic energy input rate.  {The forcing is turned on in the middle of the run at $t=18.9$~Myr and, when a finite initial velocity boost $\delta\bm u=\tau\bm a$ is applied. Here $\tau$ is a coefficient setting the finite amplitude of the one-time velocity perturbation}. 

System (\ref{mass})--(\ref{ener}) is integrated numerically using the piecewise-parabolic method on a local stencil (PPML \cite{ustyugov...09,kritsuk+11}), following the traditional implicit large eddy (ILES) approach \cite{sytine....00}. This implies that only an effective  Reynolds number $Re=u_{\rm rms}L/\nu_{\rm eff}$, Prandtl number $Pr=c_p\rho_0\nu_{\rm eff}/\kappa_{\rm eff}$ and magnetic Prandtl number $Pm=\nu_{\rm eff}/\eta_{\rm eff}$ can be defined ($\nu_{\rm eff}$, $\kappa_{\rm eff}$, and $\eta_{\rm eff}$ are the effective kinematic viscosity, thermal conductivity, and magnetic diffusivity controlled by numerical dissipation). In the molecular ISM, $Re\gg1$ and $Pm\gg1$ \cite{kritsuk+11}, while in the ILES there is no explicit control of dissipative processes and the magnetic Prandtl number is usually set by the numerical method at $Pm\sim1$. The effective Reynolds number is ultimately controlled by the nature of the numerical method and by the grid resolution $N$, with higher order accurate methods and larger grids delivering higher $Re$. Heat conduction can stabilize {thermal instability (TI)} on scales below the so-called Field length $\lambda_{\rm F}=\sqrt{\kappa T/\rho^2\Lambda(T)}$ \cite{field65} and should be ultimately included in the models. Our simulations, however, do not sufficiently resolve $\lambda_{\rm F}$ \cite{koyama.04} that varies somewhere between $0.1$~pc and $0.001$~pc, depending on local physical conditions. With such limited resolution, the heat transport will be dominated by turbulent diffusion.

Besides the effective $Re$, $Pr$, and $Pm$, two other nondimensional numbers are important for the considered system: the sonic Mach number ${\cal M}_{\rm s}=\sqrt{\langle (u/c_{\rm s})^2\rangle}$, where $c_{\rm s}$ is the sound speed, controls the degree of compressibility; and the Alfv\'en Mach number ${\cal M}_{\rm a}=\sqrt{\langle (u/v_{\rm a})^2\rangle}=\sqrt{\langle\rho u^2/B^2\rangle}$, where $v_{\rm a}=\sqrt{B^2/\rho}$ is the Alfv\'en speed, controls the mean kinetic-to-magnetic energy density ratio.

\subsection{Input parameters\label{s-param}}
\begin{table}
\caption{\label{input}Input parameters.}
\begin{indented}
\footnotesize
\item[]\begin{tabular}{@{}llccrcrr}
\br
Case& $N$ & $n_0$     & $u_{\rm rms,0}$ & $B_0$  & $\beta_{\rm th,0}$ & $\beta_{\rm turb,0}$ & ${\cal M}_{a,0}$  \\
     & & cm$^{-3}$ & km/s             & $\mu$G &           &                       &            \\
\mr
A & 512 & 5         & 16    & 9.54   & 0.2  &  3.3    & 1.3        \\
B & 512 & 5         & 16    & 3.02   &  2    &  33     & 4.0        \\
C & 512 & 5         & 16    & 0.95   & 20   &  330   & 13         \\
D & 256 & 2         & 16    & 3.02   &  2    &  13.2  & 2.6        \\
E & 256 & 5         &  7     & 3.02   &  2    &  8.3    & 2.0        \\  
\br
\end{tabular}\\
\end{indented}
\end{table}
\normalsize

Each model is defined by five control parameters: the domain size $L$;
the mean gas number density in the box $n_0$; 
the rms velocity $u_{\rm 0, rms}$ associated with the driving force amplitude; 
the uniform magnetic field strength $B_0$; 
and the grid resolution $N^3$. 

Table~\ref{input} provides a summary of parameters for the five cases discussed below. All models assume the box size $L=200$~pc, representing the scale height of the Galactic neutral H{\sc i}. Our choices for $n_0$ in the range from $2-5$~cm$^{-3}$ reflect local average conditions in a spiral arm-like setting with H{\sc i} number densities somewhat higher than the typical mean value of  1~cm$^{-3}$. Likewise, the values of $u_{\rm 0, rms}$ are chosen to bracket the realistic local ISM kinematics at the given scale $L$, matching the linewidth-size relation for local molecular clouds \cite{heyer.04}.

The table also provides the magnetic field parametrization in terms of the plasma beta
$\beta_{\rm th,0}=p_0/(B_0^2/2)$, representing the ratio of the initial thermal pressure of the gas $p_0$ to the magnetic pressure due to the applied uniform field $B_0$. The values of turbulent plasma beta (defined with the dynamic pressure $\rho u^2$ replacing thermal $p$ in the classical definition) $\beta_{\rm turb,0}= \rho_0 u^2_{\rm rms,0}/(B_0^2/2)$ and the initial Alfv\'enic Mach number ${\cal M}_{a,0}=u_{\rm rms,0}/(B_0/\sqrt{\rho_0})$ are also given for convenience. Overall, cases A, B, and C trace a transition from trans-Alfv\'enic (${\cal M}_{\rm a,0}\sim1$) to super-Alfv\'enic (${\cal M}_{\rm a,0}=4, 13$) magnetization levels, while cases D and E are only moderately super-Alfv\'enic (${\cal M}_{\rm a,0}\sim2-3$).
Finally, the total gas mass in the domain for cases A, B, C, and E is ${\mathfrak M}={9.8}\times10^5\,{\mathfrak M}_{\odot}$, while in case D the mass is 2.5 times smaller, ${\mathfrak M}={3.9}\times10^5\,{\mathfrak M}_{\odot}$. 

\begin{figure}[t]
\begin{center}
           \includegraphics[scale=0.47]{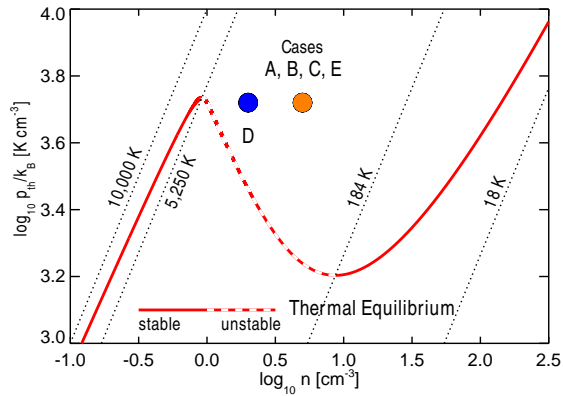}
           \end{center}
           \caption{Assumed thermal equilibrium conditions in the $p-n$ plane and initial conditions $p_0$ and $n_0$ for cases A through E. Red solid line shows stable branches of thermal equilibrium, dashed -- unstable. Tilted dotted lines are isotherms at temperatures $T=18$~K, 184~K, 5250~K, and $10^4$~K.}
           \label{phase-e}
\end{figure}
For the sake of simplicity, we adopt an analytical approximation for $\Lambda(T)$ \cite{koyama.02,koyama.06} based on a calculation \cite{wolfire....95} of the thermal equilibrium gas temperature of the diffuse interstellar medium that includes photoelectric heating rate from small grains and polycyclic aromatic hydrocarbons (PAHs) and a detailed treatment of the ionization rates and heating due to the soft X-ray background and due to cosmic rays. The cooling function approximation assumes solar metallicity. 
Radiative cooling is turned off for $T<26$~K, however there is no temperature floor and the gas temperature drops to $\sim4$~K in strong rarefactions adiabatically. The density-independent heating rate is set at $\Gamma=2\times10^{-26}$~erg/s. The high temperatures extend beyond 32,000~K in low-density gas, but the volume fraction of this `overheated' material is very low and we ignore changes in the molecular weight in the equation of state due to ionization processes. With these assumptions, the {\tt S}-shaped thermal equilibrium $\Gamma=(\rho/m_{\rm H})\Lambda(T)$ in the pressure-density plane has two extrema separating the two stable branches: a maximum at $T_{\rm max}=5250$~K and a minimum at $T_{\rm min}=184$~K (Fig.~\ref{phase-e}). These two temperatures set the upper and lower limits to the TI range of the linear isobaric mode, assuming equilibrium conditions \cite{field65}. Even though TI away from the equilibrium (i.e. where ${\cal L}(\rho,T)\ne0$) is determined by a generalized criterion \cite{hunter70}, we take a simplified approach and define the warm and cold stable phases solely based on the temperature. The warm phase thus includes all gas with $T>T_{\rm max}$ and the cold phase assumes $T<T_{\rm min}$. The range of intermediate temperatures $T_{\rm min}<T<T_{\rm max}$ roughly captures the thermally unstable regimes. With these definitions, our warm phase includes both warm neutral medium (WNM) and warm ionized medium (WIM).

    \begin{figure}[h]
\begin{center}
           \includegraphics[scale=0.55]{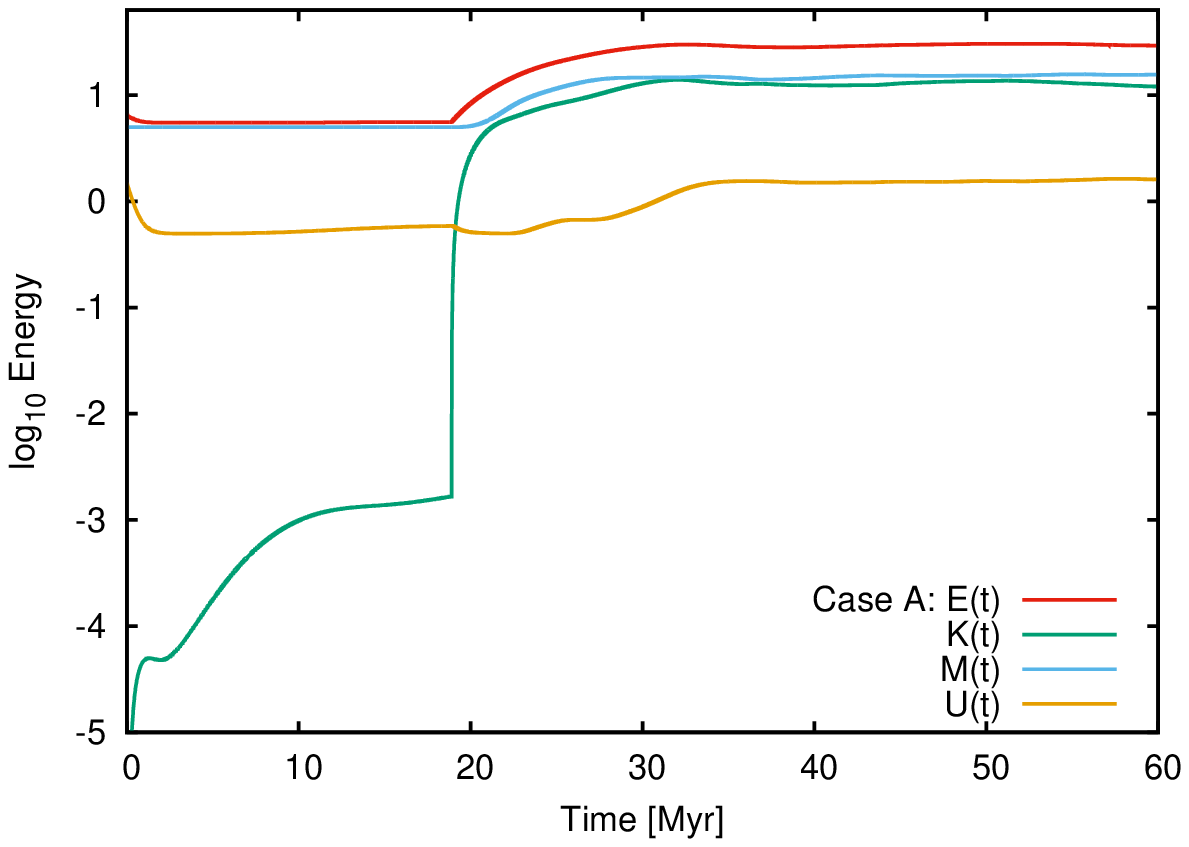}
           \includegraphics[scale=0.55]{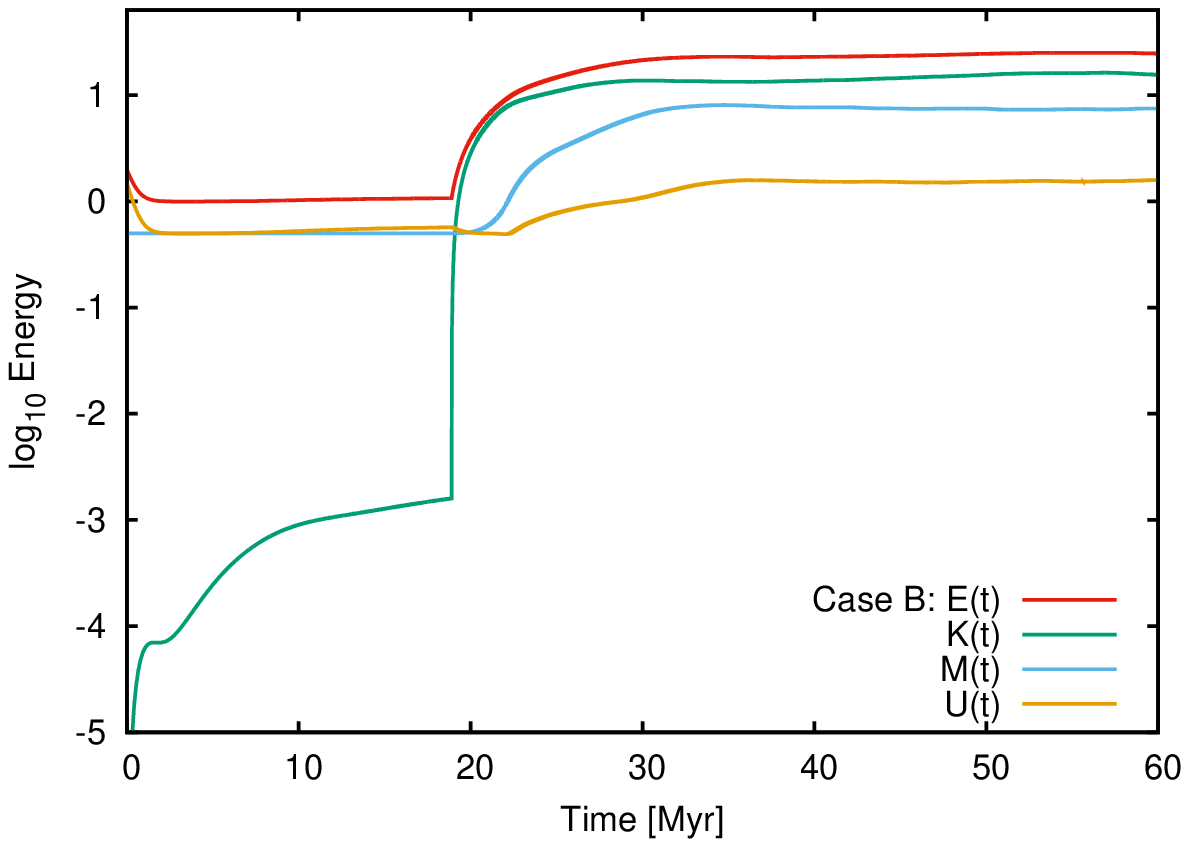}
           \includegraphics[scale=0.55]{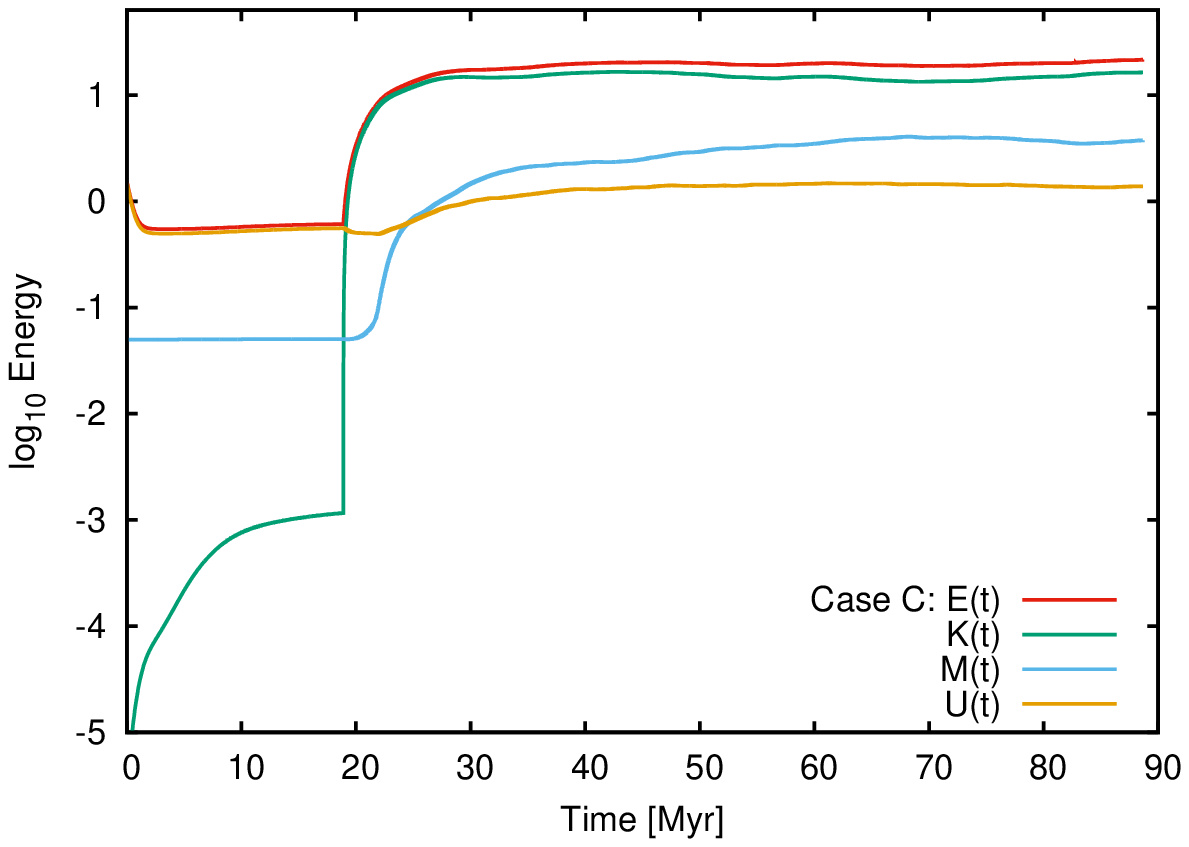}
           \includegraphics[scale=0.55]{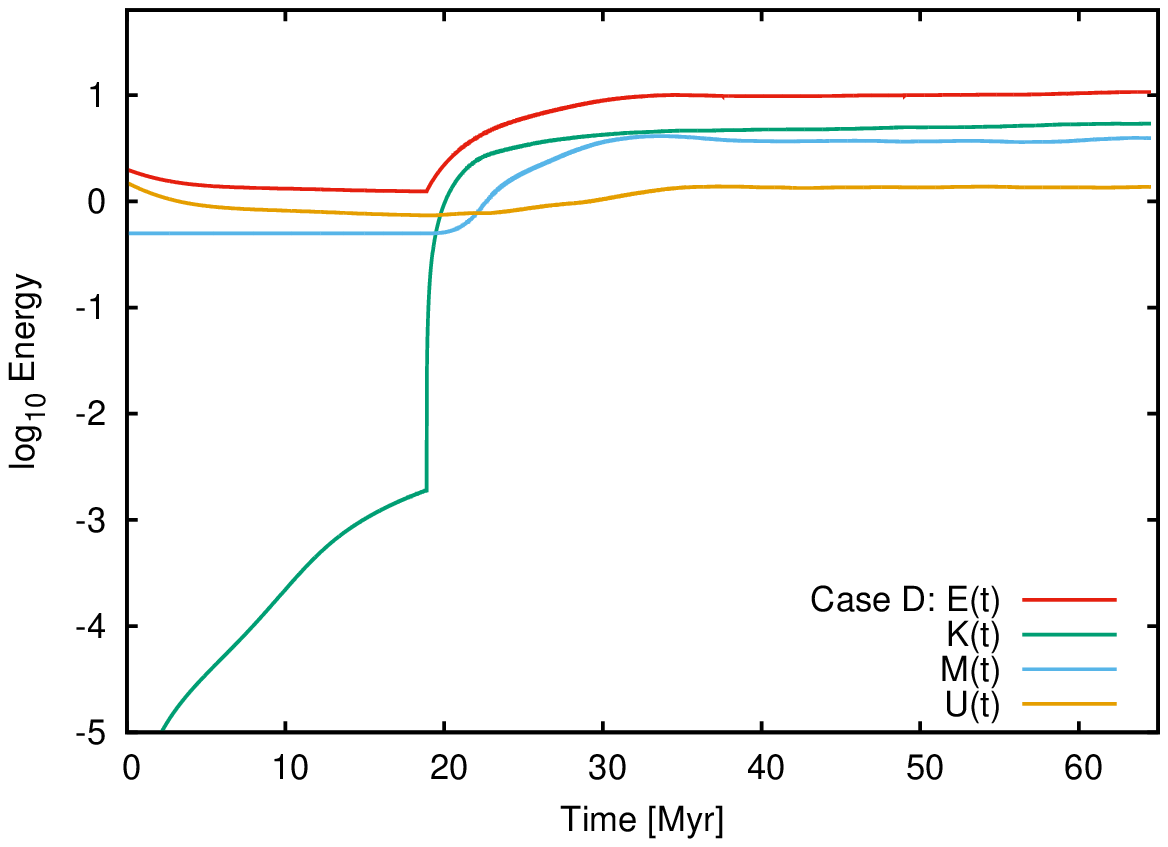}
           \includegraphics[scale=0.55]{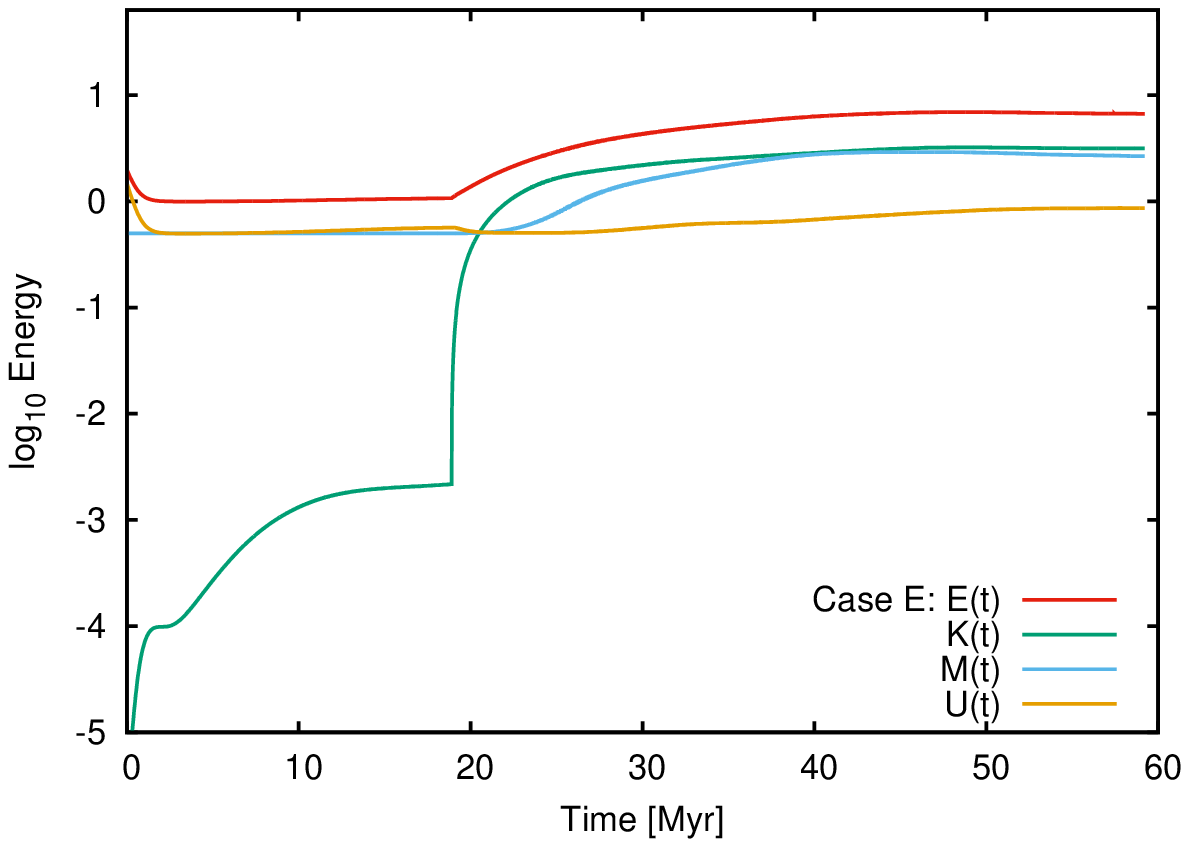}
           \caption{Time evolution of the mean kinetic ($K=\langle\rho\bm u^2/2\rangle$), magnetic ($M=\langle\bm B^2/2\rangle$), internal ($U=\langle\rho e\rangle$) and total ($E=K+M+U$) energy densities for cases A, B, C, D, and E. With forcing activated at $t=18.9$~Myr,
 all cases reach statistically stationary states $E(t)\approx const$ by $t\sim30$~Myr. Note that in case C with weak initial magnetization active exchange between $K$ and $M$ continues longer than in other cases, even though the total energy stays constant after 30~Myr. }
\label{f-energy}
\end{center}
    \end{figure}
\section{Results\label{res}}

\subsection{Energy evolution\label{s-glo}}
All cases were evolved in two stages. First, we allowed TI to develop in response to the small initial density perturbations, so that the system  undergoes a phase transition and forms cold dense clouds embedded in warm medium, generating weak subsonic turbulence \cite{kritsuk.02}. Then, as the thermal phases equilibrate and the system transitions into a stationary state, we turn on the random forcing at $t=19$~Myr. The force is normalized to reach the specific target velocity dispersion values $u_{\rm rms,0}$ assigned to each case, see Table~\ref{input}. 

Figure~\ref{f-energy} illustrates the evolution of mean energy densities in the code units for all cases. 
The kinetic energy, $K=\langle\rho\bm u^2/2\rangle$, shown in green indicates modest growth followed by saturation after the phase transition occurs. After that, a jump due to the activation of the forcing leads to a final turbulence saturation. The magnetic energy, $M=\langle\bm B^2/2\rangle$ (blue), starts at a level determined by the mean field $B_0$, which remains essentially unchanged during the phase transition stage. The magnetic energy then grows and reaches equipartition $M\sim K$ in cases A, D, and E, while remaining subdominant in cases B and C. Case D is similar to B, but has a lower kinetic energy saturation level due to $2.5\times$ lower mean density, which also helps to establish equipartition $M\sim K$. Case E is also similar to B, but has weaker forcing, causing lower energy saturation levels and equipartition $M\approx K$. The internal energy, $U=\langle\rho e\rangle$ (brown), first drops on a time scale of a few Myr due to radiative losses, then remains stationary during the post-phase-transition stage, then gets a boost from turbulence dissipation after the forcing is activated, and then finally saturates. The total energy, $E=\langle{\cal E}\rangle=K+U+M$, is shown in red. It is initially ($t<19$~Myr) dominated by $M$ in case A, by $M\sim U$ in cases B and E, and by $U$ in case C. In fully developed driven turbulence, however, the internal energy is subdominant in all cases, so that $U<M\lesssim K$. 

The particular way turbulence is initialized in the system does not really matter, as a test case (not shown) with forcing activated at $t=0$ resulted in the same stationary turbulence after a different initial transient.  The model thus keeps no memory of the initialization process and the resulting turbulence and phase structure do not depend on initial conditions, besides the generalized cooling function and global parameters listed in Table~\ref{input}.

\subsection{Magnetic field amplification and global turbulence characteristics\label{s-mag-glo}}
\begin{figure}[t]
	\begin{center}
        	   \includegraphics[scale=0.55]{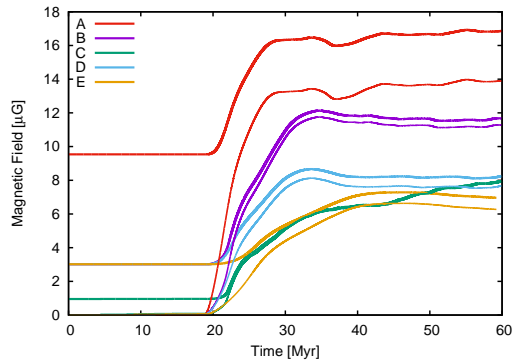}
        	   \caption{Time evolution of the magnetic field $B_{\rm rms}=\sqrt{\langle\bm B^2\rangle}$ (thick lines) and magnetic field fluctuations $b_{\rm rms}=\sqrt{\langle\bm b^2\rangle}$ (thin lines) for cases A, B, C, D, and E. In case C, $B_{\rm rms}$ continues to grow through $70$~Myr, reaching $8.5$~$\mu$G, and then by $90$~Myr the field relaxes back to $\sim8$~$\mu$G (not shown).}
        	   \label{mag}
	\end{center}
\end{figure}
The details of magnetic field growth are illustrated in Fig.~\ref{mag}, where thick and thin lines show the total field $B_{\rm rms}$ and fluctuations $b_{\rm rms}$ for each case. The magnetic field is not directly forced in these models, but it can receive energy via random compressions and stretching of the field lines through interaction with the velocity field. The nonhelical random force we use does not generate a mean field, but still leads to amplification of the small-scale magnetic energy. The mean field strength, $B_0$, 
together with the mean density, $n_0$, and the rms velocity, $u_{\rm rms,0}$, control the level of magnetic fluctuations in the developed turbulence. When $B_0$ is sufficiently strong, the saturation would imply equipartition of the kinetic and magnetic energy --- the most 
interesting and realistic situation to explore. In this case, one can identify the uniform field, $B_0$, 
with the regular field generated by {the Galactic} dynamo and spatially ordered on scales $>100$~pc 
\cite{arshakian...09,arshakian....11}. Since for the Milky Way-like galaxy 
the observed ratio of the small-scale and large-scale field strengths, $1.7\leq b_{\rm rms}/B_0\leq1.9$ \cite{ruzmaikin..88}, 
this case would fall in the range covered by cases A, D, and E (Table~\ref{mag1}). It is worth noting that in the Galactic dynamo models, the origin of the large-scale field is associated with the interplay of small-scale turbulence generated by the star formation processes {\cite{kim.15}}, differential rotation of the disk, {and the inverse cascade of small-scale helicity \cite{vishniac.01,shapovalov.11,vishniac.14}.} In our simple local ISM model, the small-scale field is instead reverse-engineered from the external mean field using external forcing. As soon as the right ratio $b_{\rm rms}/B_0$ is achieved, local magnetic field properties and cloud magnetization levels can be considered as realistic. 

\begin{table}
\caption{\label{mag1}Magnetic field properties and compressibility in developed turbulence.}
\vspace{-0.8cm}
\begin{center}
\footnotesize
\item[]\begin{tabular}{@{}crrcrccrrrrc}
\br
Case & $B_{\rm rms}$ & $b_{\rm rms}$ & $b_{\rm rms}/B_0$ & $\left<\beta_{\rm th}\right>$ & $\left<\beta_{\rm turb}\right>$ & $n_{\rm rms}/n_0$ & $u_{\rm rms}$ & $v_{\rm a, rms}$ & ${\cal M}_{\rm s,v}$  & ${\cal M}_{\rm s, m}$ & ${\cal M}_{\rm a}$\\
& $\mu$G & $\mu$G & & & &  &  km/s & km/s & & &          \\
\mr
A & 16.6 & 13.6 & 1.4 & 0.10 & 2.0& 3.7 & 15.5 & 15.4 & 4.9 & 11.6 & 1.0\\
B & 11.7 & 11.3 & 3.7 & 0.36 & 4.0& 3.5 & 15.4 & 12.6 & 5.4 & 13.1 & 1.4\\
C &  7.1 &  7.0 & 7.3 & 1.76 & 8.0&3.5 & 16.0 &  8.8 & 5.9 & 13.2 & 2.1\\
D &  8.2 &  7.6 & 2.5 & 0.77 & 2.7& 2.8 & 15.8 & 12.2 & 3.1 &  7.5 & 1.6\\
E &  7.2 &  6.5 & 2.2 & 0.37 & 2.2& 3.4 & 7.4  & 10.2 & 2.8 &  6.4 & 1.3\\  
\br
\end{tabular}
\end{center}
\end{table}
\normalsize
Table~\ref{mag1} further details magnetic field properties in quasi-stationary developed turbulence, providing average rms field values and plasma beta regimes for the whole domain. The mean $\beta_{\rm th}$ is mostly below unity, except in case C, indicating subdominant contribution of the internal energy compared to the magnetic one. The mean $\beta_{\rm turb}$ is above 2 in all cases, meaning that the dynamic pressure always dominates over magnetic pressure on average.
The table also presents diagnostics further characterizing the overall statistical properties of turbulence in numerical models. The relative rms density fluctuations $n_{\rm rms}/n_0$ in all cases fall within a narrow interval from 3 to 4, indicating very strong compressibility supported by TI and high volume- and mass-weighted sonic Mach numbers ${\cal M}_{\rm s,v}$  and ${\cal M}_{\rm s, m}$. The rms Alfv\'en speeds are comparable to the actually measured rms velocities in cases with strong magnetization, resulting in $K\sim M$ energy equipartition and ${\cal M}_{\rm a}\sim1$. Weakly magnetized cases demonstrate slightly super-Alfv\'enic rms velocities. Volume-weighted sonic Mach numbers (mostly sensitive to the warmer temperature regimes) fall roughly between 4 and 6, while mass-weighted ones (more biased to the cold phase) are largely supersonic, $6<{\cal M}_{\rm s}<13$.

\subsection{Thermodynamics of the turbulent two-phase ISM\label{s-phase}}
\begin{figure}[t]
           \includegraphics[scale=0.45]{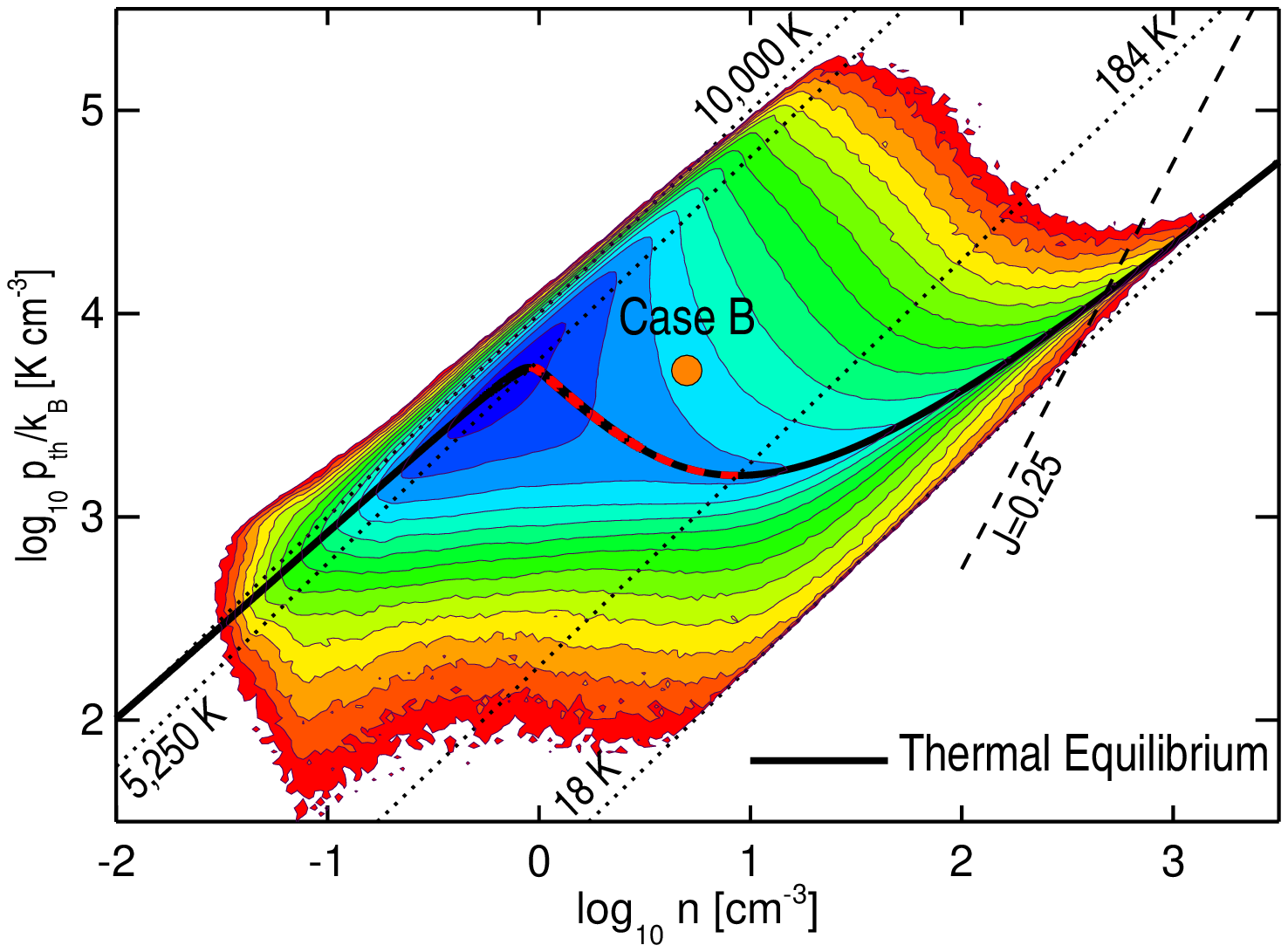} 
           \includegraphics[scale=0.45]{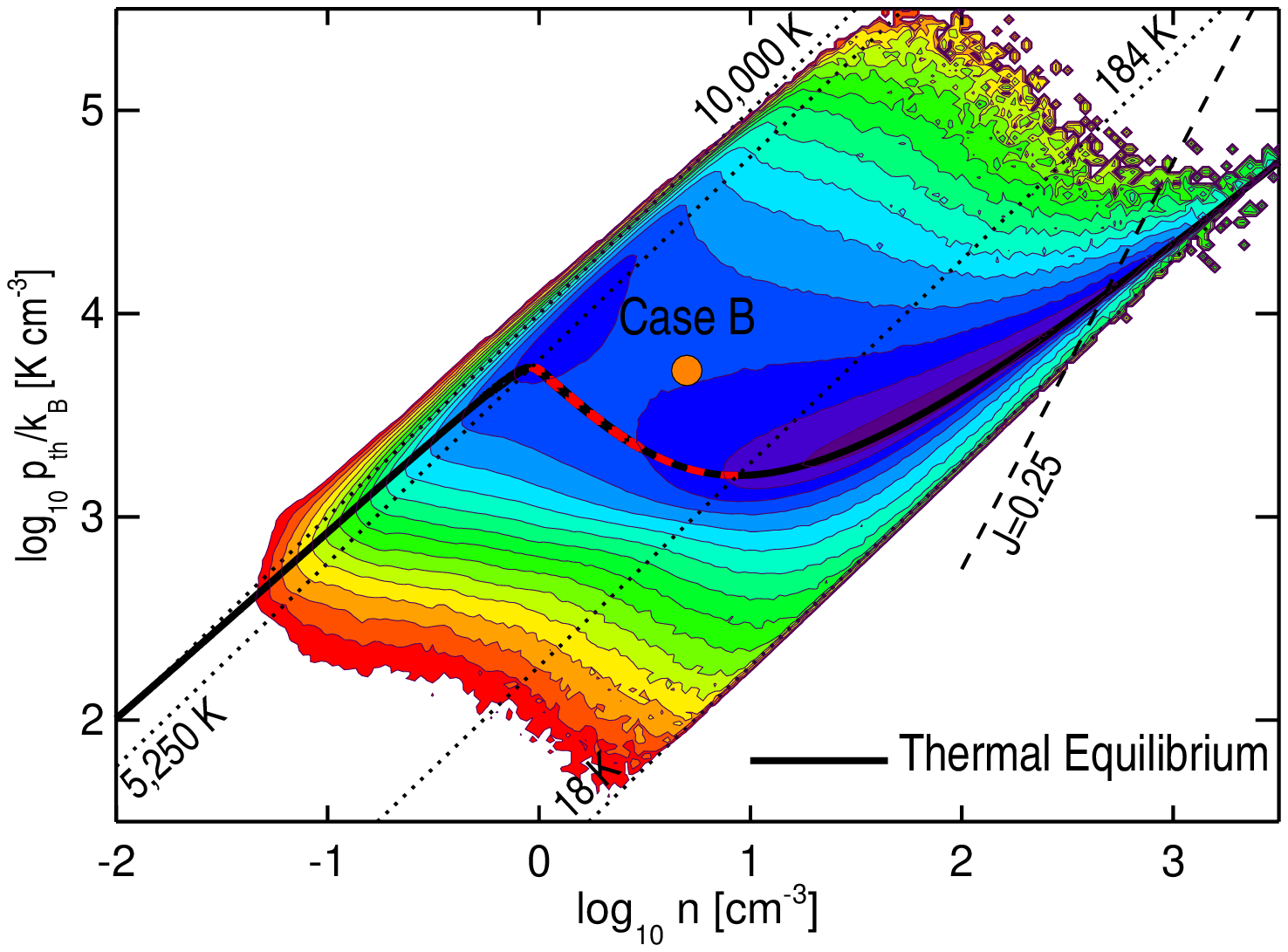} 
           \includegraphics[scale=0.45]{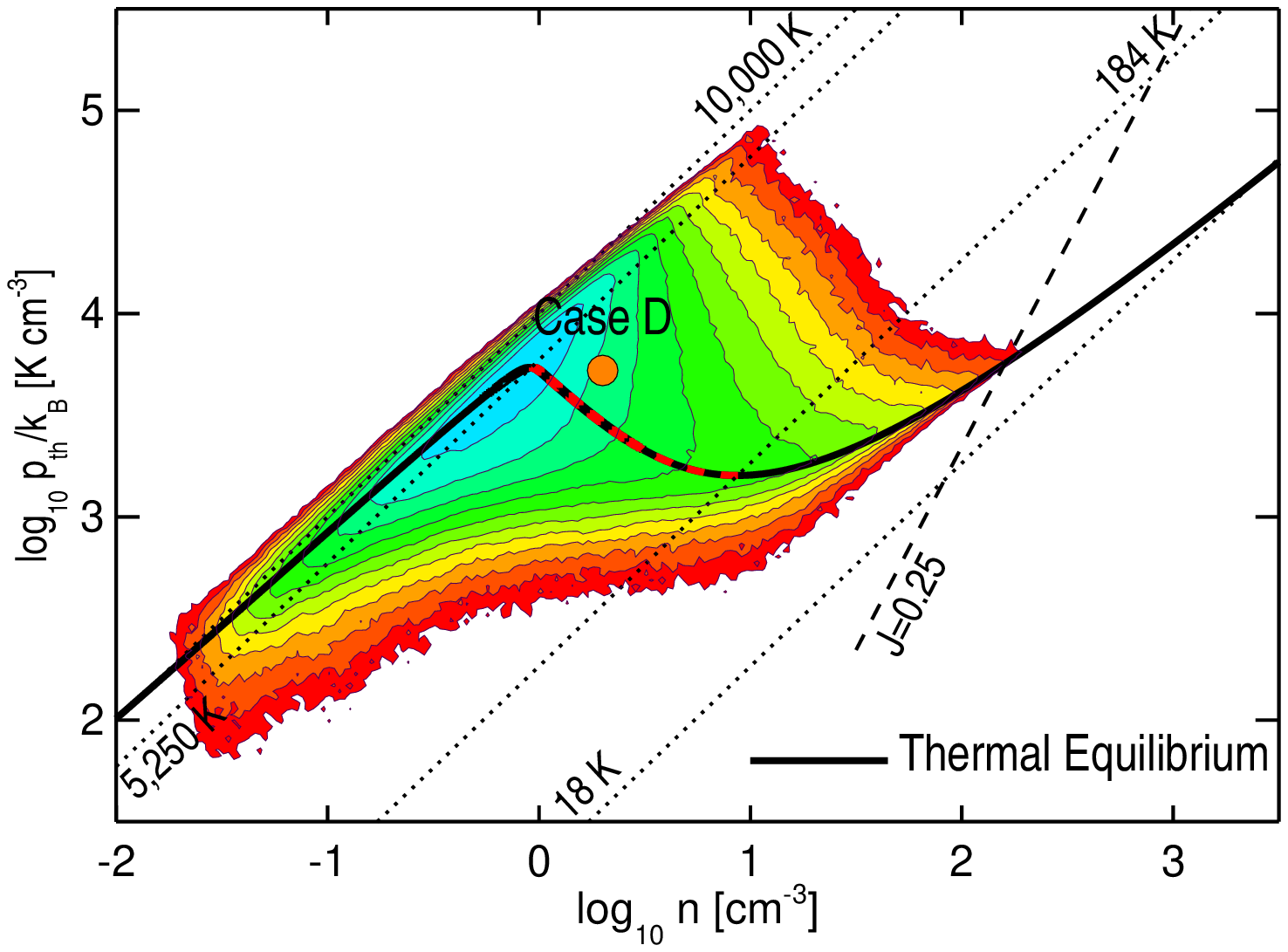} 
           \includegraphics[scale=0.45]{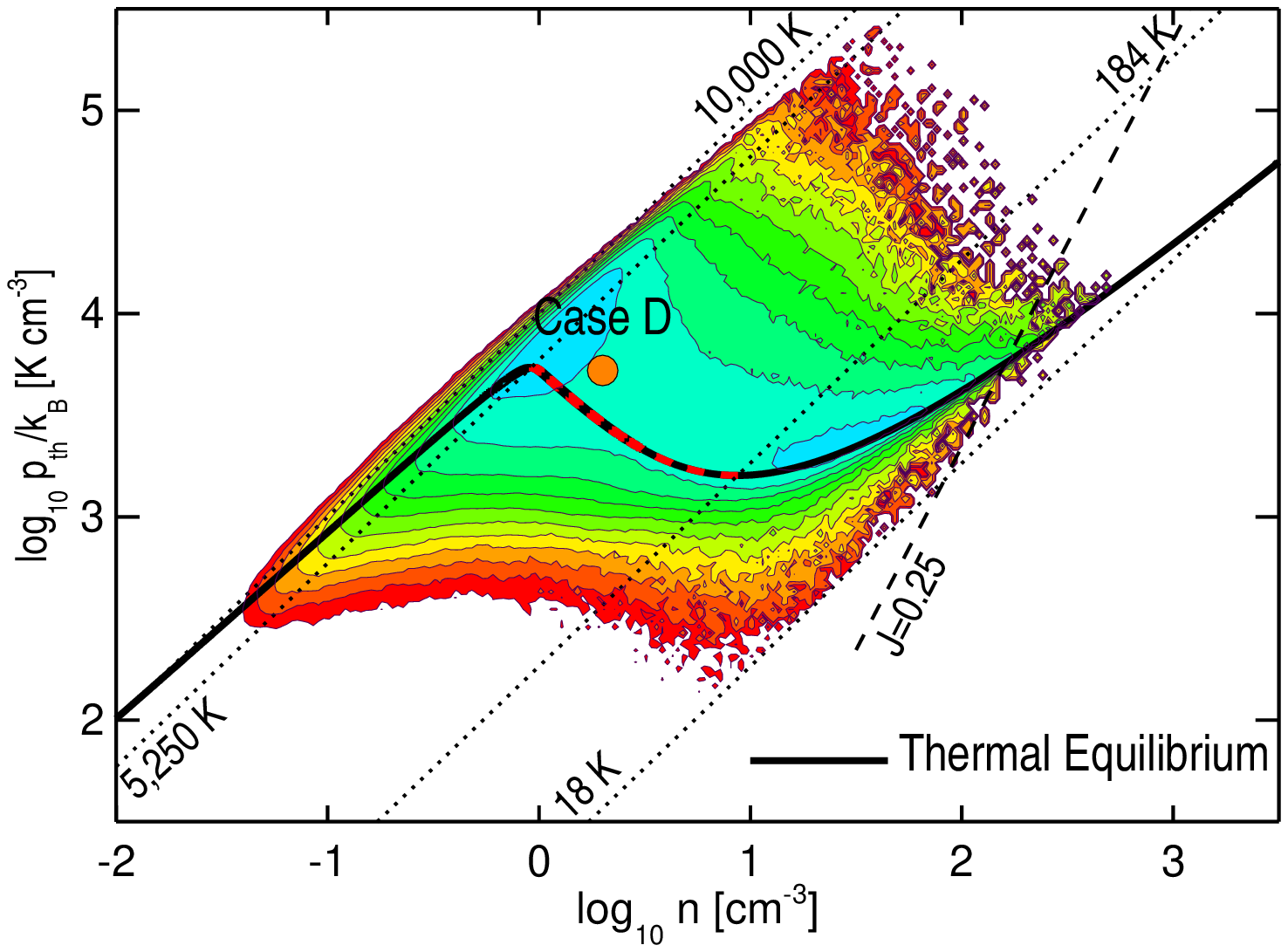} 
           \includegraphics[scale=0.45]{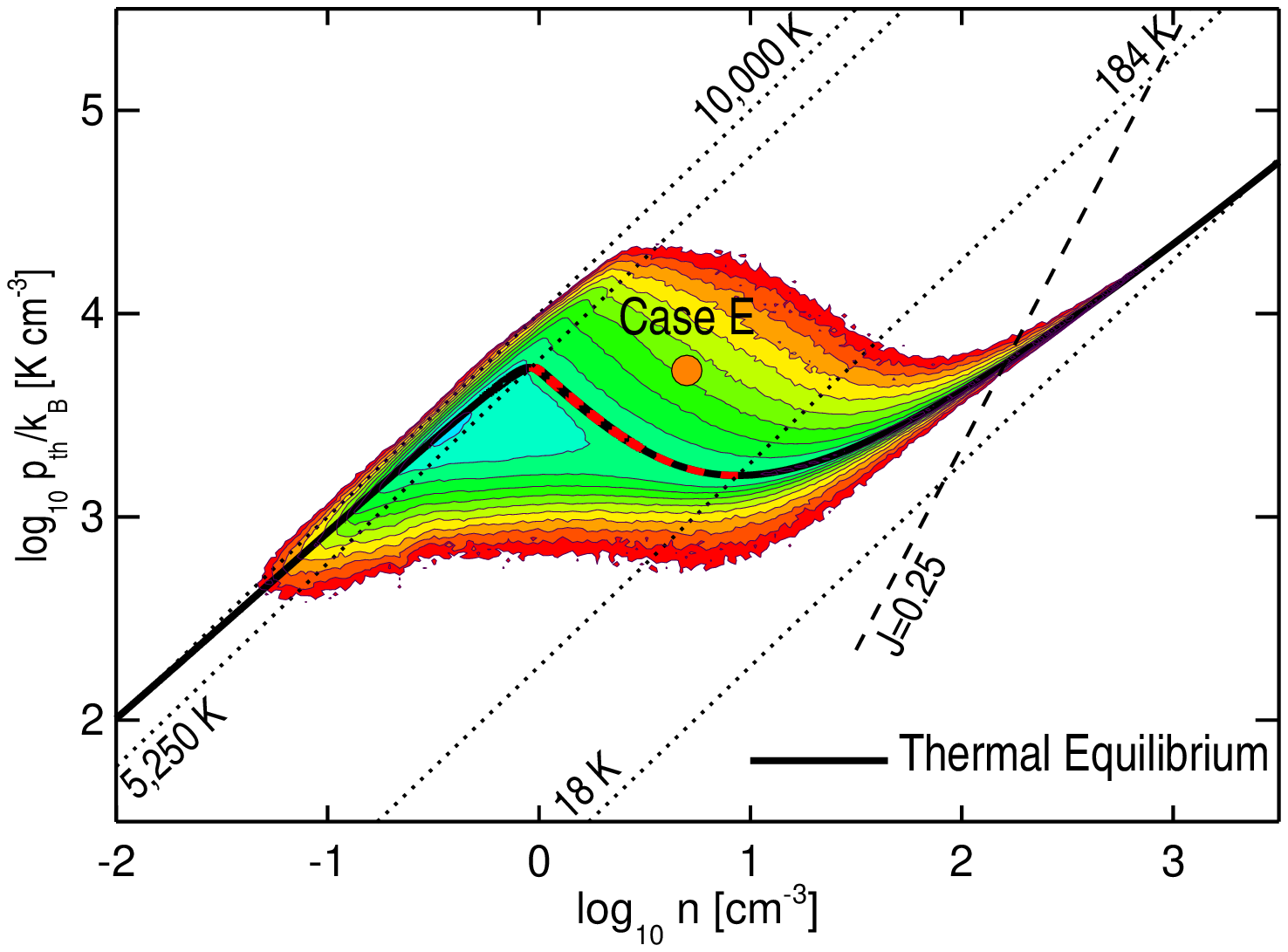} 
           \includegraphics[scale=0.45]{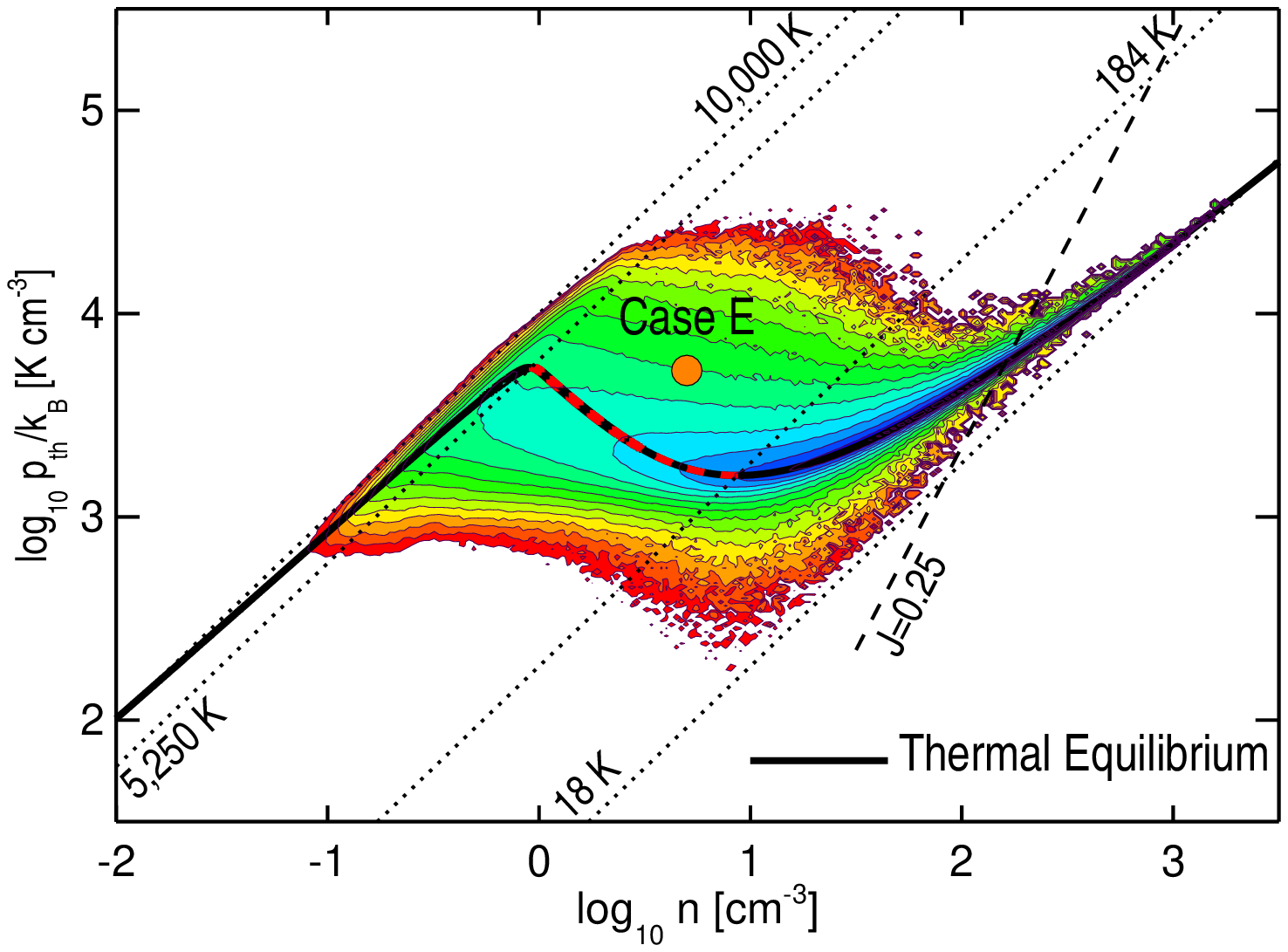} 
           \caption{Volume-weighted (left) and mass-weighted (right) phase diagrams for cases B, D, and E at $t=42.3$~Myr. Orange-filled circles indicate initial conditions in all cases selected at the maximum equilibrium thermal pressure. S-shaped black solid lines indicate stable thermal equilibrium; red dashed intervals show the unstable portion of the equilibrium. Thin dotted isotherms at $T=184$~K and 5250~K separate thermal phases. Material to the right from the straight dashed line $J=1/4$ would violate the Truelove resolution condition for Jeans-unstable self-gravitating gas \cite{truelove.....97} and thus can be considered potentially available for star formation. Note that the lower grid resolution in cases D and E limits the maximum density and shifts the Truelove-Jeans threshold line to the left. Spacing of isocontours here and in other figures below corresponds to a factor of 2 difference. {The blue-green-yellow-red sequence reflects a transition from densely to sparsely populated regimes.}}
           \label{f-phase}
\end{figure}
\begin{table}
\caption{\label{mach}Mass and volume fractions, Mach number regimes for different thermal phases.}
\begin{indented}
\footnotesize
\item[]\begin{tabular}{@{}cllrrcccccccc}
\br
Case&&$F_v$&&&$F_m$&&&${\cal M}_{\rm s}$&&&${\cal M}_{\rm a}$&\\
& W&U&C&W&U&C&W&U&C&W&U&C\\
\mr
A&25&68&7&5&44&51&1.8&4.0&13.5&0.6&0.9&2.5\\
B&23&70&7&5&44&51&1.7&4.2&15.2&1.2&1.6&4.3\\
C&16&76&8&4&44&52&1.6&4.6&15.3&2.9&3.9&8.3\\
D&35&63&2&14&63&23&1.6&3.1&12.5&1.2&1.6&4.7\\
E&13&77&10&1&35&64&0.8&2.1&6.6&0.6&1.0&3.0\\
\br
\end{tabular}\\
$F_v$ -- volume fraction (\%); $F_m$ -- mass fraction (\%); ${\cal M}_{\rm s}$ and  ${\cal M}_{\rm a}$ -- volume-weighted rms sonic and Alfv\'en Mach numbers, respectively; W, U, and C label warm, thermally unstable, and cold regimes, respectively.
\end{indented}
\label{tab2}
\end{table}
\normalsize
Scatter plots of the gas density versus thermal pressure (the so-called phase diagrams or 2D $(p,n)$-PDFs) illustrate imprints of the cooling and heating processes on the overall dynamics of multiphase turbulence. 
Figure~\ref{f-phase} shows volume-weighted (left) and mass-weighted phase diagrams for data snapshots taken at $t=42.3$~Myr in cases B, D, and E. As in Fig.~\ref{phase-e}, thick black lines show thermal equilibrium and dotted lines are isotherms at given $T$-values. The contours indicate constant levels of the volume (mass) fraction for different regimes of the thermal pressure, $p$, and density, $n$, separated by factors of 2. An orange circle corresponds to initial conditions $(p_0,n_0)$ for each case. 

{The volume-weighted diagrams show that the warm phase together with thermally unstable gas have large filling factors, while the cold phase occupies a tiny fraction of the domain volume. The mass-weighted diagrams instead highlight the distribution of the mass between the warm/unstable gas and the cold phase  (Table~\ref{tab2}).}

The phase diagrams for case B show that turbulence supports a wide range of thermal pressures and also that $p$ in the molecular gas ($n>100$~cm$^{-3}$) is higher than that in the diffuse ISM, even though self-gravity is ignored in the model. Most of the volume is filled with the warm gas, while the mass is roughly equally distributed between warm (including considerable abundance of thermally unstable component \cite{heiles01,gazol...01,kritsuk.02,gazol.16}) and cold components. Volume ($F_{\rm v}$) and mass ($F_{\rm m}$) fractions of thermal phases in simulations given in Table~\ref{mach} are generally consistent with locally observed values \cite{heiles.03}, {e.g. $\gtrsim48$\% by mass of the warm neutral medium (WNM, $>500$~K) lies at temperatures that belong to the thermally unstable regime; about 60\% of all H{\sc i} is WNM, which fills very roughly $50$\% of the volume.
}

{Phase diagrams for cases D and E illustrate the dependence of the resulting multi-phase structure on the mean density of the gas $n_0$ and on the rms velocity $u_{\rm rms}$ (Table~\ref{mag1}). With a $2.5\times$ smaller density, case D retains less than a half of the cold phase mass fraction, compared to case B. The maximum pressure and the maximum density of the cold phase also drop. Note that the rate of kinetic energy supply required to support $u_{\rm rms}$ at the same level as in case B, is lower because the mean density is lower. Hence case D illustrates the combined effects of  a reduced density and an effectively $\sim2\times$ weaker forcing in a low-density environment. Case E has the same $n_0$ as B, but $2\times$ smaller $u_{\rm rms}$, which corresponds to $\sim4\times$ smaller energy injection rate. Since the mean density remains moderately high, but the turbulence is weaker, the cold phase mass fraction is higher compared to B, but the range of thermal pressures  visibly shrinks.}

{The effects of magnetization can be studied using the A-B-C sequence of cases, which have the same $n_0$ and $u_{\rm rms}$, but different $B_0$. However, as far as the $p_{\rm th}-n$ phase diagrams are concerned, the differences are quite subtle. Hence we defer the discussion to \S~\ref{s-sfreg} with a primary focus on control of star formation.
}

\subsection{Mach number regimes of thermal phases\label{s-mach}}
\begin{figure}[t]
\centering
           \includegraphics[scale=0.45]{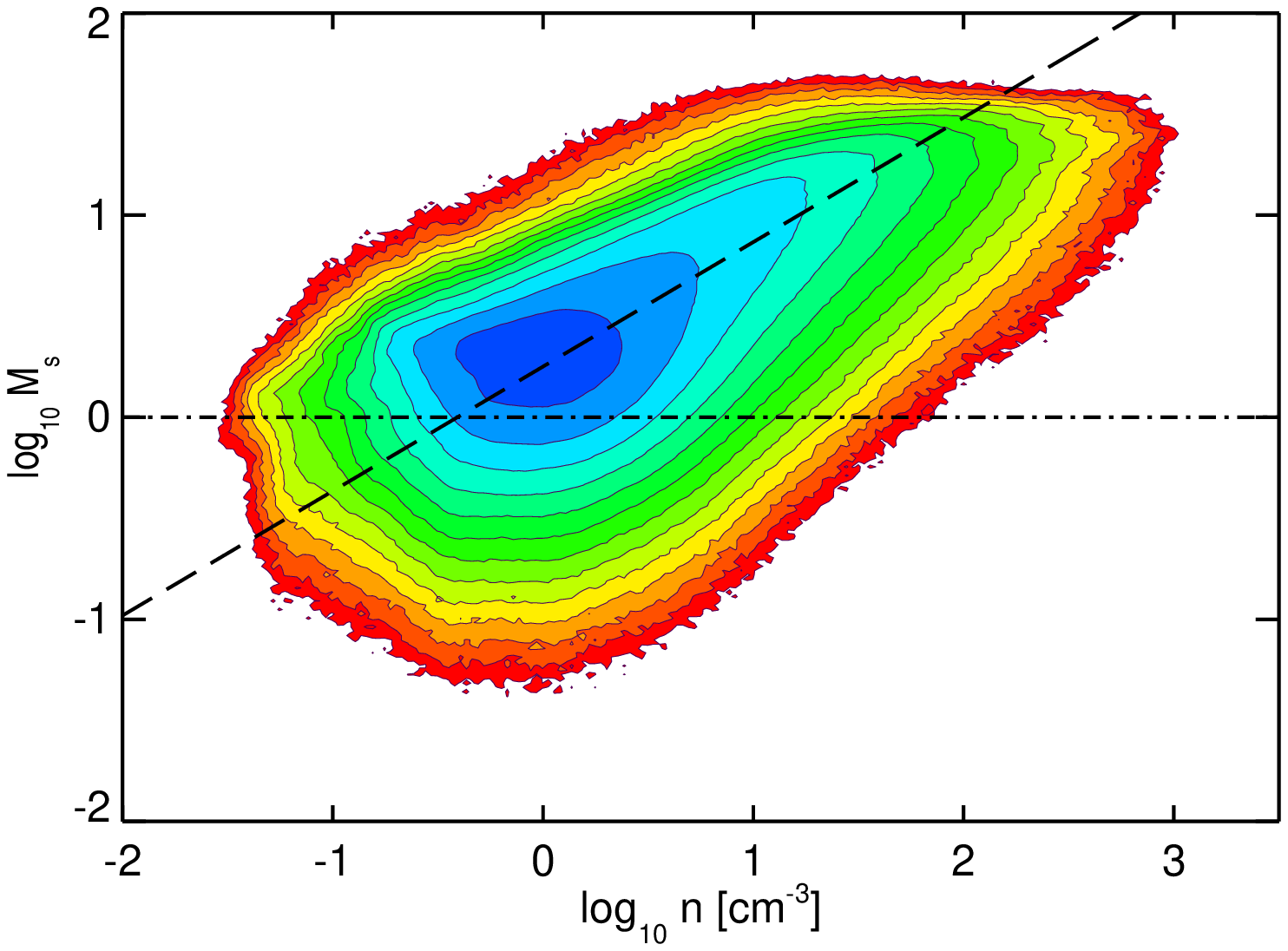}
           \includegraphics[scale=0.45]{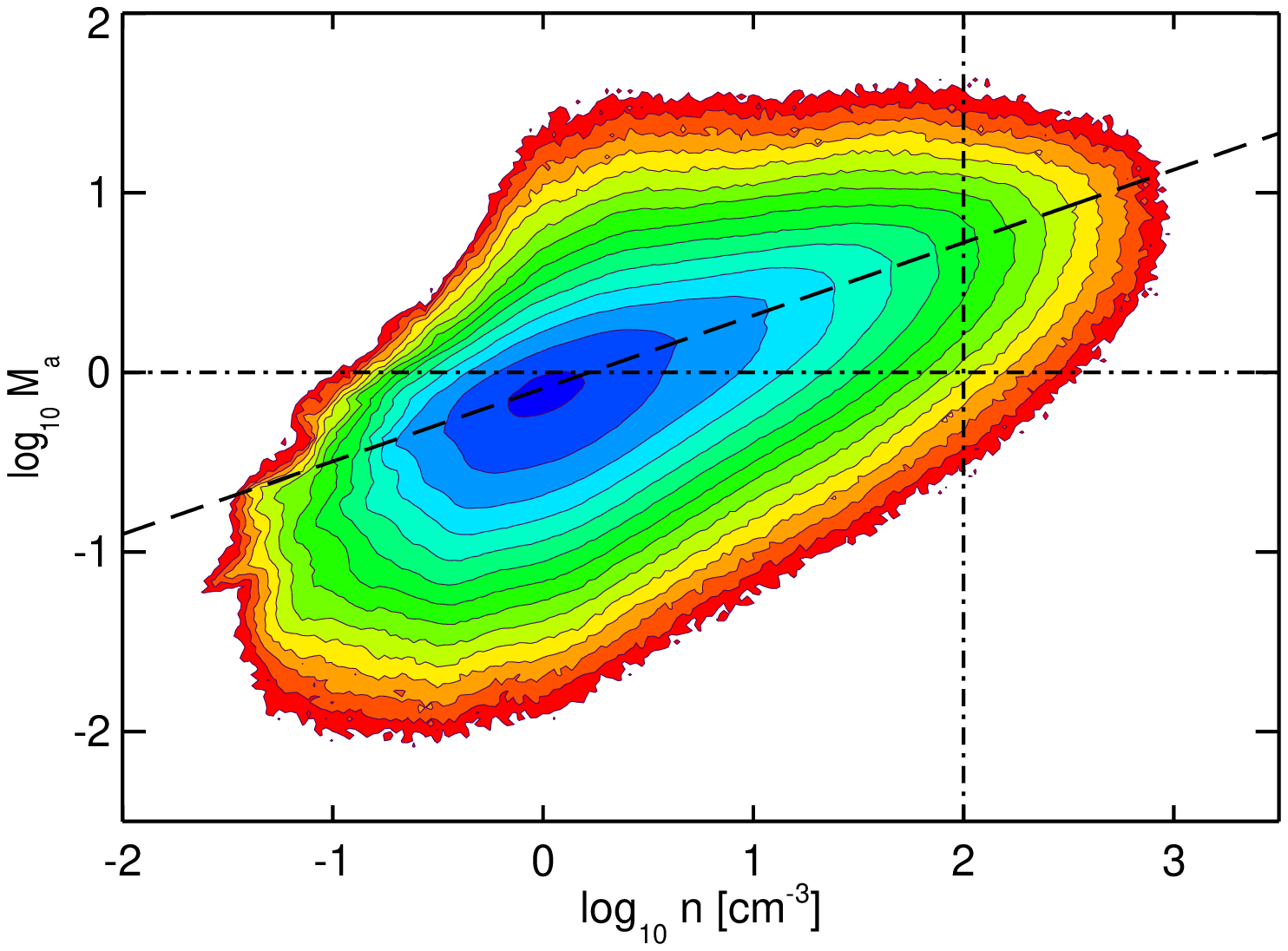}
           \caption{{Sonic (left) and Alfv\'en (right) Mach numbers versus} density for case B, $t=42.3$~Myr. Dashed lines show least-square fits ${\cal M}_{\rm s}\propto n^{0.62\pm0.01}$ ($\log(n)\in[0.1,1.2]$) and ${\cal M}_{\rm a}\propto n^{0.41\pm0.01}$ ($\log(n)\in[0,2]$).}
           \label{f-mach}
\end{figure}

Table~\ref{mach} also lists average Mach numbers for the cold, warm, and unstable phases. The sonic Mach number ${\cal M}_{\rm s}$ (a measure of compressibility) is {transonic} in the warm phase (weakly compressible), is {supersonic} in the unstable regime (mild-to-strong compressibility) and is {highly supersonic} in the cold phase (very strong compressibility). Hence, the velocity scaling is expected to be Kolmogorov-like in the warm gas and Burgers-like in the cold phase (\S~\ref{larson}). The ${\cal M}_{\rm s}-n$ correlation, shown in Fig.~\ref{f-mach} (left), follows the prediction ${\cal M}_{\rm s}\propto n^{1/2}$ for isobaric conditions at intermediate densities $n\in[1,10]$~cm$^{-3}$ \cite{kritsuk.04} reasonably well and flattens into ${\cal M}_{\rm s}\approx const$ in quasi-isothermal conditions at low and high density extremes (\S~\ref{s-dpdf}).

The Alfv\'en Mach number ${\cal M}_{\rm a}$ (a measure of relative strength of magnetic and kinetic effects) is mostly trans-Alfv\'enic in the warm phase, except in case C, indicating approximate energy equipartition $K\sim M$. The unstable phase is transient from trans-Alfv\'enic to super-Alfv\'enic ($K\gtrsim M$), and the cold phase is mildly (cases A and D) or strongly (case C) super-Alfv\'enic ($K\gg M$). This creates a curious physical situation, where nonlinear relaxation of the system results in a statistically stationary state with moderately magnetized turbulence across most of the volume, enforcing overall kinetic-to-magnetic energy equipartition. At the same time, the cold, high density material potentially available for star formation remains in supersonic, super-Alfv\'enic regime with kinetic energy dominating magnetic by a large margin. Fig.~\ref{f-mach} illustrates a clear positive correlation between the Alfv\'en Mach number and density for case B with the bulk of the gas mass at $n>10^2$~cm$^{-3}$ having local ${\cal M}_{\rm a}>1$. {A similar trend is seen in isothermal MHD simulations \cite{burkhart...09} in supersonic and weakly super-Alfv\'enic regimes, which can be explained by the dependence ${\cal M}_{\rm a}(B,\rho)$ and the relationship between $B$ and $\rho$. Our multiphase cases combine a transonic, sub-Alfv\'enic behavior at low densities with a supersonic and super-Alfv\'enic one at high densities.}

\subsection{Control of star formation\label{s-sfreg}}
Our simulations do not include gravitational effects and all cases represent globally Jeans-stable settings. We can, however, extrapolate our results to cases with self-gravity by picking a sample of evolved snapshots with stationary statistics and asking a simple question: What happens if we turn on self-gravity? To address this, we have to return to the discussion of phase diagrams in Fig.~\ref{f-phase}, where each plot contains a dashed line in the upper-right corner, separating locally Jeans-unstable material.

All gas to the right of the dashed line {($J\equiv\Delta x/\lambda_{\rm J}=0.25$, where $\lambda_{\rm J}=\sqrt{\pi c_{\rm s}^2/G\rho}$ is the Jeans length)} would fail the Truelove resolution constraint \cite{truelove.....97}, {which requires the Jeans length to be resolved with at least four grid zones to avoid artificial fragmentation. If we were to follow local gravitational collapses with adaptive mesh refinement in these simulations, the grid zones failing Truelove's condition would be flagged. The Truelove condition is also used to determine whether sink particles should be introduced in models which rely on this technology.} In practice, the condition implies that (with some efficiency factor $\lesssim1$) this material would be converted into stars within one free-fall time $t_{\rm ff}=\sqrt{3\pi/32G\rho}\approx\sqrt{(10^3\;{\rm cm}^{-3})/n }$~Myr, where $G$ is the gravitational constant \cite{kritsuk..11a}. 

This Jeans-unstable gas occupies a tiny fraction (0.01\%) of the computational volume and {its mass $\Delta{\mathfrak M}$  comprises 1.7\%} of the mass in case B at $t=42.3$~Myr. The cold phase in this case constitutes on average 7\% of the volume and 51\% of the mass (Table~\ref{tab2}). In this case, the star formation rate per free-fall time \cite{krumholz.05,krumholz14} $\epsilon_{\rm ff}\equiv t_{\rm ff}/t_{\rm dep}{=\Delta{\mathfrak M}/{\mathfrak M}}\lesssim 0.017$, where the gas depletion time $t_{\rm dep}\equiv{\mathfrak  M}/\dot{\mathfrak M}$, is consistent with the recent {\em Spitzer}  legacy survey of low-mass star-forming clouds near the Sun, which gives $\epsilon_{\rm ff}\sim0.01-0.1$ for clouds with mean densities $n_{\rm H_2}\sim10^3$~cm$^{-3}$ \cite{evans+09}. 

In case D with $2.5\times$ lower mean density compared to B (and proportionally lower kinetic energy injection at large scales), the cold phase constitutes only 2\% by volume and 23\% by mass. In this case, star formation is very inefficient. It is active in just 0.002\% of the domain volume and involves 0.22\% of the mass per $t_{\rm ff}$. In case E with a $2\times$ lower rms velocity,  the cold phase constitutes  10\% of the volume, 64\% of the mass, and the upper limit for the star formation rate is set at 7.1\% of the mass per $t_{\rm ff}$ in 0.14\% of the volume. In this case, relatively low level of turbulence results in a high star formation rate. {Note that cases D and E have a $2\times$ lower grid resolution compared to B, which could to some degree affect the phase fractions. The estimates of $\epsilon_{\rm ff}$ include the resolution dependence, which enters the Truelove constraint \cite{truelove.....97}.}

{We have traced the dependence of $\epsilon_{\rm ff}$ on the mean density and rms velocity in the box, but more subtle magnetic effects
remained unexplored. In order to see how magnetization changes the expected star formation efficiency, we calculated the time-averaged star formation rates per free-fall time for cases A, B, and C, which have the same $n_0$ and $u_{\rm rms}$, but different $B_0$. The results: $\langle\epsilon_{\rm ff}\rangle=0.021\pm0.005$ (A), $0.013\pm0.002$ (B), and $0.011\pm0.002$ (C) indicate an overall $\sim2\times$ drop in $\langle\epsilon_{\rm ff}\rangle$ from A to C. These estimates, however, are based on the  purely thermal Truelove condition \cite{truelove.....97}, which does not fully take into account the magnetic effects even though the pressure and density fields that enter the Truelove condition are affected. Since strong magnetization is able to suppress fragmentation, higher densities would be required to trigger the collapse, relaxing the critical $J$ value $J_{\rm M}=J/\sqrt{1+0.74/\beta_{\rm th}}$ \cite{myers....13}. If we apply this $\beta_{\rm th}$-dependent correction, the average rates drop by a factor of $\sim10$ in all three cases, but the statistics become poor because the maximum density is limited by the grid resolution. Thus, higher resolution modeling is required to measure how magnetization changes the expected star formation efficiency.}

{We note that the star formation rates discussed here differ from those in the context of idealized isothermal simulations of MC turbulence, where $\epsilon_{\rm ff}$ is determined primarily by the virial parameter of an individual cloud \cite{padoan..12}. These rates may vary in a wide range of values, depending on cloud's dynamical state \cite{padoan...17}. Since our multiphase models contain large statistical ensembles of realistic clouds, $\epsilon_{\rm ff}$ measures ensemble-average rates. Naturally, the scatter between different snapshots from a given run under statistically stationary conditions is reasonably small, with standard deviations of order $(15-20)$\%. A better understanding of the functional form of $\epsilon_{\rm ff}(L,n_0, u_{\rm rms}, B_0)$  will help to design robust subgrid-scale prescriptions for star formation and feedback in galaxy formation simulations \cite{semenov..16,semenov..17}.}

Thus the two-phase turbulence has a potential to regulate the supply chain of star formation, permitting only so much cold gas, dripping through the upper-right `nose' of the phase diagram per $t_{\rm ff}$, to undergo gravitational phase transition. Star formation rates per free-fall time $\epsilon_{\rm ff}\sim(0.001-0.1)$  we measured in cases B, E, and D are controlled by the problem parameters $(L, n_0, u_{\rm rms,0}, B_0,\Gamma)$, but this simple model does not explicitly include the feedback processes, {nor does it include the mass source associated with the cosmological gas infall onto the galaxy}. Natural further steps in development of the model would include closing the energy feedback loop and allowing the mass flux through the system (with external gas accretion on the disk as a source and star formation as a sink). This would be an advanced variant of the galactic {\em bathtub} model \cite{dekel.14}. Remarkably, $\varepsilon_{\rm ff}\sim0.01$ that we get in case B is a reasonable estimate for the star formation rate on scales from entire galaxies to single clouds \cite{krumholz14}. This suggests that closing the feedback loop might help to nail down a stable, statistically steady regime for the recycling of gas into stars by disk-like galaxies between major merger events.

\begin{figure}[t]
           \includegraphics[scale=0.6]{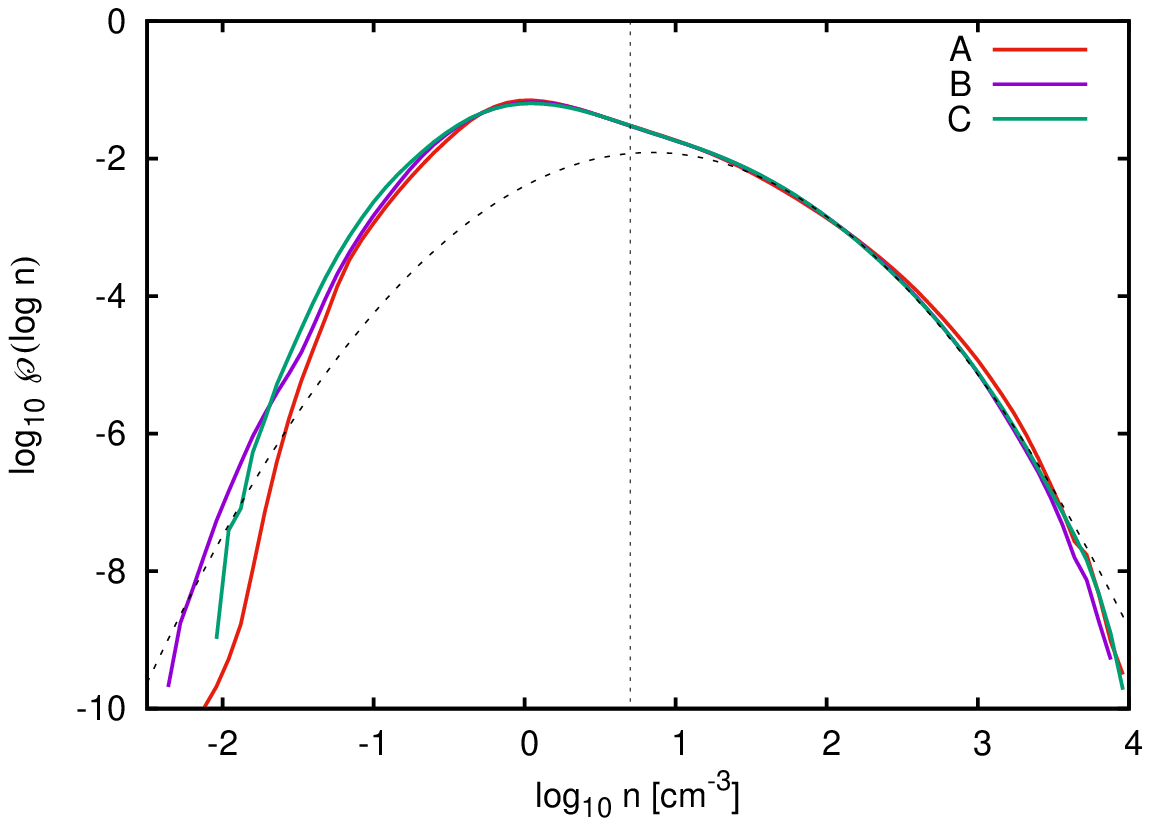}
           \includegraphics[scale=0.6]{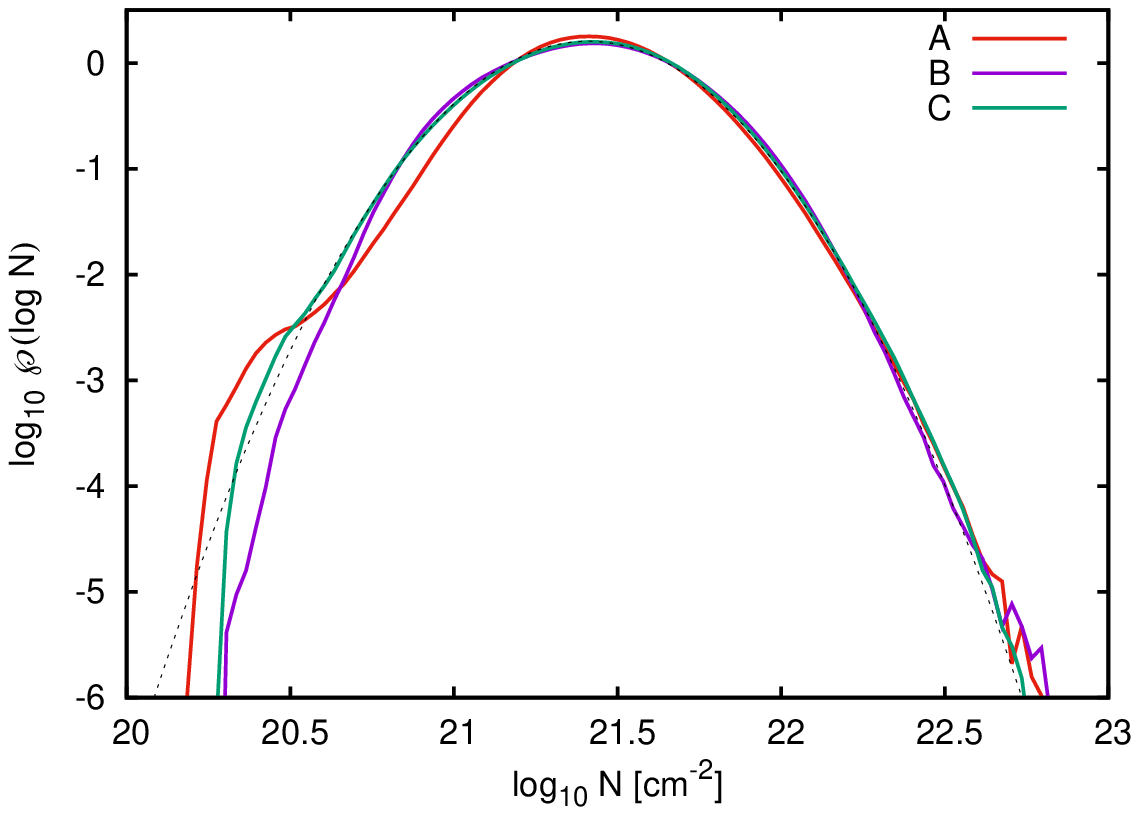}
           \caption{Time-averaged PDFs of density ($\wp(\log n)$, left) and column density ($\wp(\log N)$, right) for cases A, B, and C. Vertical dotted line shows the the mean density $n_0=5$~cm$^{-3}$. Dashed line in both cases shows a lognormal approximation to the high density tail. PDFs for cases B and C with moderate and weak magnetization follow the lognormal shape at high densities ($n>10$~cm$^{-3}$) and at high-to-moderately low column densities ($N>10^{20.7}$~cm$^{-2}$).}
\label{f-dpdf}
\end{figure}

\subsection{Density and column density PDFs\label{s-dpdf}}
The gas density PDF is one of the basic statistics of compressible turbulence that bears unique signatures of gas thermodynamics  \cite{passot.98}, gravitational instability \cite{kritsuk..11a}, and the presence of shocks. For instance, in isothermal homogeneous supersonic turbulence, the PDF is lognormal \cite{blaisdell..93,vsemadeni94,kritsuk...07}. In two-phase ISM turbulence with moderate velocity dispersions, however, the PDF is expected to have two peaks  corresponding to thermally stable phases and a power-law excess at high densities because the cold phase thermal equilibrium can be approximated by an effective soft polytropic relation $p_{\rm eq}\propto\rho_{\rm eq}^{\gamma_{\rm eff}}$ with $\gamma_{\rm eff}\approx0.7<1$ \cite{kritsuk.02,seifried..11,gazol.13}. Cases A through D assume a moderately strong turbulence driving at an injection scale $\lambda_f\sim100$~pc, which effectively merges the usual double-peak distribution obtained at  $u_{\rm rms}\sim5-10$~{km} s$^{-1}$ into an asymmetric single-peak PDF. Density PDFs for cases A, B, and C shown in Fig.~\ref{f-dpdf} (left) display a minor tendency toward weaker rarefactions at higher magnetization levels moderated by stronger magnetic tension. Otherwise, the three PDFs are very similar and have characteristic lognormal shapes at the high end with no signature of any power-law extension. The lack of a thermal power-law tail is due to: (i) high levels of turbulence, supporting a wide range of thermal pressures and populating the unstable phase, which smear the soft effective polytropic law (Fig.~\ref{f-phase}) and (ii) flattening of the Mach number--density correlation ${\cal M}_{\rm s}\propto\rho^{1/2}$, consistent with quasi-isobaric conditions \cite{kritsuk.04}, above $n\sim30$~cm$^{-3}$ (Fig.~\ref{f-mach}). It is ultimately the lack of ${\cal M}_{\rm s}-n$ correlation that supports the lognormal shape of the PDF high end populated primarily with cold material under supersonic conditions. Note that the moderately strong forcing responsible for the lognormal shape here is not merely a random choice of a free model parameter, but is required to match the local observed distribution of thermal pressure (\S~\ref{s-pth}) and local linewidth--size relationship (\S~\ref{larson}).

Figure~\ref{f-dpdf} (right) shows PDFs of column densities (or N-PDFs) obtained by averaging projected density distributions in three coordinate directions. Interestingly, the convolutions involved in the projection operation completely remove the signature of two-phase density structure and return nearly perfect lognormal N-PDFs, particularly in case B, however with noticeable statistical variations at low column densities in cases A and C. Such lognormal N-PDFs should be observable with temperature-blind tracers 
in clouds with no signs of active star formation, where the effects of self-gravity are too weak to build a power-law tail at high column densities {(e.g. in the Draco cloud \cite{schneider+17}). Note that the sonic Mach numbers for cases A, B, and C are generally higher than 5 (Table~\ref{mag1}). At ${\cal M}_{\rm s,v}\lesssim2$, the lognormal property of the low tail of N-PDFs is lost \cite{gazol.13}.}

\begin{figure}[t]
\begin{minipage}{.46\textwidth}
\vskip-0.47cm
           \includegraphics[scale=0.364]{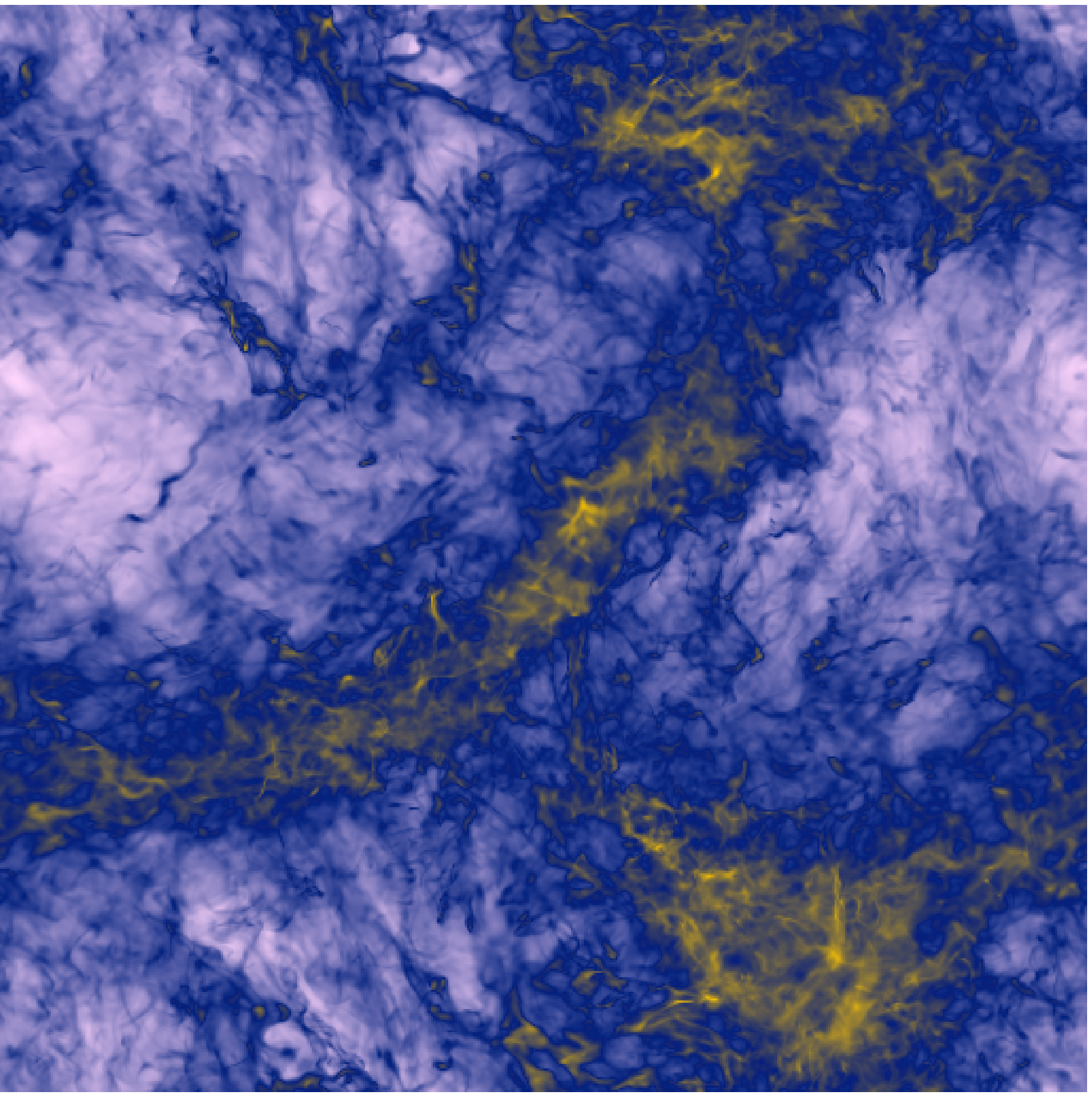}
\end{minipage}%
\begin{minipage}{.46\textwidth}
           \includegraphics[scale=0.448]{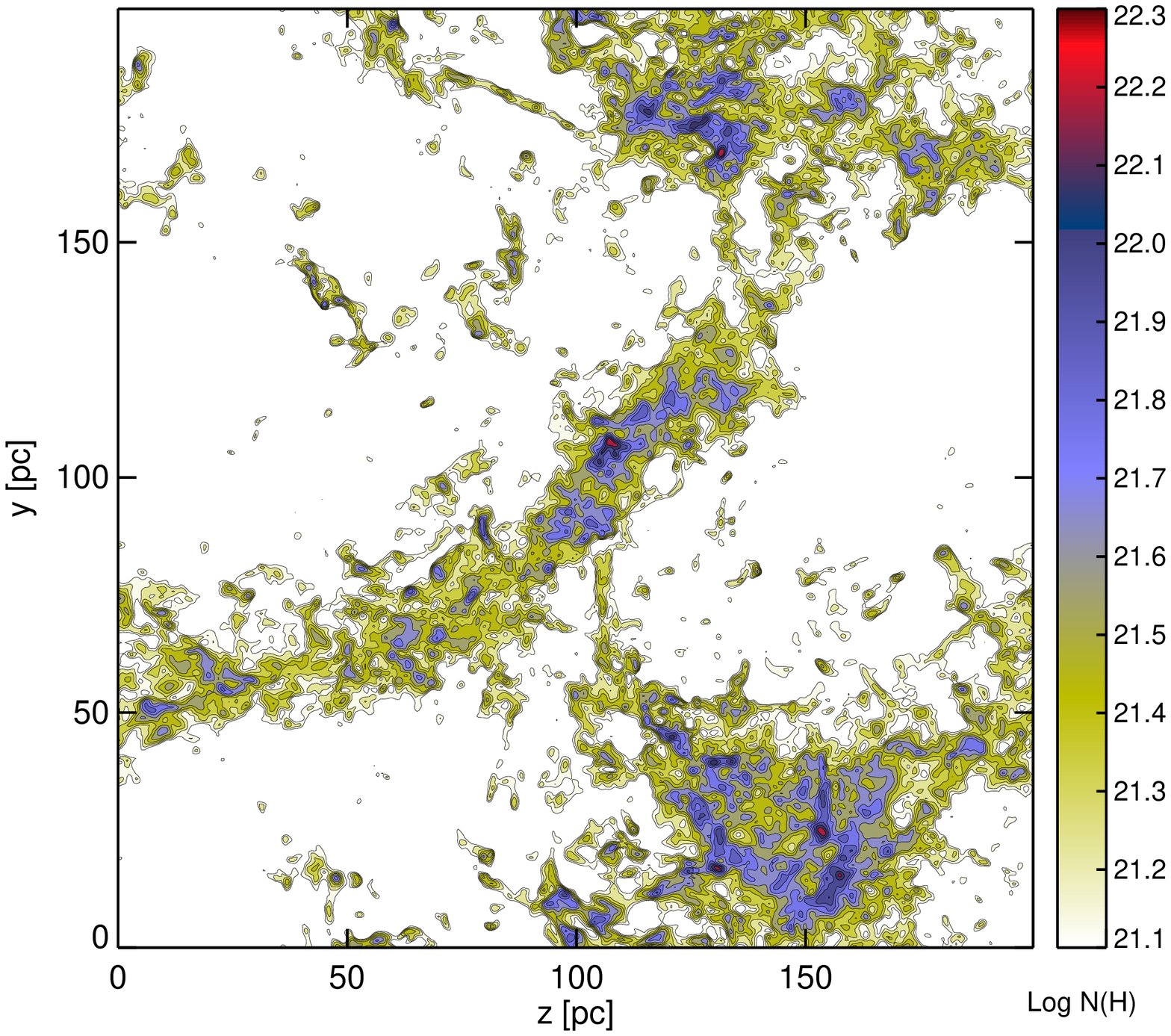}
\end{minipage}%
           \caption{Projected density image (left, logarithmic color sequence white-blue-brown shows low-to-high total projected densities) and a synthetic contour map of the cold phase ($T<184$~K) column density (right) for a case B snapshot taken at $t=46$~Myr, representing a simulated equivalent of a molecular cloud complex. To create the map, the original grid resolution of the snapshot, $\Delta=0.39$~pc, has been coarsened by a factor of $\sim3$ to match the $0.5^{\circ}$ angular resolution of the
historic $^{12}$CO map of molecular clouds in Perseus, Taurus, and Auriga \cite{ungerechts.87}. The morphology of contours
closely resembles the observations. Color bar indicates the logarithm of column density of the cold material in cm$^{-2}$. {\em A movie is available to illustrate the evolution of projected density in case B.}} 
\label{f-map}
\end{figure}
\begin{figure}[t]
           \includegraphics[scale=0.6]{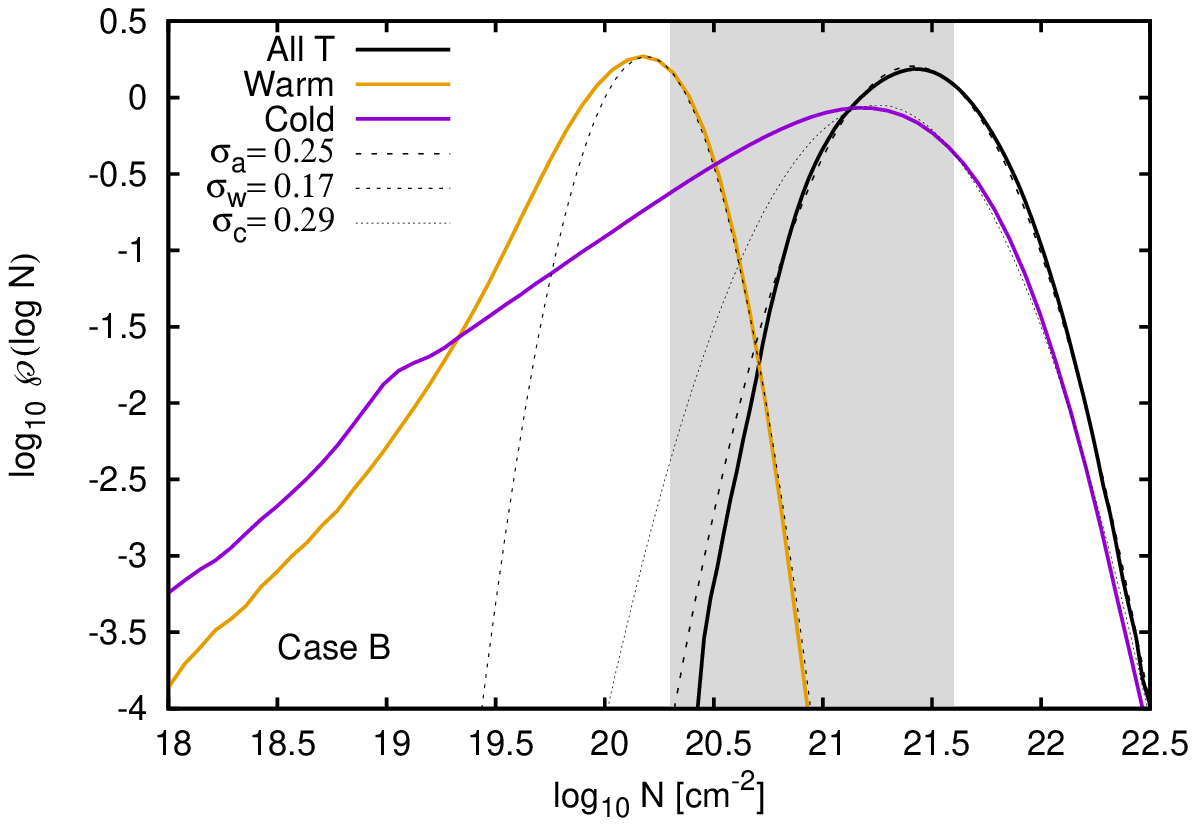}
           \includegraphics[scale=0.6]{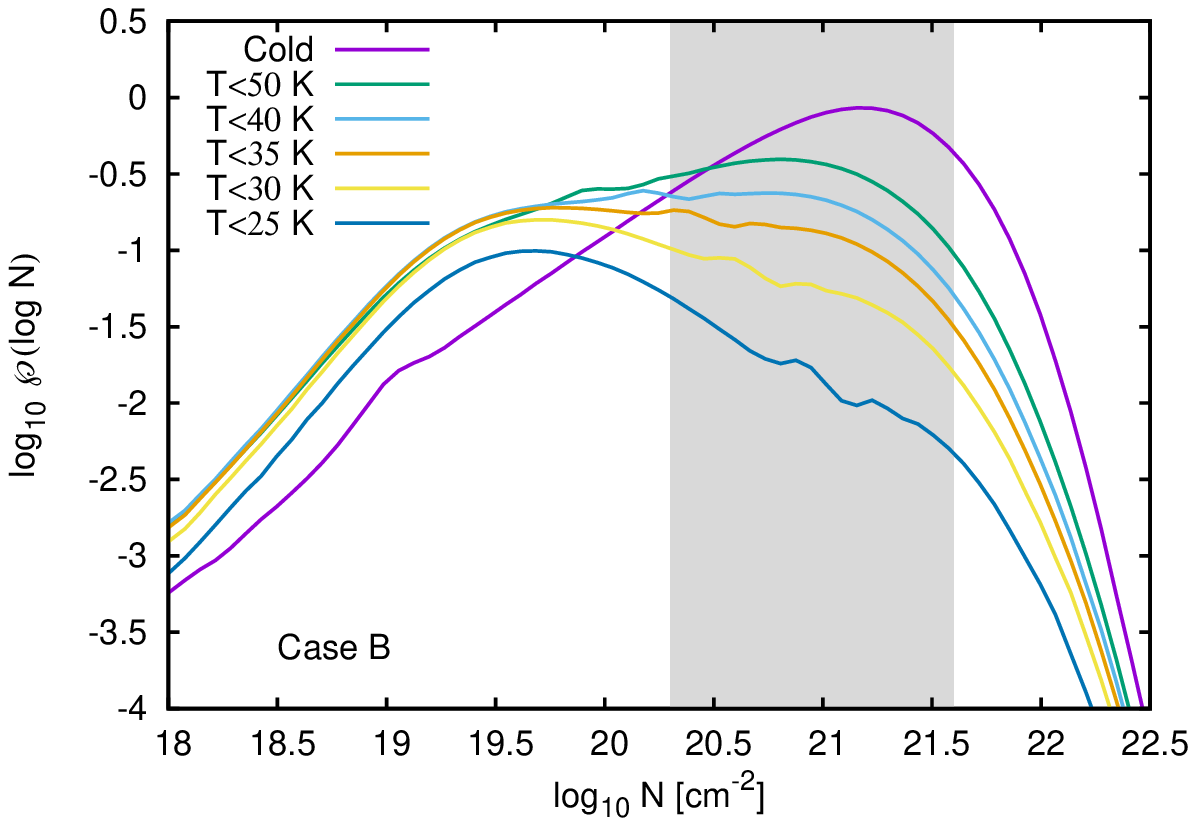}
           \caption{Time-averaged PDFs of column density for case B. {\em Left:} separate PDFs conditioned on the gas temperature are shown for the cold ($T<184$~K) and warm ($T>5,250$~K) phases. Black solid line shows PDF for all temperatures (same as line B in previous figure). 
Dashed lines show best-fit lognormal representations for the high column density tails; $\sigma_{\rm a,w,c}$ are standard deviations for the lognormals. {\em Right:} PDFs for the cold gas conditioned on the temperature, e.g. $\wp(\log N\, | \, T<50 {\rm K})$.
Grey-shaded areas indicate column densities $N\in[2\times10^{20},4\times10^{21}]$~cm$^{-2}$ favorable for `CO-dark H$_2$' formation. 
}
\label{f-npdf}
\end{figure}

Since temperature plays the role of an important third dimension in column-density mapping of star forming clouds \cite{marsh..15}, we proceed by computing N-PDFs conditioned on the temperature $\wp(\log N|\Delta T)$, such that $\int\wp(\log N|\Delta T)d\log N=1$ and $\Delta T$ defines the allowed temperature interval. As an example, Fig.~\ref{f-map} shows a projected density image (left) and a synthetic map of the column density of the cold phase gas (right), showing a plethora of multi-scale filamentary structures created by turbulence with the aid of TI. Even though our models do not include  any particular tracer's chemistry, simple conditioning on the temperature can yield new predictions for temperature-dependent N-PDF shapes. Figure~\ref{f-npdf}  (left) shows case-B N-PDFs for the cold (purple, $4<T[{\rm K}]<184$) and warm (red, $T>5250$~K) phases, as well as the distribution including all temperature regimes (black solid line, $\forall\; T$). 
Three best-fit lognormal distributions are shown with black dashed and dotted lines together with their corresponding standard deviations: $\sigma_{\rm c}=0.29$, $\sigma_{\rm w}=0.17$, and $\sigma_{\rm a}=0.25$. Remarkably, both warm and cold phases bear extended low-end tails (that would be difficult measuring observationally), while the full N-PDF is purely lognormal. None of the lognormal distribution widths can be accurately predicted using the variance--Mach number relation $\sigma_N^2=0.11\ln(1+b^2{\cal M}_{\rm s}^2$) based on simulations of isothermal turbulence \cite{burkhart.12}, if we assume ${\cal M}_{\rm s}$ values from Table~\ref{mach} and $b\approx1/3$ due to a divergence-free nature of the forcing \cite{federrath..08}. Similar discrepancy of the widths of fitted lognormal distributions with isothermal predictions was seen earlier in multiphase ISM simulations without a magnetic field \cite{gazol.13}, in GALFA-H{\sc i} data for the Perseus molecular cloud \cite{burkhart...15}, and in data for seven molecular clouds from the Leiden/Argentine/Bonn Galactic H{\sc i} survey \cite{imara.16}. In all cases, the widths of observational H{\sc i} N-PDFs are similar to our model predictions. 

%

Self-shielding of H$_2$ becomes effective around $N_{\rm H}\sim 2\times10^{20}$~cm$^{-2}$, while column densities in excess of $N_{\rm H}\sim 4\times10^{21}$~cm$^{-2}$ are required to form CO \cite{hollenbach.79,lee....96}. The range of column densities in between is populated by the `CO-dark H$_2$' gas that may account on average for $(30-40)$\% of the Milky Way molecular mass \cite{pineda...13,smith....14}. The shaded areas in both panels of Fig.~\ref{f-npdf} show the column density interval, where CO-dark H$_2$ is expected to reside. Fig.~\ref{f-npdf} (right) details the temperature dependence of N-PDFs for the cold gas in case B. In all cases, except perhaps for the $T<25$~K case that hits our resolution limits, the high tails of these distributions can be reasonably well represented as lognormal. {Similar plots for cases A and C (not shown) display only subtle differences in the N-PDFs with case B, mostly limited to the low column density tails. This is consistent with Fig.~\ref{f-dpdf} and indicates that corresponding observational diagnostics are not very sensitive to the ISM magnetization levels.}

\subsection{Thermal pressure PDFs\label{s-pth}}
\begin{figure}[t]
           \includegraphics[scale=0.6]{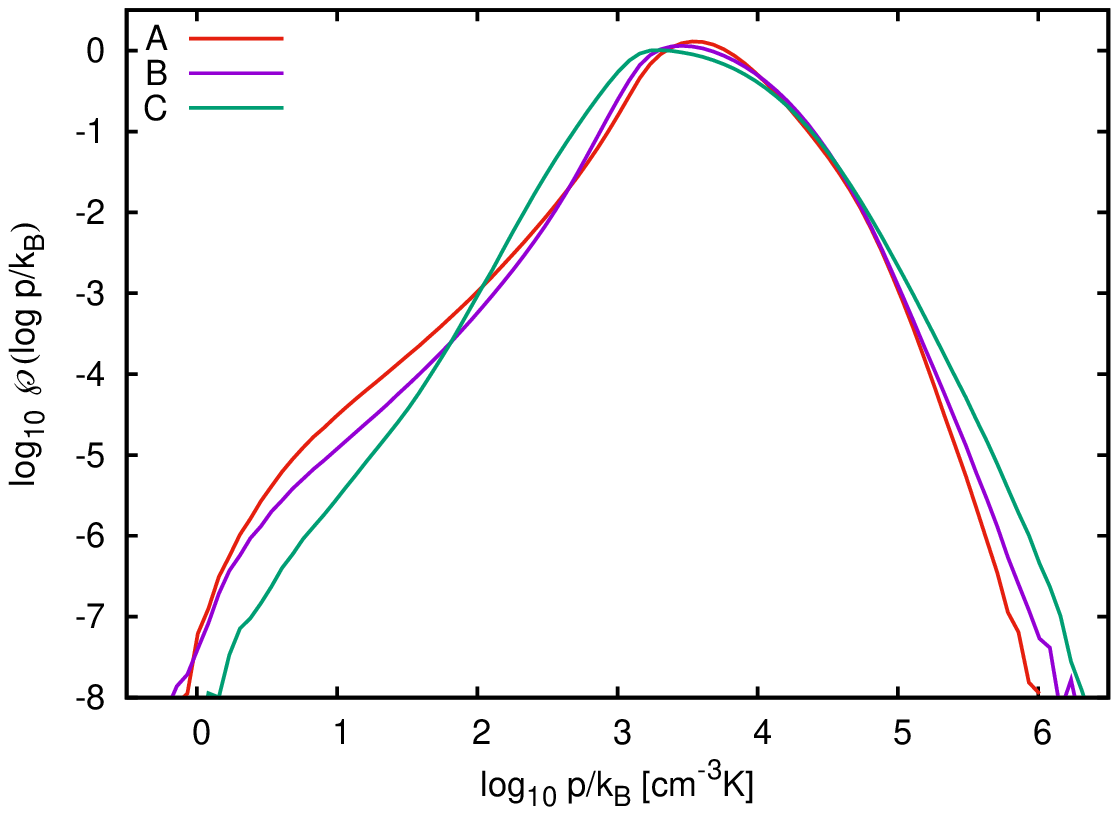}
           \includegraphics[scale=0.6]{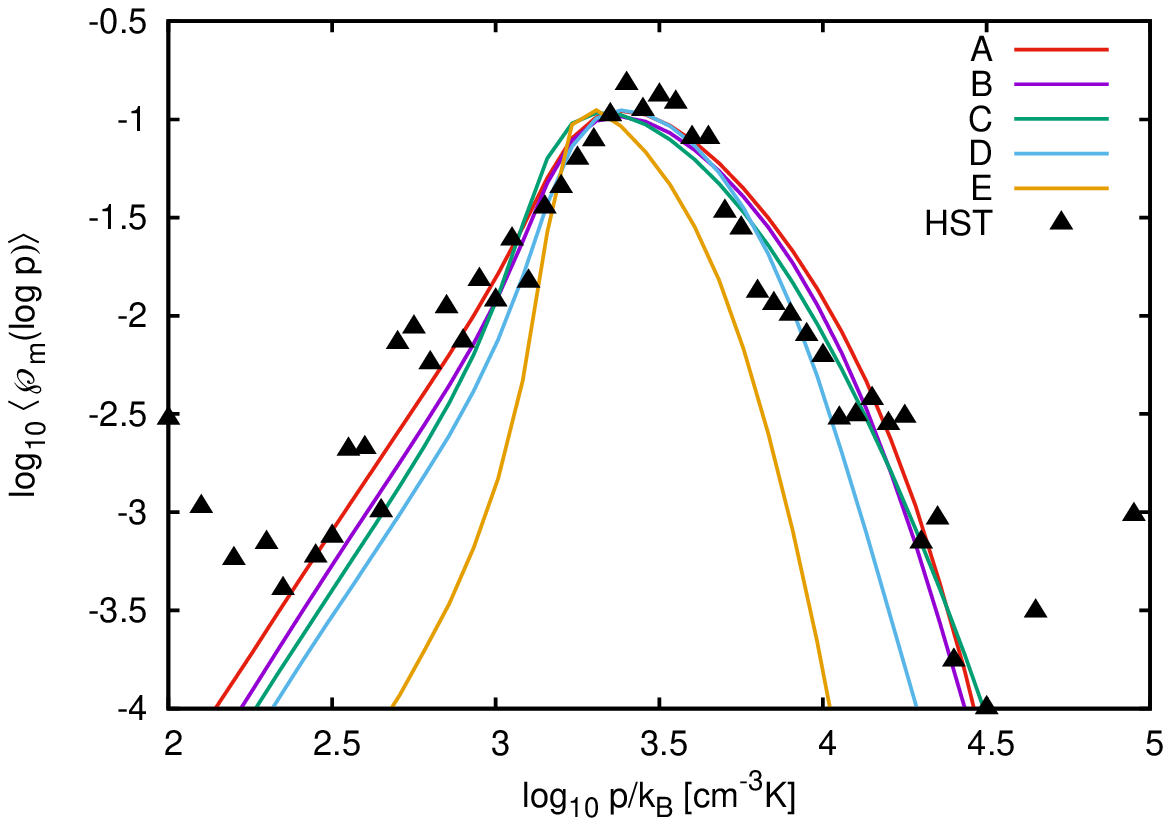}
           \caption{Time-average PDFs of the logarithm of thermal pressure for cases A, B, and C (left). Mass-weighted PDFs of $\log(p)$ conditioned on lines of sight with column densities below a threshold $N_{\rm c}({\rm HI})=2.5\times10^{21}$~cm$^{-2}$ and on temperature of the cold phase ($40$~K$<T<183$~K) for all cases (right). Black triangles show HST data for the diffuse CNM in the Galaxy based on UV spectra of 89 stars \cite{jenkins.11}. }
\label{f-pthpdf}
\end{figure}

At high levels of turbulence, the distribution of thermal pressure for cases A, B, and C spans about 6 decades leaving no room for the classical isobaric pressure-supported cloud picture in the violent ISM (Fig.~\ref{f-pthpdf}, left). The PDFs  are asymmetric in a very characteristic way, showing a positive skewness, reported in simulations earlier \cite{kritsuk..11,hill......12,gent....13,gazol14}. The origin of this asymmetry can be traced back to the turbulence-regulated and radiative cooling-controlled distribution of thermal phases in the phase diagram (Fig.~\ref{f-phase}). The peak tends to shift slightly to higher pressures as the magnetization becomes stronger, approaching the mean ISM pressure of 3700~K~cm$^{-3}$ reported in \cite{jenkins.11} in the range of $B_0$ bracketed by cases A and B. 
{The peak location is sensitive to pressure in the warm phase and in the thermally unstable regime, which together make up $\approx93$\% of the volume in cases A, B, and C (Table~\ref{tab2}). Models with higher magnetization, however, tend to have a progressively more abundant warm phase and less populated thermally unstable regime. Since the stable warm phase has higher $p_{\rm th}$ on average, compared to the unstable gas (Figure~\ref{f-phase}), the PDF's peak shifts to the right along the C-B-A sequence. The physics behind the shift apparently has to do with the stabilizing effect of the magnetic field on the TI \cite{field65}.
}
Note that the overall shape of the PDF is not lognormal because at high turbulence levels each pressure bin includes contributions from a wide range of temperatures, while some parts of the high tail can still be approximated by a lognormal with reasonable accuracy.

In order to compare the pressure PDF with observations that include line-of-sight convolutions, we use density-weighed PDFs compiled using sightlines parallel to the orthogonal coordinate directions of $\sim70$ data snapshots from each model. In addition, conditioning is used to match observations that can be blind to certain temperature regimes and are limited to sightlines below some extinction threshold. The synthetic PDFs shown are conditioned to mimic the HST archive data for the Solar Neighborhood  obtained from fine structure C{\sc i} lines in UV stellar spectra sensitive to the cold neutral gas \cite{jenkins.11}.\footnote{The HST data points provided by Dr. Ed Jenkins in 2010 reflect an earlier version of the analysis that relied on Wolfire et el. (1995) description of thermal processes \cite{wolfire....95}, which is consistent with our simulations (\S~\ref{s-param}) and differs from \cite{jenkins.11}, where the Wolfire et al. (2003) \cite{wolfire...03} version was used.} We thus exclude lines of sight with $N({\rm H}{\rm\sc I})>2.5\times10^{21}$~cm$^{-2}$ and mask out the gas with temperatures outside $40$~K$<T<183$~K. Besides these limits, there are no other free parameters involved, although we use arbitrary normalization vertically to fit the data. The effects of the column density and temperature  thresholds can be seen in the synthetic map shown in Fig.~\ref{f-map}.

Distributions for cases A through C with $u_{\rm rms,0} = 16$~km/s match the characteristic mean pressure typical for the Milky Way disk at the solar radius, reproduce its overall asymmetry, and show only weak dependence on $B_0$. The width of the distribution, however, is sensitive to $u_{\rm rms}$ and $n_0$. Case D marginally reproduces the shape and the width of the observed distribution with some deficiency in the high tail. A lower turbulence level in case E yields too narrow a distribution, and hence should be rejected. {We also formally applied the Kolmogorov-Smirnov (KS) test to compare the HST sample with our modelled cumulative distribution functions. Based on the KS test results, case B provides the best match to the HST data with the goodness of fit $D=0.066$ and the significance level of this KS statistic $\wp=0.95$. Case B is only slightly better than A, but other cases returned poor results, as the rough visual comparison suggested.}
Overall, due to the conditioning by temperature and opacity, the mass-weighted PDF is substantially narrower than the actual PDF shown in the left panel. While both low and high ends are suppressed, the effects of conditioning are stronger at high thermal pressures.

\subsection{Magnetic, dynamic, and thermal pressure\label{s-pre}}
\begin{figure}[t]
       \begin{center}
           \includegraphics[scale=0.45]{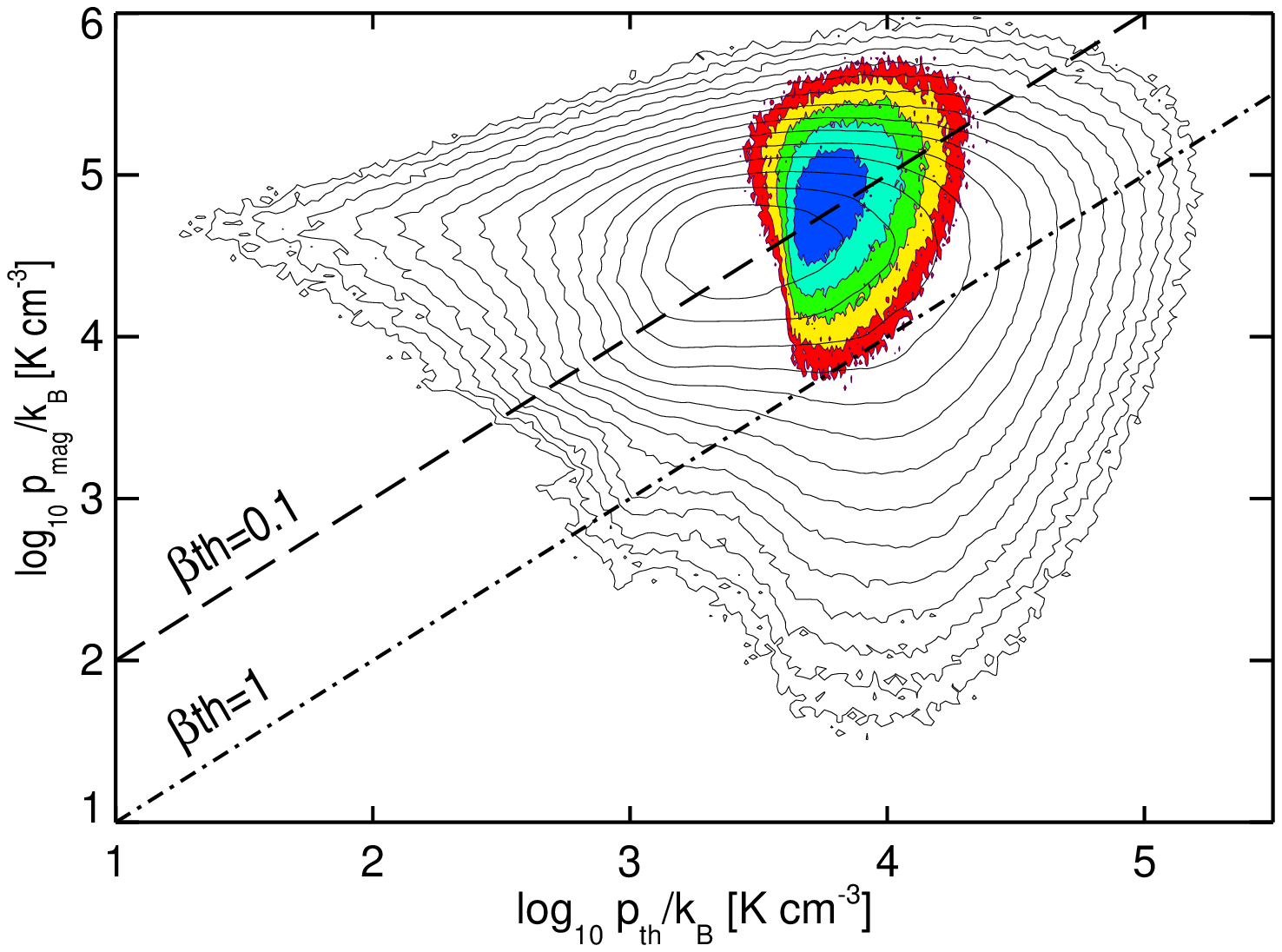}
           \includegraphics[scale=0.45]{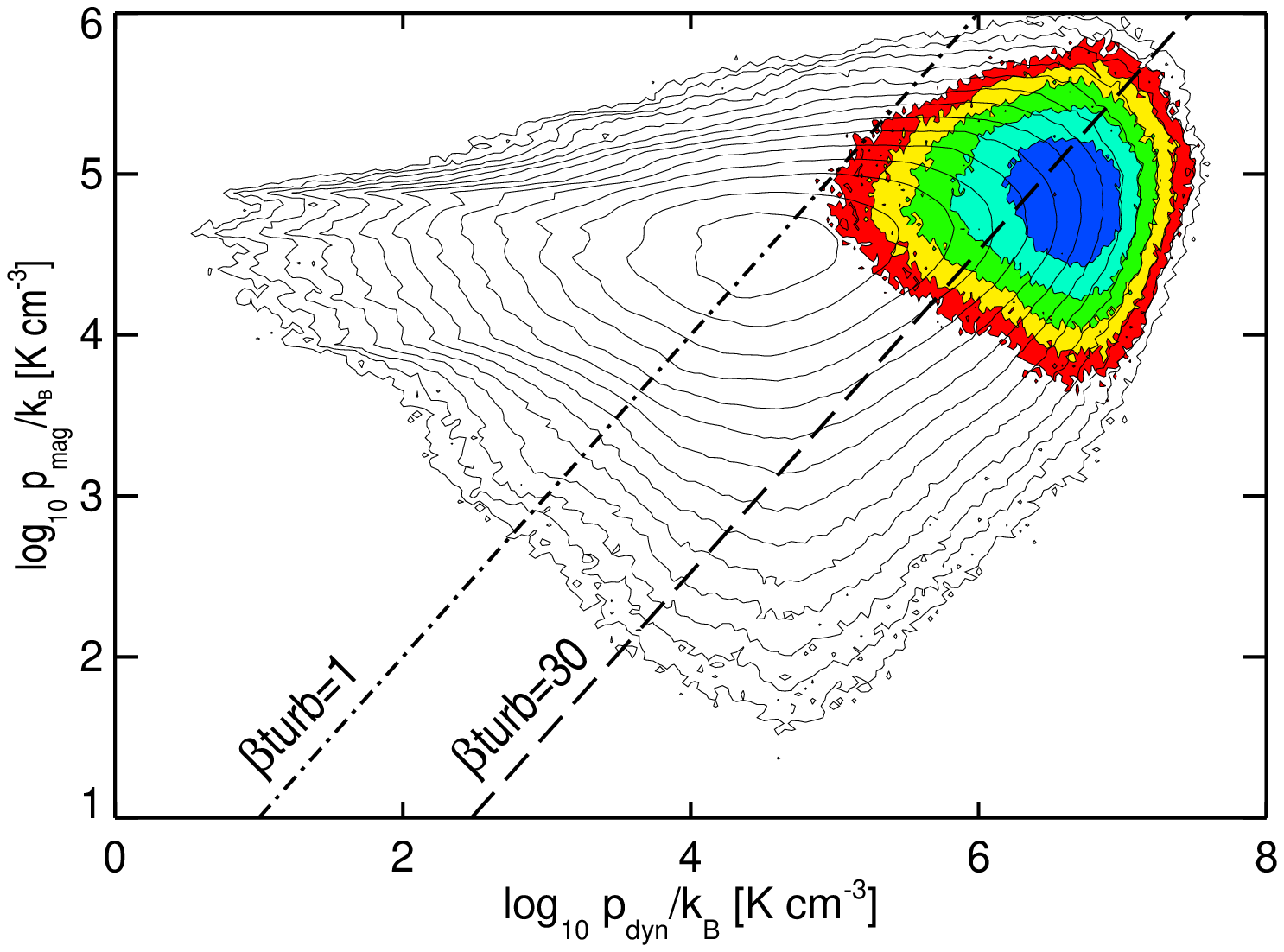}
                 \end{center}
           \caption{Magnetic pressure vs. thermal pressure (left) and vs. dynamic pressure (right) for a snapshot from case B at $t=42.3$~Myr. Black contours show a scatter plot sampled from the whole computational domain, while contours shown in color sample only the cold molecular gas, which is conveniently defined as all material with $n>100$~cm$^{-3}$ and $T<100$~K. {The contours are logarithmically spaced, reflecting factors of 2 difference in number density; the blue-green-yellow-red sequence corresponds to a factor of 16 drop in the number of grid zones with given physical conditions.}}
           \label{f-pmdt}
\end{figure}
These numerical experiments probe the levels of magnetic field strength in molecular clouds that form self-consistently in the magnetized turbulent diffuse ISM. Figure~\ref{f-pmdt} shows scatter plots of the magnetic pressure $p_{\rm mag}\equiv B^2/2$ versus the thermal pressure $p_{\rm th}=(\gamma-1)\rho e$ (left) and versus the dynamic pressure $p_{\rm dyn}\equiv\rho u^2$ (right) for case B at 42.3~Myr. Since $\beta_{\rm th}\equiv p_{\rm th}/p_{\rm mag}=(\gamma-1)U/M$, $\beta_{\rm turb}\equiv p_{\rm dyn}/p_{\rm mag}=2K/M=2{\cal M}_{\rm a}^2$, and $\beta_{\rm turb}/\beta_{\rm th}=\gamma{\cal M}_{\rm s}^2=2/(\gamma-1)K/U$, one can use the two plasma beta parameters to access pairwise energy equipartition and Mach number regimes at different locations on these diagrams. For instance, the dashed line $\beta_{\rm th}=0.1$ in the left panel, crossing right through the `summit' of the black contours, indicates that magnetic pressure is an order of magnitude higher than thermal in the bulk of the volume (same applies to magnetic and internal energies, $M\gg U$). Interestingly, the same remains true for the cold molecular gas ($T<100$~K and $n>100$~cm$^{-3}$), as indicated by the color contour map in Fig.~\ref{f-pmdt}.

The dynamic pressure, however, is of the same order as the magnetic one in the bulk of the volume, as the dash-dot line $\beta_{\rm turb}=1$ in the right panel shows. This also means that $K\sim M$ and ${\cal M}_{\rm a}\sim1$ in the bulk of the volume, consistent with data in Table~\ref{mach}. It is worth noting, however, that these conditions do not extend to the cold molecular gas, as the color contours centering around $\beta_{\rm turb}=30$ indicate. The cold molecular gas in our simulations is instead super-Alfv\'enic with ${\cal M}_{\rm a}\sim8$ (see also Fig.~\ref{f-mach}, right) and $K\sim10M$. Qualitatively, this holds even in the strongly magnetized case A.

A key to understanding the origin of this super-Alfv\`enic regime of strong MHD turbulence in the cold and dense molecular gas lies in the process of self-organization {caused by stochastic flux freezing, which is in turn mediated by the reconnection diffusion process \cite{eyink11,lazarian...15}.} 
 Magnetic flux conservation in turbulent fluids at high magnetic Reynolds numbers is not valid in the classical sense, nor is it entirely broken. Instead it holds in a statistical sense associated with the `spontaneous stochasticity' caused by turbulent Richardson-like diffusion of magnetic field lines, which leads to a breakdown of the classical flux freezing concept. This subject is still under development for compressible MHD turbulence and our simulations may not completely capture the small-scale dynamo effects in the cold dense gas due to limited resolution and because of the implicit numerical nature of dissipation resulting in $Pm\sim1$ and uniform $\nu_{\rm eff}$ that do not properly approximate conditions in the multiphase ISM.

\subsection{Filaments, striations, and the alignment of magnetic and velocity fields\label{align}}
\begin{figure}[t]
       \begin{center}
           \includegraphics[scale=0.56]{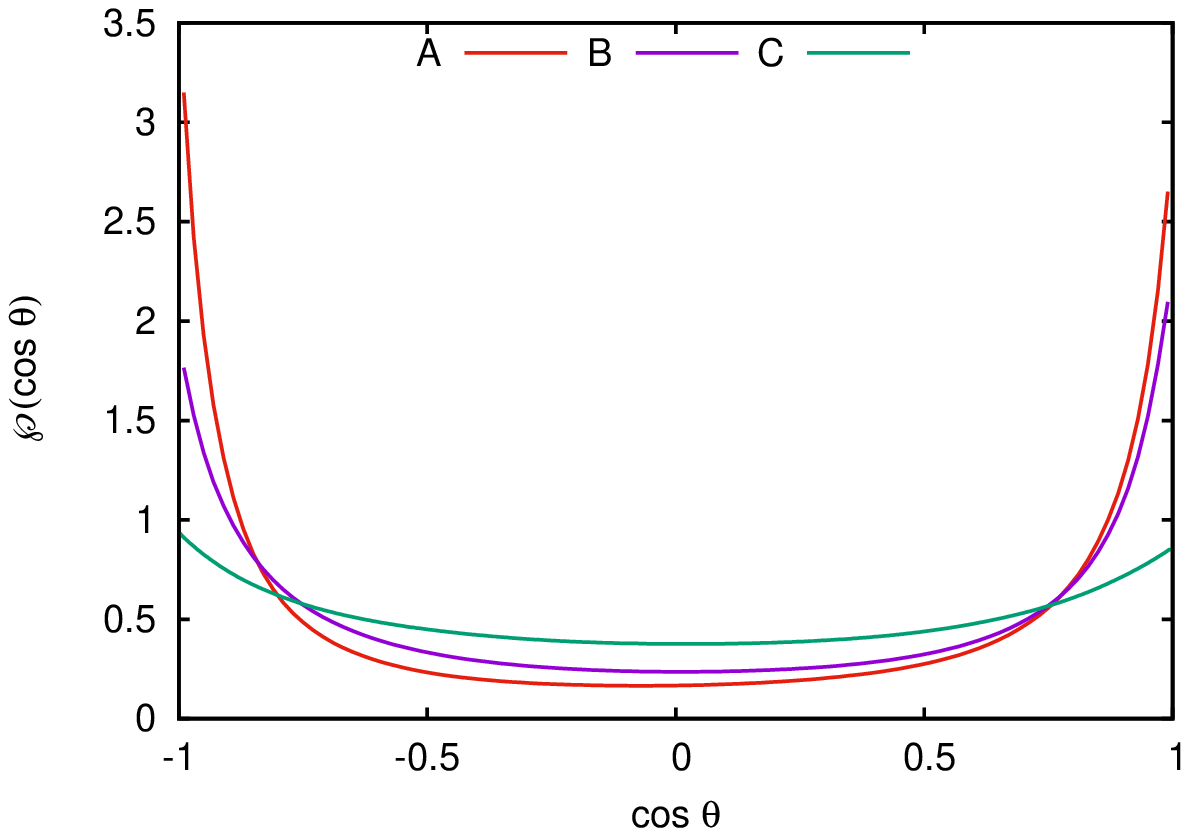}
           \includegraphics[scale=0.42]{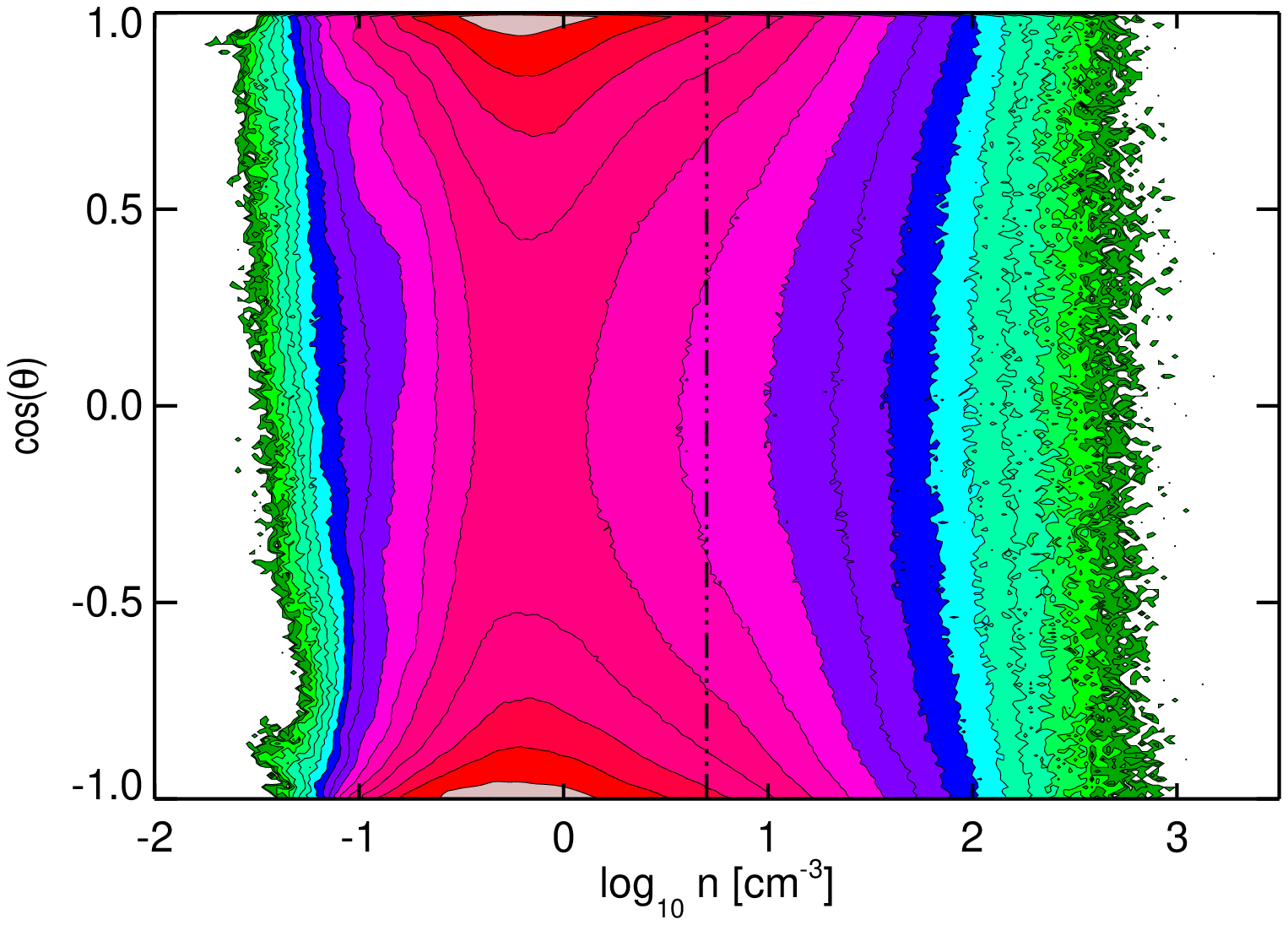}
                 \end{center}
           \caption{Time-average PDFs of the cosine of the velocity-magnetic field alignment angle $\theta=\arccos(\bm B\cdot\bm u/\sqrt{B^2u^2})$ for cases A, B, and C (left) and scatter plot of  $\cos{\theta}$ vs. gas density for case A at $t=42.3$~Myr (right).
Isocontour spacing corresponds to a factor of 2 in likelihood. Vertical dash-dotted line in the bottom-right plot indicates the mean density $n_0=5$~cm$^{-3}$. }
           \label{ali}
\end{figure}
The statistical steady state explored by our simulations is characterized by a certain degree of alignment between the velocity and magnetic field lines. If molecular clouds form in the turbulent ISM via large-scale compression of the warm diffuse trans-Alfv\'enic H{\sc i} primarily along the field lines, then turbulence in the clouds will naturally be super-Alfv\`enic \cite{padoan......10}. A closely related subject is the relative orientation of the magnetic field direction with respect to filamentary density structures in simulated turbulent molecular clouds \cite{soler+13,chen..16}. Observations of striations and filaments in local molecular clouds and their close environments \cite{goldsmith.....08,palmerim+13,alves+14,kalberla......16,cox+16,soler+16} suggest that the filaments are preferentially oriented perpendicular to the magnetic field lines, while the low-density subcritical striations are parallel. Moreover, it is believed that the background cloud material is funneled along the magnetic field lines onto star-forming filaments through striations, consistent with the kinematic constraints
from CO observations \cite{schneider+10}.
Here we check the likelihood of such a scenario by looking at the alignment of local magnetic field and velocity vectors in different density regimes. 

Figure~\ref{ali} shows the PDFs of the cosine of the magnetic field--velocity angle $\theta=\arccos(\bm B\cdot\bm u/\sqrt{B^2u^2})$ for cases A, B, and C (left) and details the alignment properties as a function of density for case A (right).  The $\bm B-\bm u$ alignment is most pronounced in the strongly magnetized case A and in the bulk of the volume at densities slightly below the mean $n_0$ (shown by the dash-dot line), where the turbulence is sub-Alfv\'enic. However, the alignment is very weak at both extreme ends of the density distribution: in rarefactions and in shocked material at very high densities (Fig.~\ref{ali}, right). 
Indeed, in the magnetically dominated (sub-Alfv\'enic) warm phase, the gas structure is expected to be preferentially aligned with the local magnetic field. 

For the dense cold phase with super-Alfve\'nic conditions, however, this {\em dynamic} alignment mechanism does not work, and the filament elongation becomes preferentially perpendicular to the local magnetic field.
Case C, with supersonic and super-Alfv\'enic turbulence in all phases (Table~\ref{tab2}) shows no significant alignment between the velocity and magnetic field vectors. In this case, however, the {\em kinematic} alignment (between $\bm B$ and the eigenvectors $\bm \lambda$ of a symmetric part of the rate-of-strain tensor $s_{ij}=(\partial_j u_i+\partial_iu_j)/2$) is actually expected instead \cite{tsinober04}.
Filament orientation with respect to the magnetic field depends on the shear strain alignment with respect to the field lines. If the strain is strong and (anti)parallel to the magnetic field, the filament will be aligned with the field \cite{inoue.16}. Gravitational effects can further complicate the alignment pattern in dense filaments through amplification of the potential (curl-free) component of the gas velocity.

\subsection{Magnetic field PDFs\label{s-mpdf}}
\begin{figure}[t]
           \includegraphics[scale=0.6]{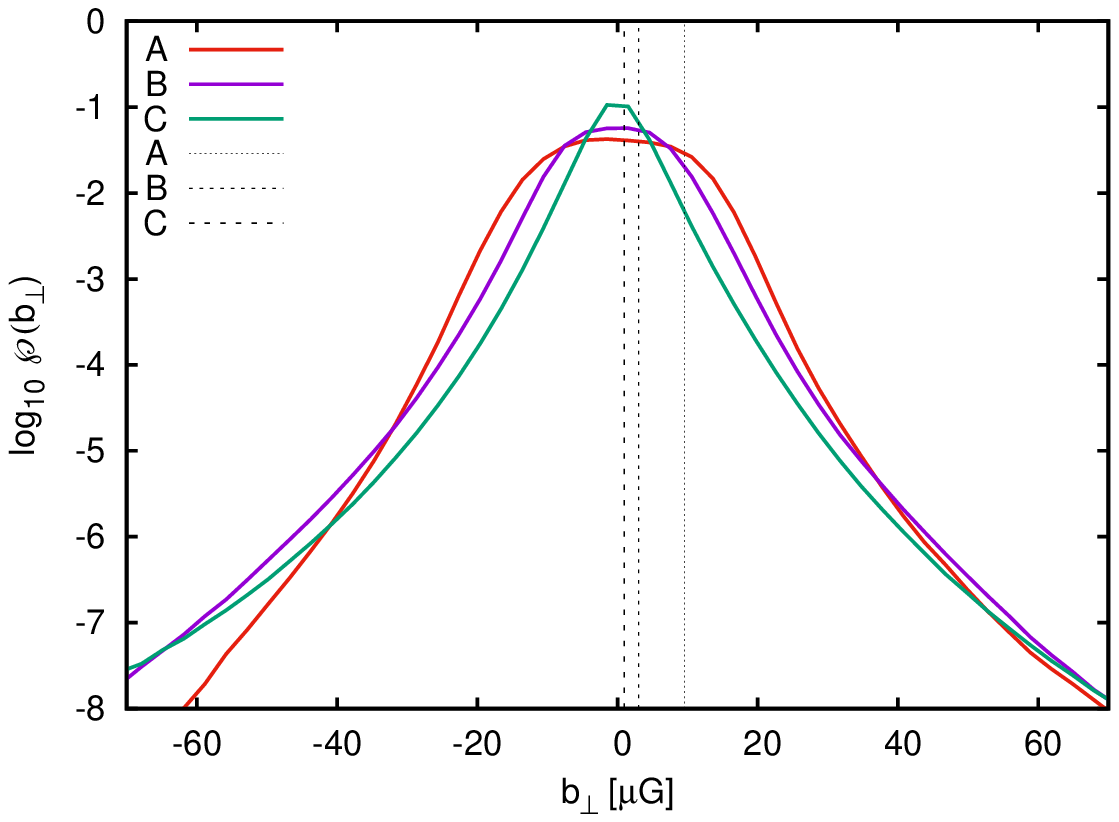}
           \includegraphics[scale=0.6]{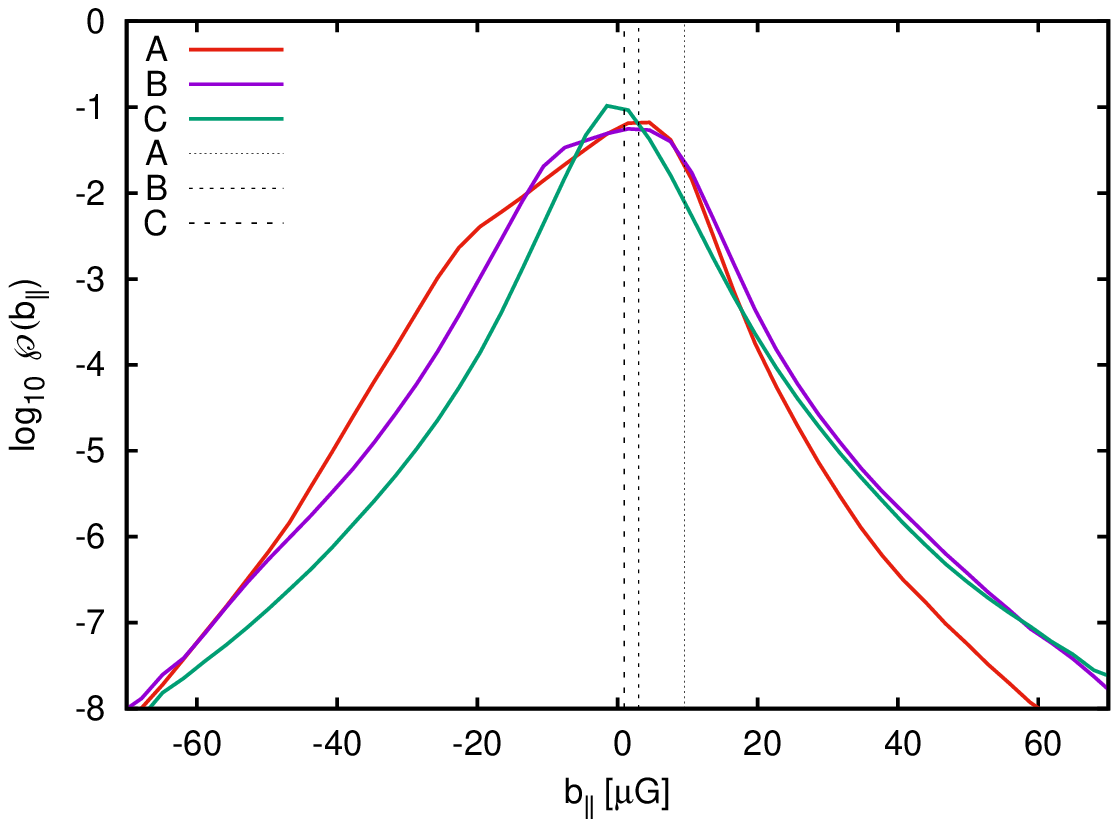}
           \caption{Time-average PDFs of magnetic field fluctuations $\bm b=\bm B-\bm B_0$ perpendicular (left) and parallel (right) to the mean field $B_0$ for cases A, B, and C. Vertical straight lines show the mean field values.}
\label{kurtic}
\end{figure}
Most of what we know about the interstellar magnetic fields comes from observations that usually involve complex convolutions along the observed sightlines \cite{beck16}. In order to decipher that convoluted information, additional assumptions must be made concerning possible symmetries involved (e.g., whether the turbulent fluctuations of the field are isotropic) and the shape of probability distributions of various magnetic field components (e.g., if they are Gaussian). Figure~\ref{kurtic} shows PDFs of magnetic field fluctuations $\bm b=\bm B-\bm B_0$ perpendicular ($b_{\perp}$) and parallel ($b_{\parallel}$) to the mean field $\bm B_0$ for cases A, B, and C. All PDFs represent averages over $70-100$ flow snapshots, uniformly covering the statistically stationary state. In the case of $b_{\perp}$, the results include both $b_y$ and $b_z$ components normal to $\bm B_0$ directed along the positive $x$-coordinate direction.

The plots show that the PDFs are strongly non-Gaussian in all cases and that the $B$-field fluctuations in the direction of $\bm B_0$ are strongly asymmetric, i.e. they know fairly well about the direction of the mean field. In the strongly magnetized case A, the PDFs of perpendicular components show a plateau around zero with a width of order $B_0$. The case-A PDF is {\em platykurtic}  with an excess kurtosis $\gamma_2\equiv\mu_4/\sigma^4-3=-0.4$. Moderately and weakly magnetized cases B and C instead show {\em leptokurtic} (i.e. more peaked than the normal distribution) PDFs with $\gamma_2=0.3$ and 3.6, respectively.

The PDF of the parallel component $b_{\parallel}$ is skewed to negative values with the skewness $\gamma_1\equiv\mu_3/\sigma^3=-1.0$ and $-0.2$ in cases A and B, respectively. Case C shows modest positive skewness $\gamma_1=0.2$. The models thus predict strongly non-Gaussian distributions for the magnetic field components both parallel and perpendicular to the mean field in cases with sufficiently strong magnetization.

\begin{figure}[t]
\centering
           \includegraphics[scale=0.6]{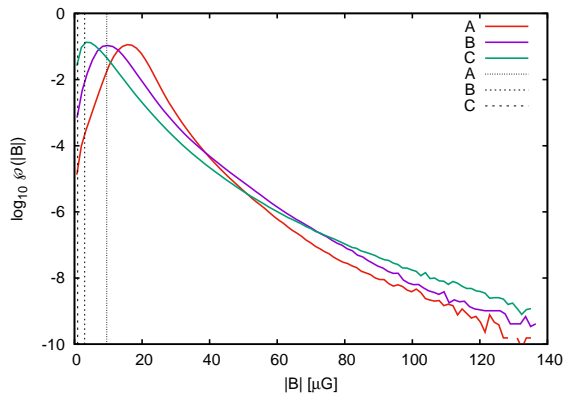}
           \caption{Time-average PDFs of the magnetic field strength for cases A, B, and C. Vertical straight lines show the mean field values.}
\label{f-absb}
\end{figure}

{Figure~\ref{f-absb}} shows PDFs of $|\bm B|$ for cases A, B, and C; vertical black dashed lines indicate the values of $B_0$. All three PDFs have fat stretched-exponential tails extending to the field strengths exceeding dozens or even hundreds of the mean field value. Curiously, the likelihood to find extreme values in the range $B\in[100,120]$~$\mu$G in the weakly magnetized case C (green line) is approximately eight times higher than in the strongly magnetized case A (red line). The mean field $B_0$ thus controls the abundance of cold dense gas with extreme field values produced in strong intermittent (both in space and in time) 3D compressions which are naturally more likely under weak magnetization.
\begin{figure}[t]
\centering
           \includegraphics[scale=0.6]{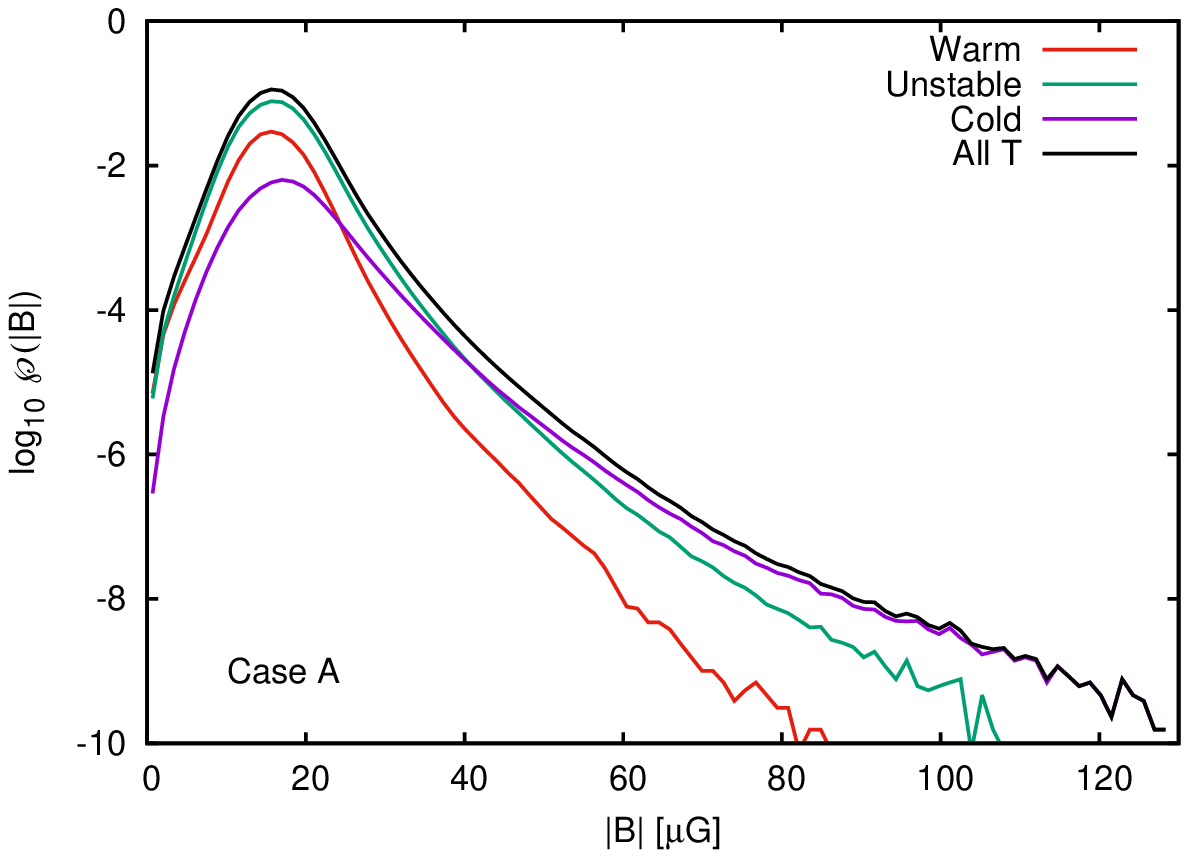}
           \includegraphics[scale=0.6]{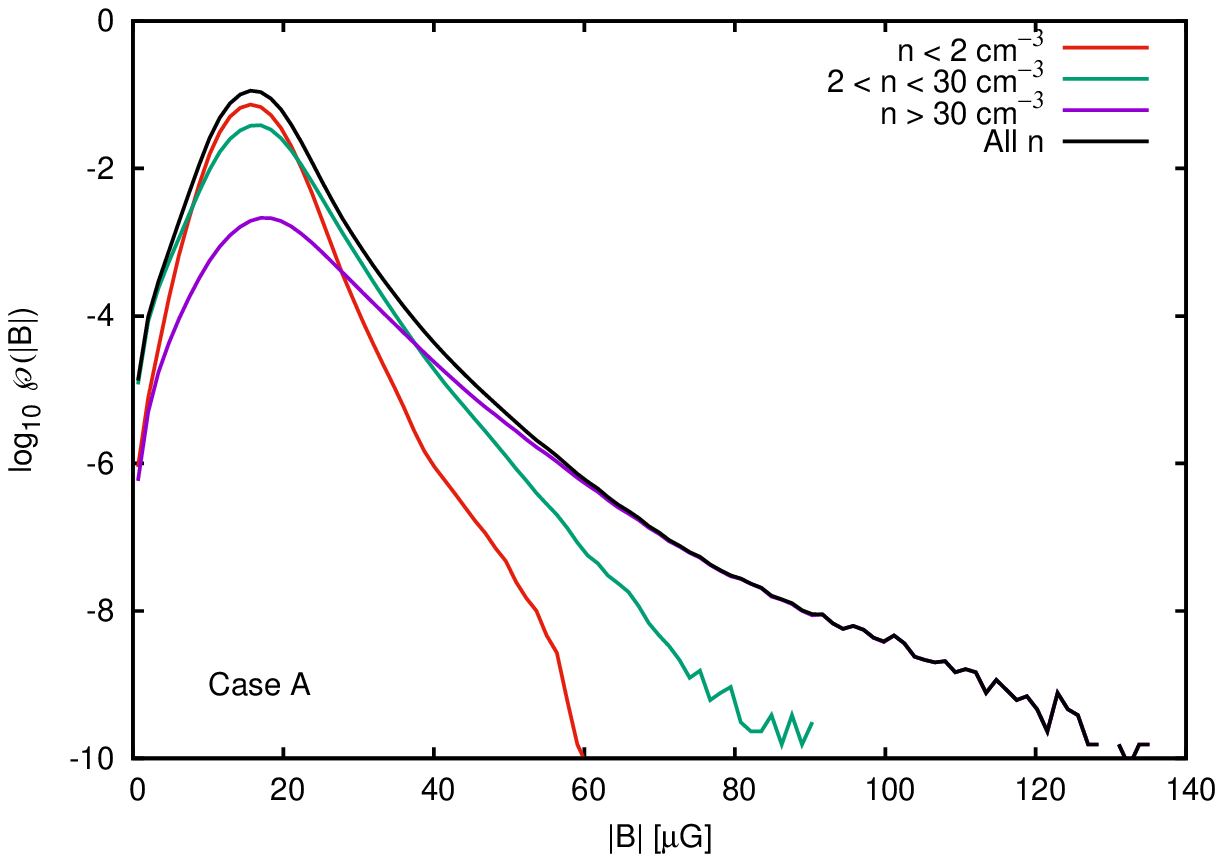}
           \caption{Time-average PDF of magnetic field strength (black) and contributions from different thermal phases (left) and density regimes (right) for case A.}
\label{f-absbnt}
\end{figure}
Using the gas temperature and density as additional variables to condition the distributions, one can check which physical regimes contribute most to the fat tails or to the peak of the $|\bm B|$-PDFs. Figure~\ref{f-absbnt} shows that in the strongly magnetized case A the extreme tail is due to the cold phase (left) and contains exclusively dense ($n>30$~cm$^{-3}$, most likely molecular) gas (right), whereas the peak of the distribution is mostly due to the gas at intermediate (thermally unstable) temperatures and at lower densities ($n<2$~cm$^{-3}$). The most likely rms $B$-field values for the cold phase (or for the dense gas) are only slightly higher than those for the full distribution, whereas the peak of the warm phase (or low density gas) is only slightly shifted to weaker field values.
These distributions can predict general trends in expected magnetic field regimes accessible with observational tracers populating parts of molecular clouds with different temperatures and densities.

\subsection{Velocity structure functions\label{larson}}
\begin{figure}[h]
\centering
           \includegraphics[scale=0.6]{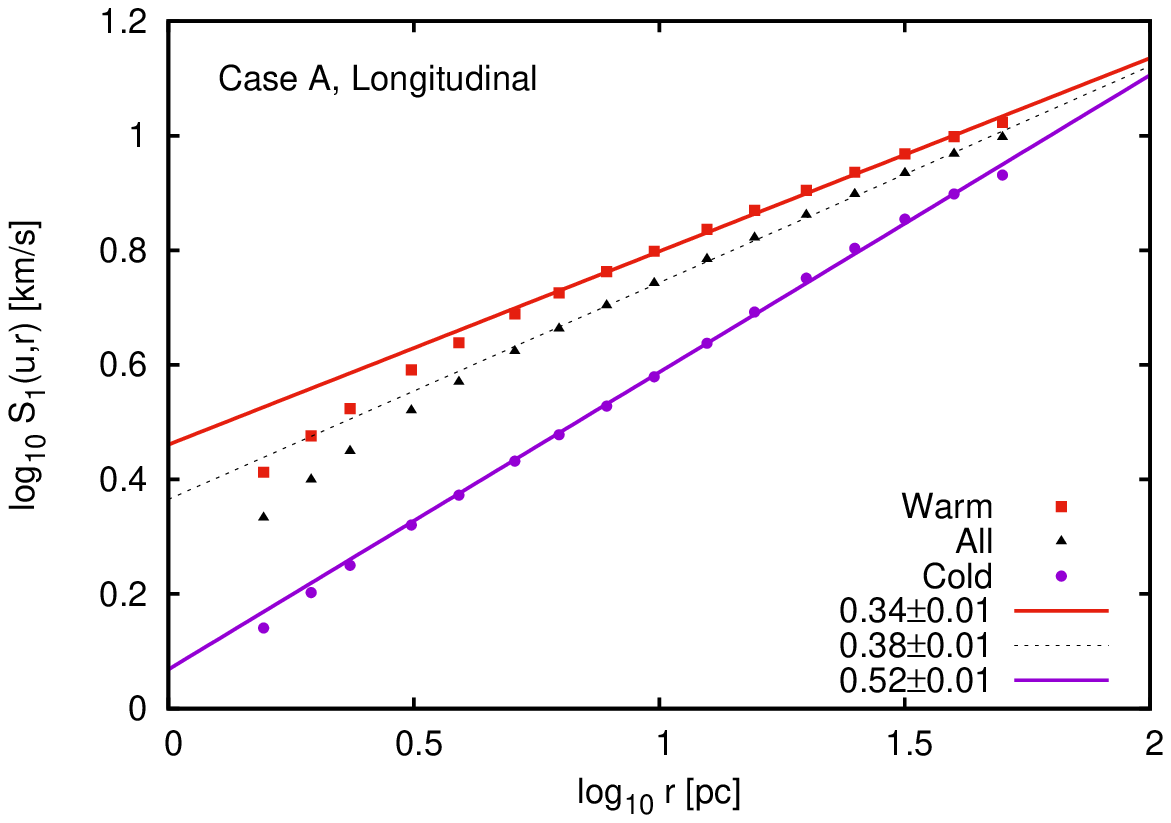}
           \includegraphics[scale=0.6]{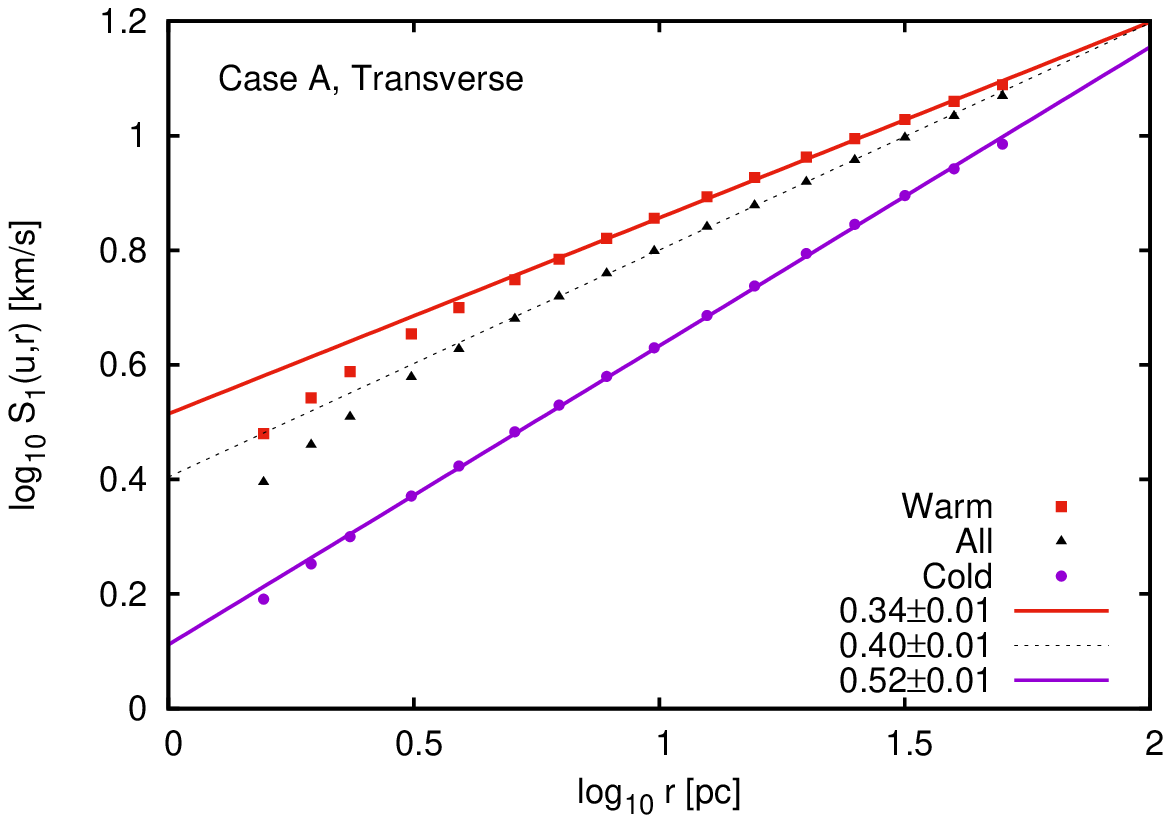}
           \includegraphics[scale=0.6]{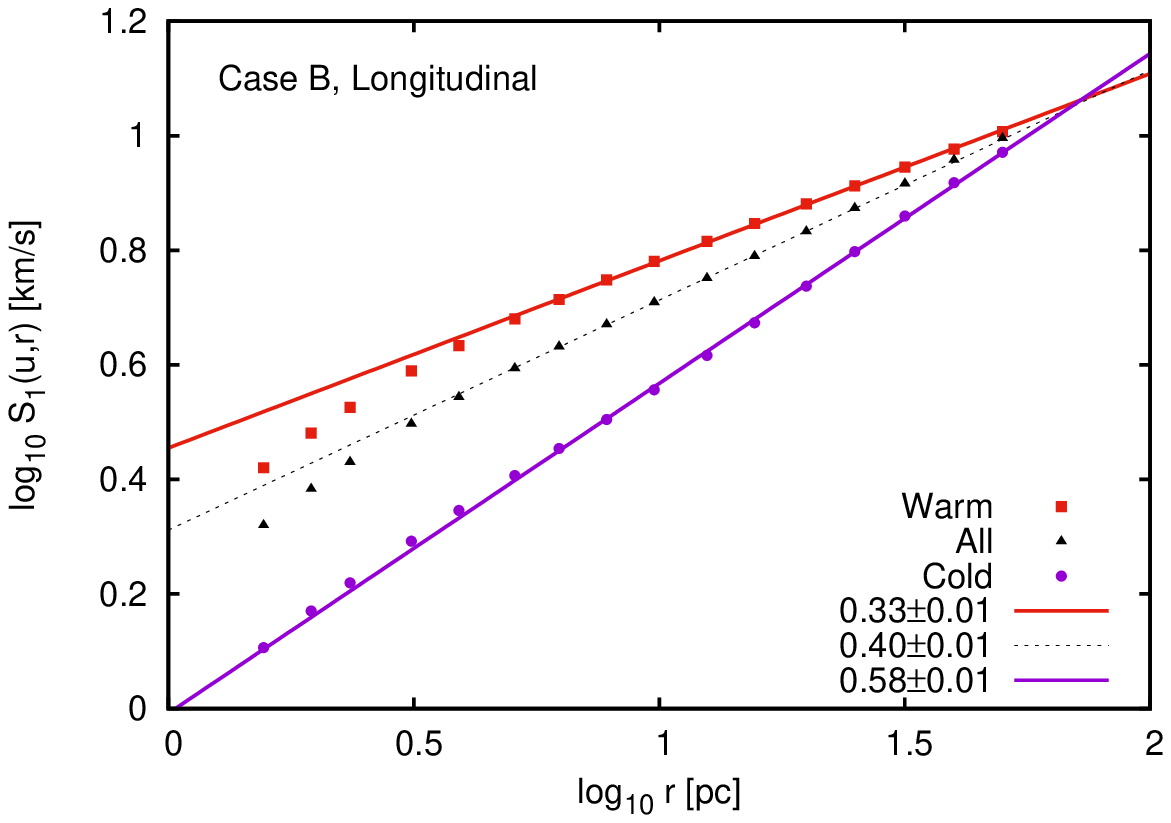}
           \includegraphics[scale=0.6]{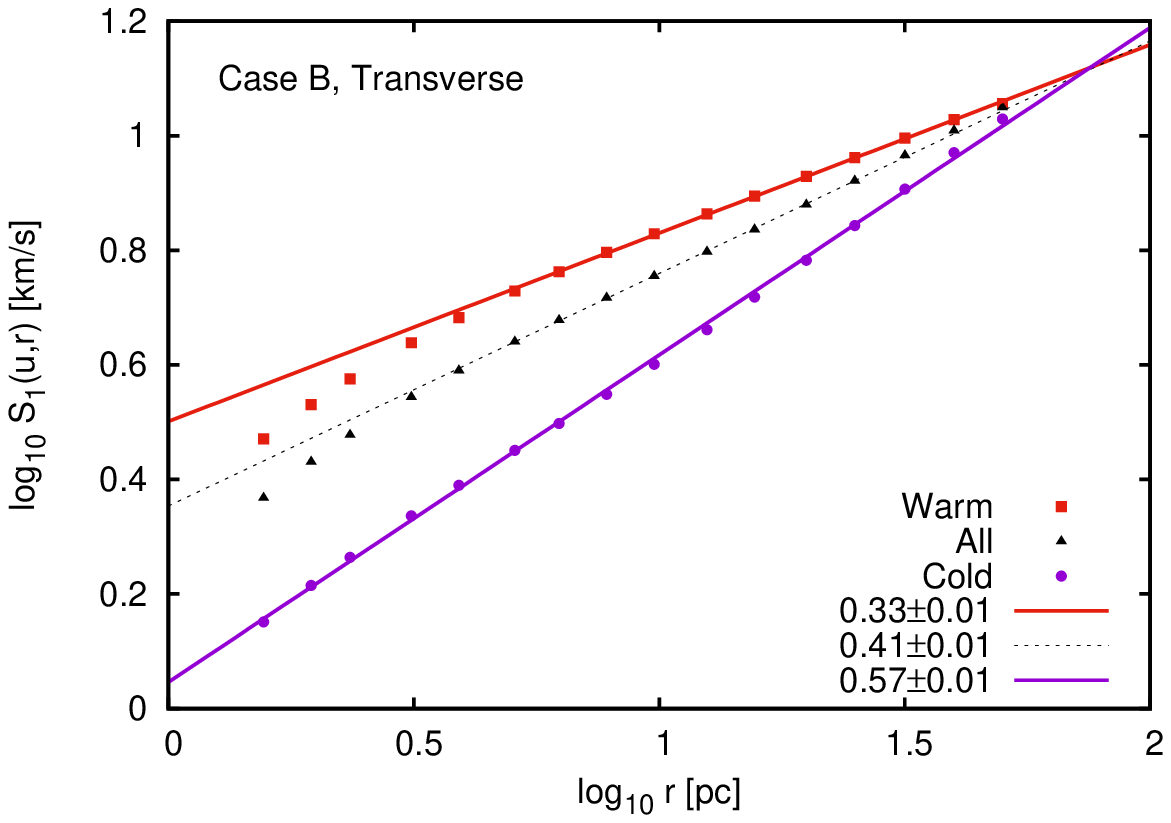}
           \includegraphics[scale=0.6]{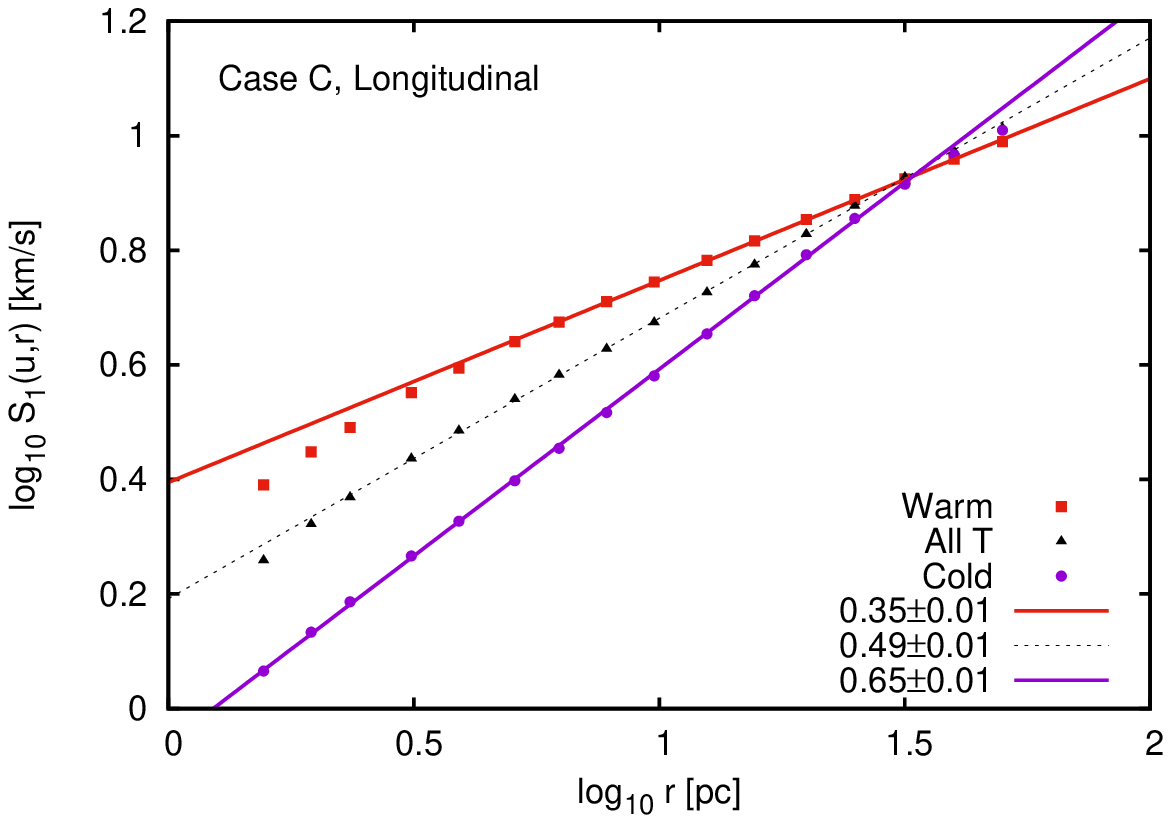}
           \includegraphics[scale=0.6]{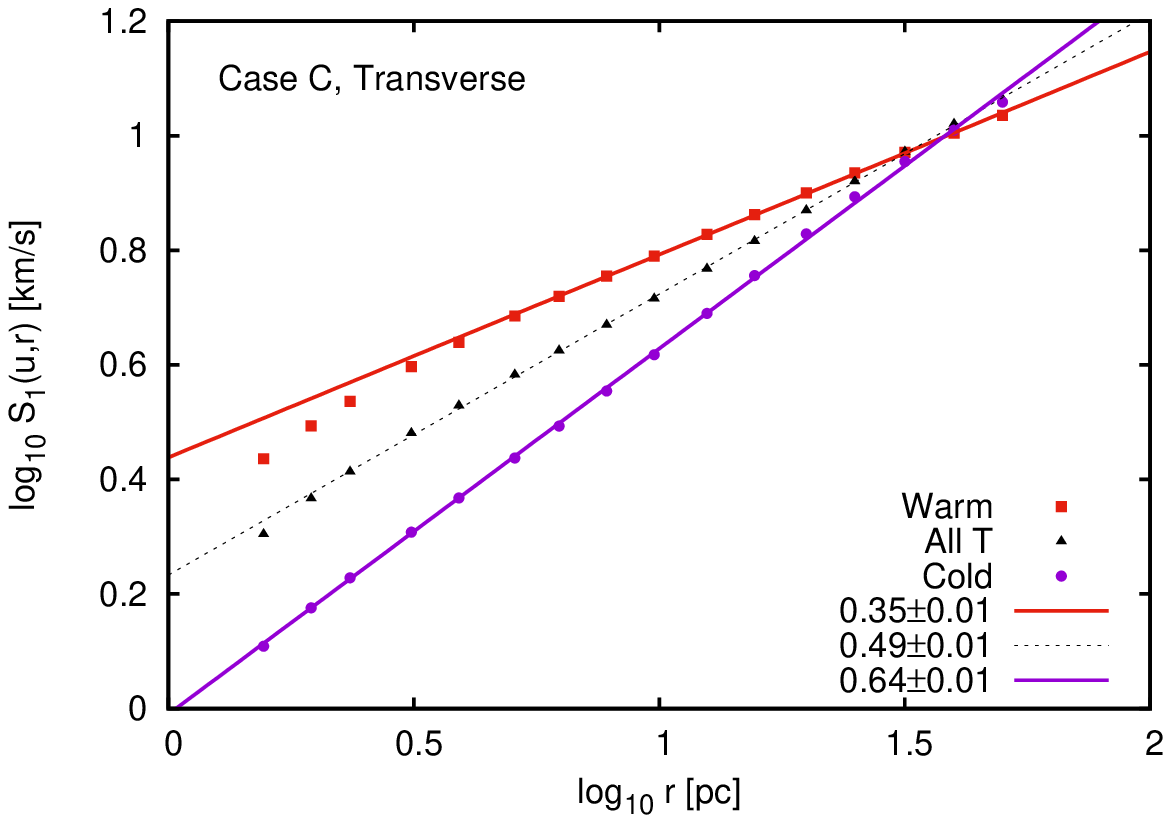}
           \caption{Time-average first order velocity structure functions $S_{1,\parallel}(r)$ and $S_{1,\perp}(r)$ for cases A, B, and C.  Results for all temperature regimes are shown in black; ensemble averages conditioned on the warm ($T>5,250$~K) and cold ($T<184$~K) phases are shown in red and purple, respectively.}
\label{f-sf1}
\end{figure}
Larson's first law \cite{larson81,kritsuk..13}, also known as the linewidth--size relation, is an observable surrogate for first order ($m=1$) structure functions of the velocity field that are defined by 
\begin{equation}
S_m(r)=\langle[ u(\bm x+\bm r)-u(\bm x)]^m\rangle\propto r^{\zeta_m},
\label{struc}
\end{equation}
where $\langle\cdot\rangle$ denotes the ensemble average over all possible point pairs separated by the lag $r=|\bm r|$. The velocity components here can either be parallel or perpendicular to vector $\bm r$, corresponding to longitudinal $S_{m,\parallel}(r)\propto r^{\zeta_m^{\parallel}}$ and transverse $S_{m,\perp}(r)\propto r^{\zeta_m^{\perp}}$ structure functions, respectively. 

Figure~\ref{f-sf1} shows scaling results for $S_{1,\parallel}(r)$ and $S_{1,\perp}(r)$ based on 16 lag values in the range of $r\in[1.5,50]$~pc for cases A, B, and C. Black triangles and black dashed lines show structure functions obtained from uniform sampling of the whole computational domain, where all temperature regimes are represented. In cases A and B with strong-medium magnetization, scaling exponents $\zeta_1^{\parallel}\approx\zeta_1^{\perp}=0.4$, while in weakly magnetized case C $\zeta_1^{\parallel}\approx\zeta_1^{\perp}=0.5$. The fact that case C shows an exponent close to that found in isothermal hydrodynamic turbulence simulations at high Mach numbers is not surprising since ${\cal M}_{\rm s}\approx13$ and ${\cal M}_{\rm a}\approx2$ (Table~\ref{tab2}), i.e. the turbulence is supersonic and super-Alfv\'enic on average in the whole volume. In all cases, the scaling range is about one dex, which implies the exponents are reasonably accurate.

Observational measurements are usually done using a single molecular hydrogen tracer (e.g. CO) available only in certain density and temperature regimes. To mimic this situation, one can condition the ensemble averaging operator in (\ref{struc}) to only include point pairs in the density and temperature regimes appropriate for the given tracer. Technically, simulations with sufficient resolution will provide reasonably accurate statistics for the density and temperature regimes that are not excessively restrictive. This would help study trends in the tracer-specific bias in observed velocity statistics. For instance, Figure~\ref{f-sf1} shows structure functions conditioned for the cold phase ($T<184$~K, purple) and for the warm phase ($T>5,250$~K, red). The warm phase shows Kolmogorov scaling $\zeta_1^{\parallel}=\zeta_1^{\perp}\approx1/3$ in all three cases. Indeed, the Mach number regimes for the warm phase are only mildly transonic (Table~\ref{tab2}) and thus imply $S_1(r)\propto r^{1/3}$ \cite{porter..02}. In contrast, the scaling of structure functions for the cold phase is steeper:
$\zeta_1\approx0.5$ in case A, 0.6 in case B, and 0.65 in case C. This is not surprising since ${\cal M}_{\rm s}\approx15$ for the cold phase  (Table~\ref{tab2}) and $\zeta_1\sim1/2$ are common for isothermal turbulence at ${\cal M}_{\rm s}\sim10$ \cite{kritsuk...07}. Our simulations, thus, reproduce the observed linewidth--size relations for molecular clouds (including both the slope and the offset) within the uncertainty introduced by the projection effects.

\subsection{Velocity and density power spectra\label{s-rvspec}}
\begin{figure}[t]
           \includegraphics[scale=0.6]{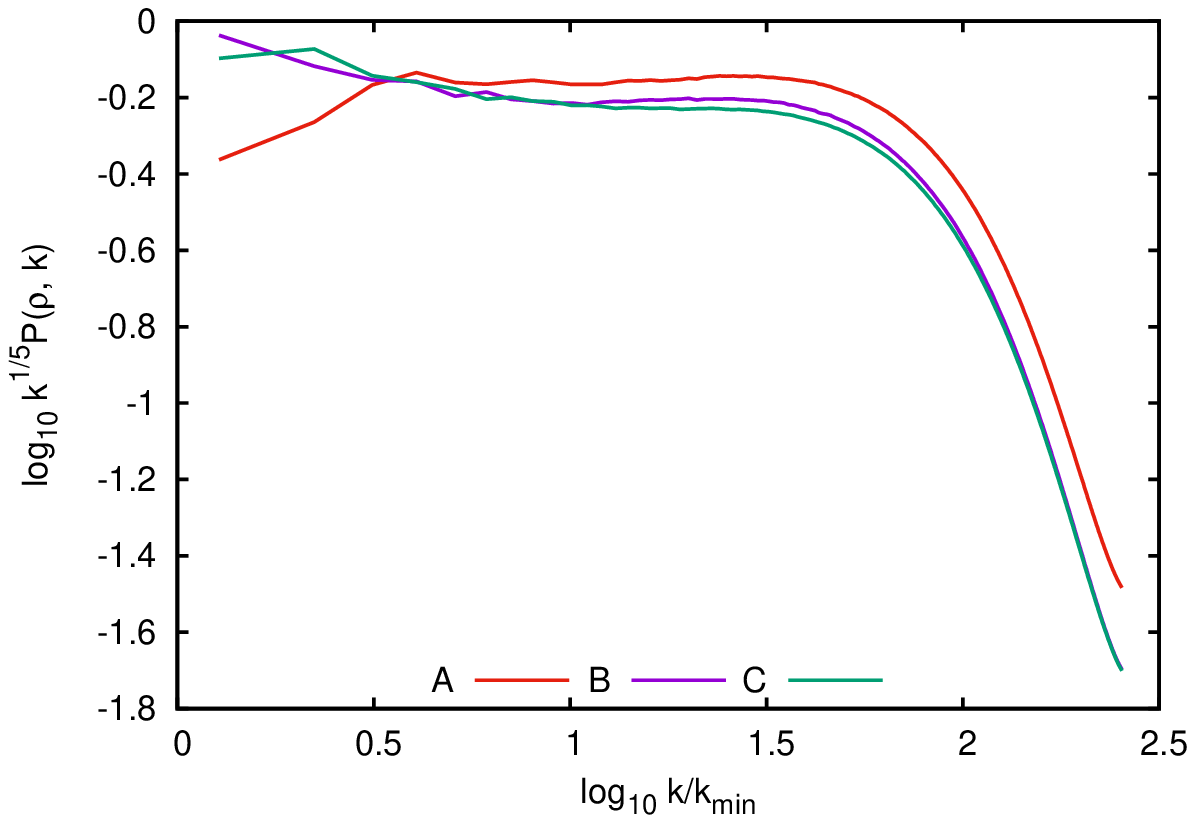}
           \includegraphics[scale=0.6]{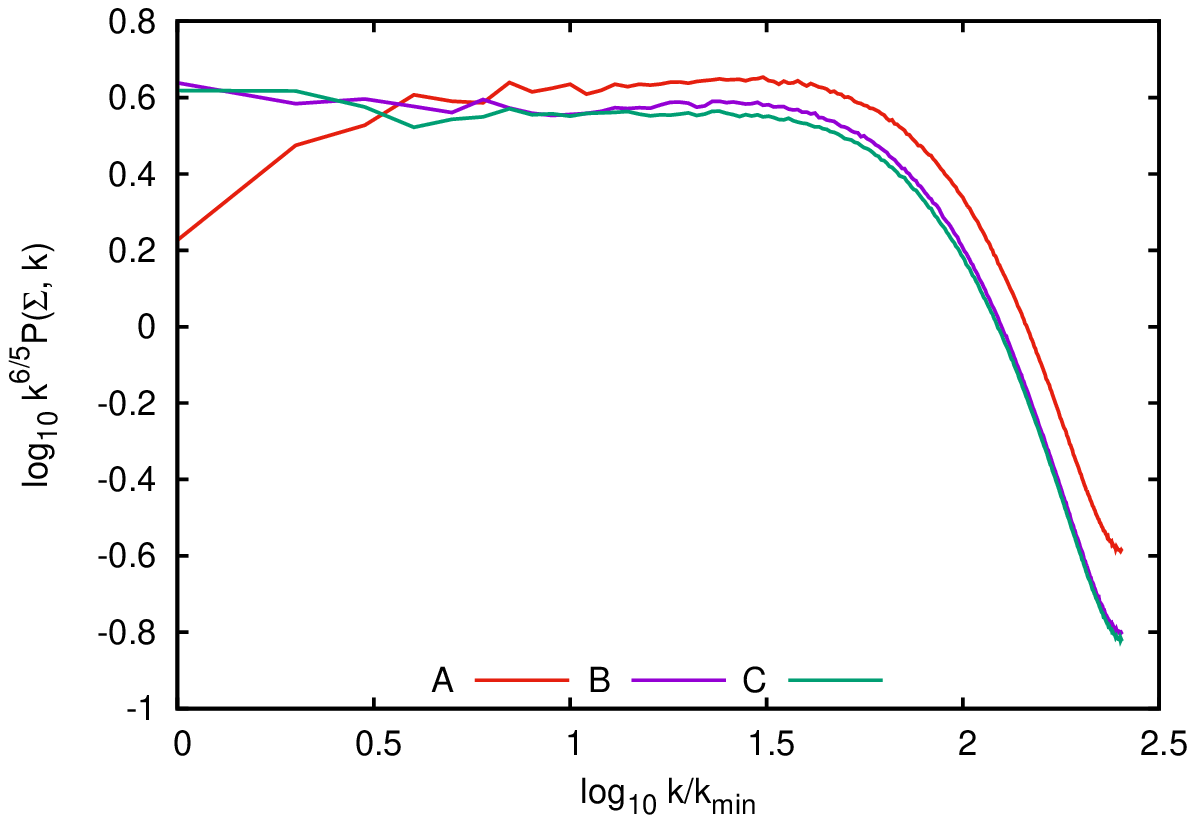}
           \caption{Compensated power spectra of density (left) and column density (right) for cases A, B, and C. {The column density spectra are averages over three orthogonal projection directions. We found no significant differences between spectra for individual projection directions in all three cases, including the strongly magnetized case A.}}
\label{f-dpow}
\end{figure}
We now turn to second order, two-point statistics and discuss the power spectra of primitive variables, such as density and velocities.
The power spectrum is defined here in a usual way, including integration over spherical shells, e.g., for a scalar field $q(\bm x)$, the spectrum $P(q,k)\equiv\int|\widehat{q}(\bm \kappa)|^2\delta(k-|\bm \kappa|)d\bm \kappa$, where $k$ is the wavenumber, $\widehat{q}(\bm\kappa)$ denotes the Fourier transform of $q(\bm x)$; $\delta(k)$ is the Dirac delta function.

Figure~\ref{f-dpow} shows power spectra of the density ($P(\rho,k)$, left) and projected density ($P(\Sigma,k)$, right) for cases A, B, and C compensated with $k^{0.2}$ and  $k^{1.2}$, respectively, so that the scaling range is approximately flat. Unlike in isothermal turbulence at sonic Mach numbers $5-10$, where $P(\rho,k)\propto k^{-1}$, the spectra in our multiphase simulations are substantially more shallow. This is in part due to the modest resolution of the simulations, which is responsible for very fragmented under-resolved cold phase at high densities. The spectra for moderately and weakly magnetized cases B and C are nearly identical, and the case-A spectrum has only a mild power offset and the same slope. Remarkably, the projected density spectra, which involve 2D Fourier transforms and integration over annuli in $k$-space, are very similar in shape and have a slope offset of $-1$, i.e. $P(\Sigma, k)\approx L^2k^{-1}P(\rho,k)/2$. Thus the $\Sigma$-spectrum is a very reliable proxy to the $\rho$-spectrum, assuming isotropy. This important property of the column density spectrum provides direct access to the scale-dependent gravitational potential energy distribution in self-gravitating turbulence (\S~\ref{s-espec} and \cite{banerjee.17}). {We also note that the measured slope of the column density spectrum $P(\Sigma,k)\propto k^{-1.2}$ implies a very weak dependence of $\Sigma$ on scale, $\Sigma\propto r^{0.1}$ for $r\in[5,60]$~pc.}

\begin{figure}[t]
\centering
           \includegraphics[scale=0.6]{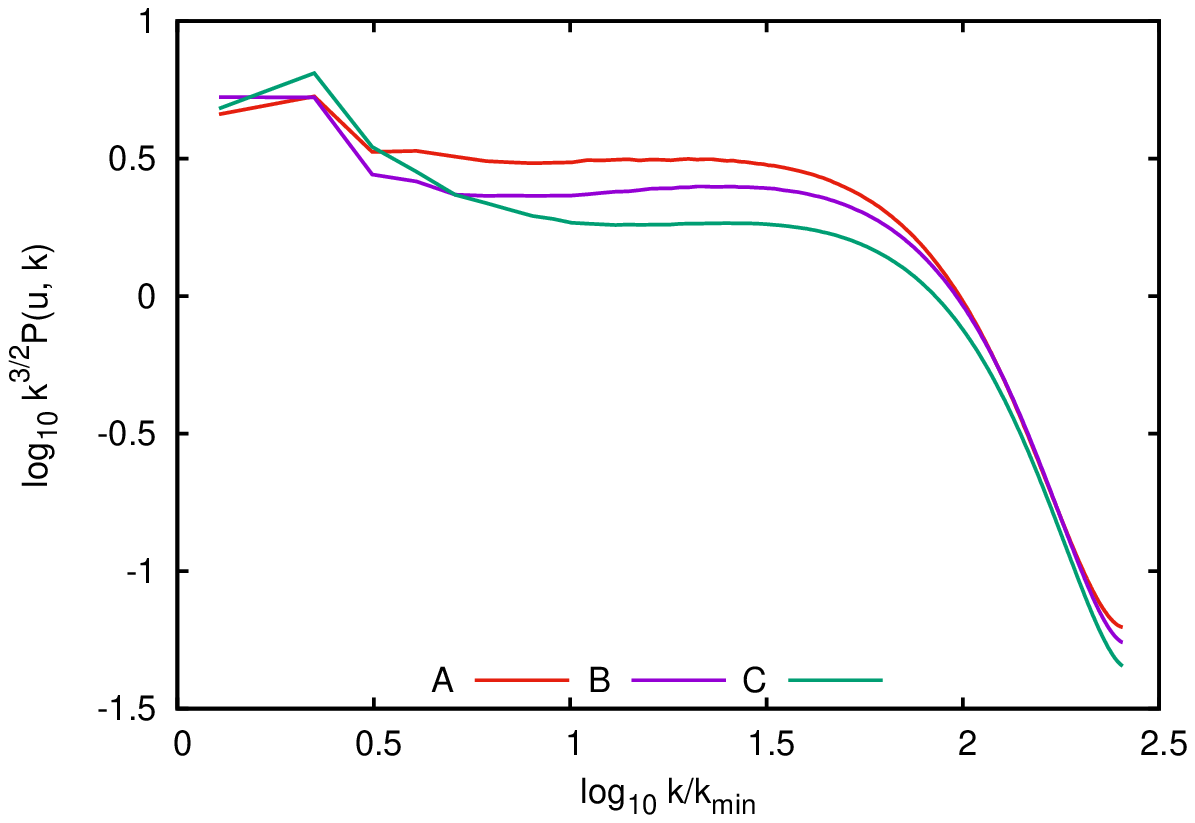}
           \includegraphics[scale=0.6]{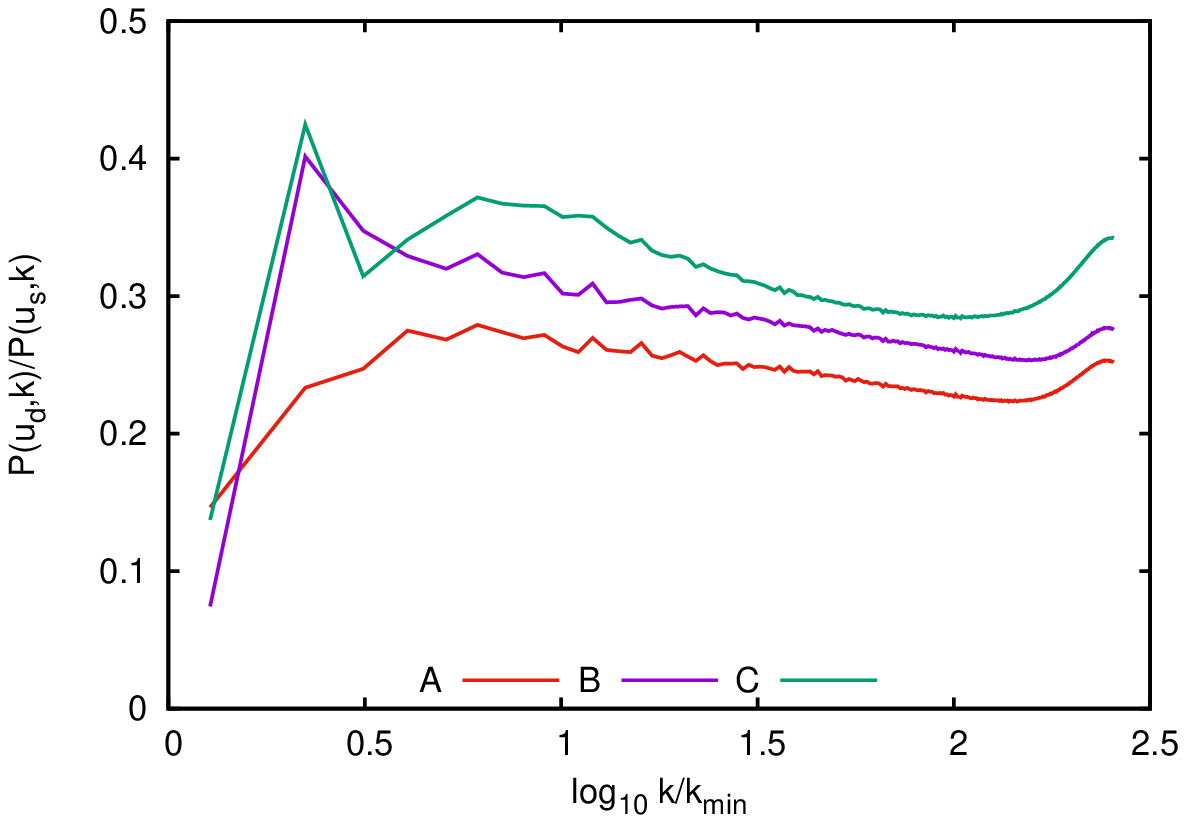}
           \caption{Compensated velocity power spectra (left) and the ratio of dilatational-to-solenoidal power (right) for cases A, B, and C.}
\label{f-upow}
\end{figure}
Figure~\ref{f-upow} (left) shows compensated velocity power spectra $P(\bm u,k)$ for the same three cases. The spectra show similar scaling properties $P\propto k^{-3/2}$ at $k/k_{\rm min}\in[10, 30]$, which most likely does not reflect the inertial range scaling in the limit of $Re\gg1$ due to the limited resolution. The bump in the spectrum at $k<3k_{\rm min}$ reflects the random forcing operating at these scales. The `true' inertial range is most likely contaminated by overlapping effects of the forcing and numerical dissipation.  

Figure~\ref{f-upow} (right) shows results of the Helmholtz decomposition of the velocity field into dilatational (curl-free) and solenoidal (divergence-free) components, $\bm u=\bm u_{\rm d}+\bm u_{\rm s}$. The dilatational-to-solenoidal ratio $P(\bm u_{\rm d},k)/P(\bm u_{\rm s},k)$ starts below $\sim0.3$ in case A and increases to $\sim0.4$ in case C with decreasing magnetization, consistent with expectations for solenoidally driven supersonic isothermal MHD turbulence \cite{kritsuk...10} and for supernovae-driven multiphase ISM turbulence \cite{pan...16}. Since our case includes the baroclinic effect otherwise missing in isothermal simulations, we quantified its contribution to the production of enstrophy in stationary multiphase turbulence, using the formalism based on the enstrophy equation \cite{porter..15}
\begin{equation}
\partial_t\Omega=S_{\rm s}+S_{\rm b}+S_{\rm c}+S_{\rm m}+S_{\rm d}.
\end{equation}
Here  $\Omega=\langle\omega^2/2\rangle$ is the enstrophy, $\langle\cdot\rangle$ denotes averaging over the triply periodic domain, $\bm\omega=\bm\nabla\bm\times\bm u$ is the vorticity, and the dissipative term $S_{\rm d}$ cannot be evaluated explicitly in ILES simulations. Following \cite{porter..15},  we also dropped the contribution from random forcing, which is small since the large-scale acceleration $\bm a$ is mostly uncorrelated with small-scale vorticity $\bm\omega$. 
The remaining four terms of interest here include the vortex stretching term $S_{\rm s}=\langle\bm\omega\bm\cdot(\bm\omega\bm\cdot\bm\nabla)\bm u\rangle$, the baroclinic term $S_{\rm b}=\langle\bm\omega\bm\cdot(\bm\nabla\rho\bm\times\bm\nabla p)/\rho^2\rangle$, the compression term $S_{\rm c}=-\langle\omega^2\bm\nabla\bm\cdot\bm u/2\rangle$, and the magnetic term $S_{\rm m}=\langle\bm\omega\bm\cdot(\bm\nabla\rho\bm\times\bm\nabla p_{\rm mag})/\rho^2+\bm\omega\bm\cdot\bm\nabla\bm\times(\bm B\bm\cdot\bm\nabla\bm B/\rho)\rangle$. 

		\begin{table}
		\caption{\label{enst}Enstrophy, source terms, and associated time scales.}
		\begin{indented}
		\footnotesize
		\item[]\begin{tabular}{@{}ccccccc}
		\br
		Case& $\Omega$ & $S_{\rm s}$ & $S_{\rm b}$ &$S_{\rm c}$ &$S_{\rm m}$ & $\tau$ \\
		& (Myr$^{-2}$) & (\%) & (\%) & (\%) & (\%) & (Myr)\\
		\mr
A & 31.7 & 12.1 & 2.4  & 11.0 & 74.5  & 0.23\\
B & 28.1 & 14.7 & 2.7 & 11.1  & 71.5  & 0.26\\
C & 21.8 & 18.3 & 5.4 & 11.9  & 64.4  & 0.28\\
		\br
		\end{tabular}\\
		\end{indented}
		\end{table}
		\normalsize
Table~\ref{enst} summarizes the relative contributions of these four sources in statistically stationary conditions for cases A, B, and C. The magnetic term $S_{\rm m}$ strongly dominates enstrophy production in all cases, but contributes slightly less under weaker magnetization.  The baroclinic effect $S_{\rm b}$ is generally weak and contributes even less in presence of a stronger mean field. Vortex stretching $S_{\rm s}$ is always the second largest contributor, but magnetic tension progressively reduces its effect in cases with stronger magnetization. Vorticity amplification by compression $S_{\rm c}$ (e.g. in shocks, where a net alignment of the enstrophy gradient and velocity vectors is common \cite{porter..15}) is not sensitive to the magnetization level. The overall levels of enstrophy given in the first column of Table~\ref{enst} correlate positively with the magnetization level, being 1.5 times larger in case A, compared to C. Meanwhile, the characteristic time of enstrophy production by all four sources, $\tau=\Omega/(S_{\rm s}+S_{\rm b}+S_{\rm c}+S_{\rm m})$,
 is roughly the same in all three cases. Since this time scale $\tau\approx0.25$~Myr is short compared to the system lifetime, it naturally has to do with the small-scale processes in the turbulent ISM.

\subsection{Energy spectral densities\label{s-espec}}
\begin{figure}[t]
\centering
           \includegraphics[scale=0.6]{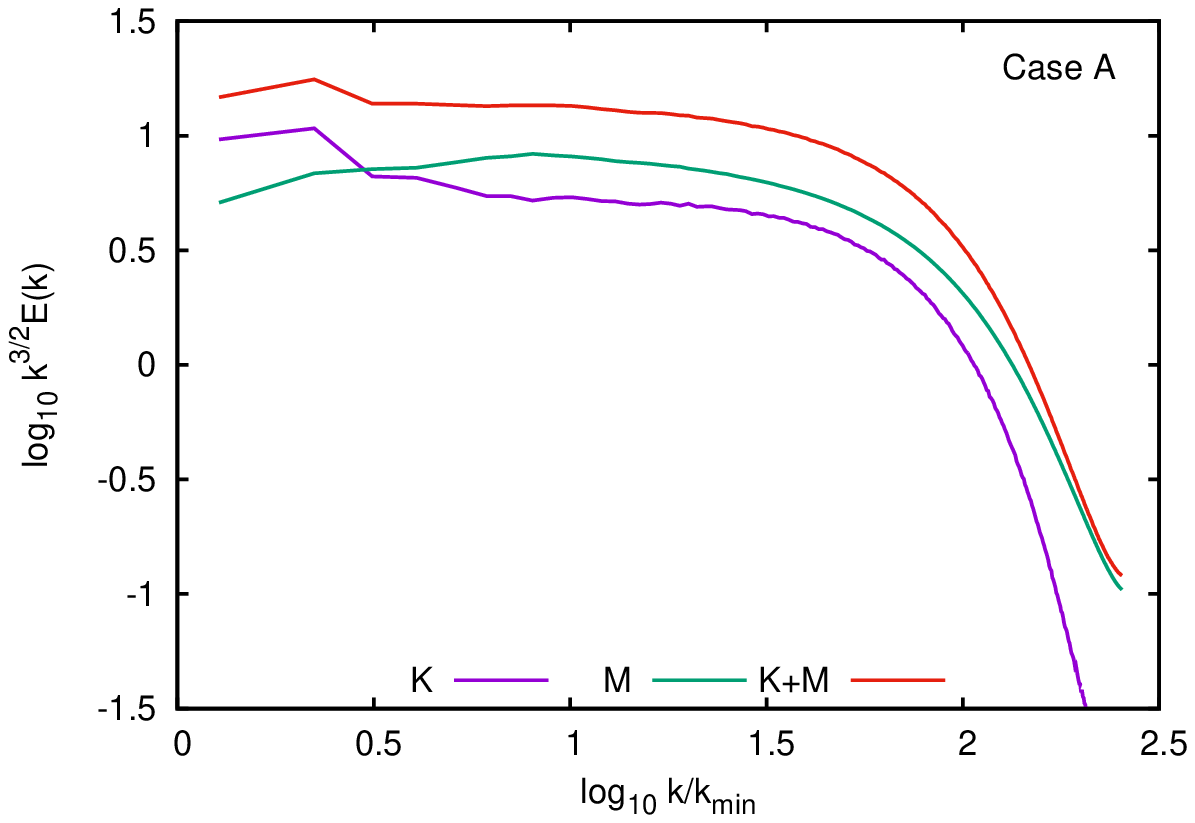}
           \includegraphics[scale=0.6]{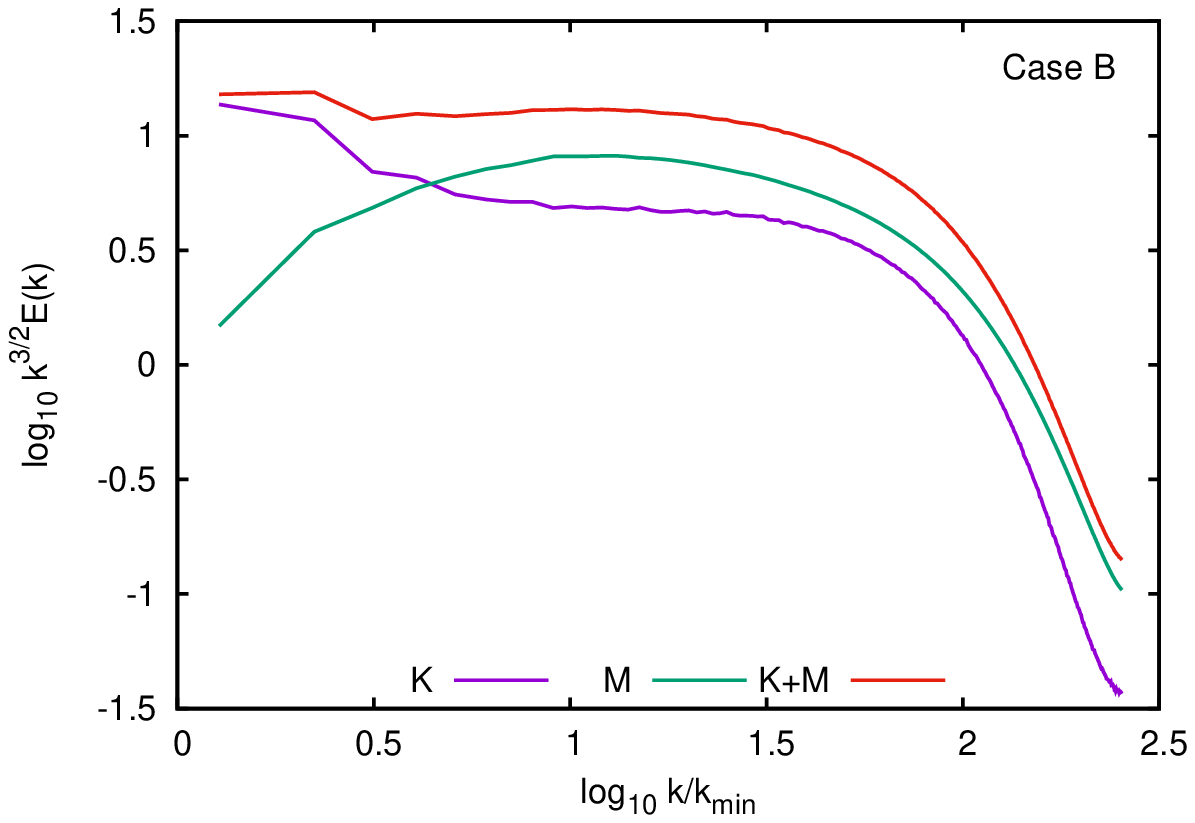}
           \includegraphics[scale=0.6]{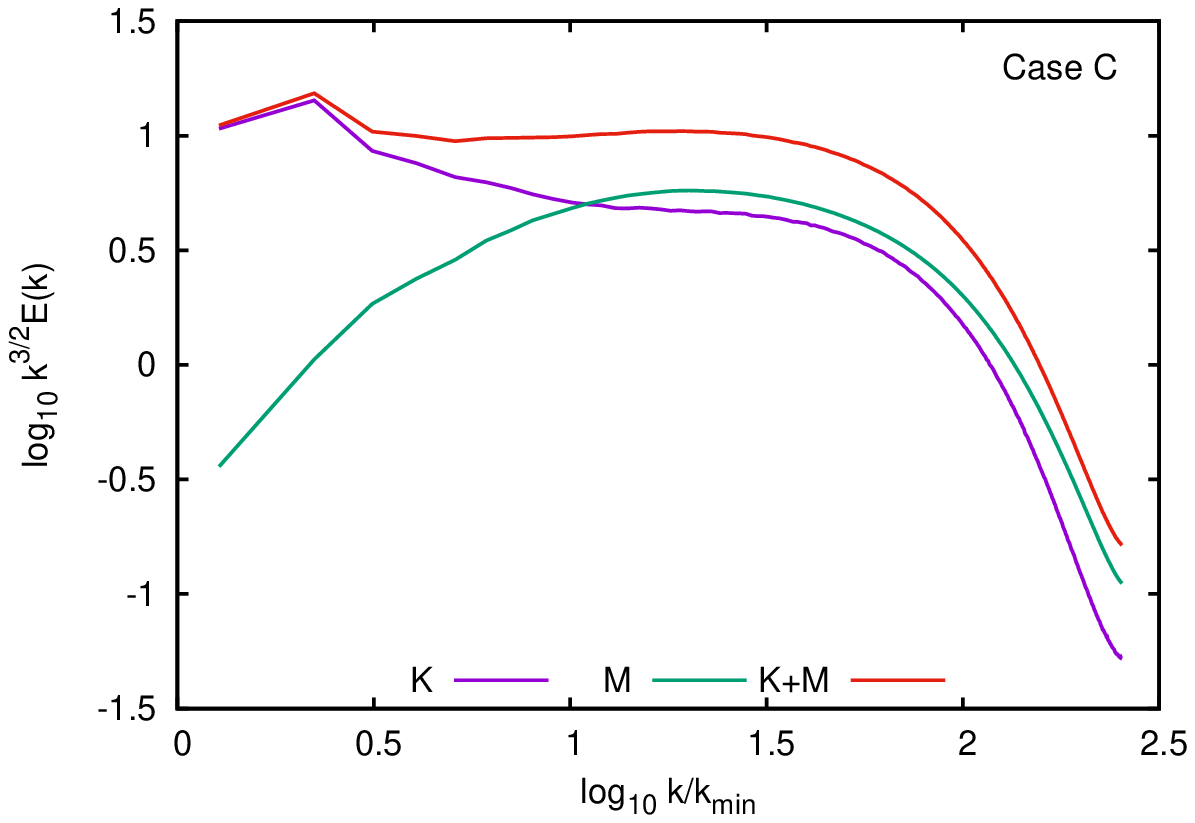}
           \caption{Compensated kinetic and magnetic energy spectra for cases A, B, and C. In all cases, the sum $K(k)+M(k)\propto k^{-3/2}$ at this resolution and the scaling range for the sum is longer than that for any of the two components.
           }
\label{f-cosp}
\end{figure}
The total energy conserved by the ideal system without dissipation, heating, cooling, and forcing includes the kinetic, internal, and magnetic energy contributions: $E=K+U+M$, respectively. Realistic ISM turbulence at length scales $\sim100$~pc, is dominated by the kinetic and magnetic energy constituents, hence $U\ll K\sim M$. The individual spectral densities are defined by $K(k)=P(\bm j,\bm u;k)/2$ and  $M(k)=P(\bm b,k)/2+\delta(k)B_0^2/2$. Here the kinetic energy spectral density is represented by a cospectrum\footnote{The cospectrum $P(\bm j,\bm u;k)$ is the Fourier transform of the symmetric part of the cross-covariance function, ${\cal C}_{ju}(\bm r)=\langle\bm j\bm\cdot\bm u'+\bm j'\bm\cdot\bm u\rangle/2$, integrated over spherical shells, i.e. $P(\bm j,\bm u;k)\equiv\int\widehat{{\cal C}_{ju}}(\bm \kappa)\delta(k-|\bm \kappa|)d\bm \kappa$.} of the momentum density $\bm j\equiv\rho\bm u$ and velocity $\bm u$. Both spectral densities satisfy Parseval's theorem: $K=\int_0^{\infty}K(k)dk$ and $M=\int_0^{\infty}M(k)dk$ \cite{banerjee.17}. {An alternative approach to calculating $K(k)$ for compressible turbulence, using the power spectrum of $\sqrt{\rho}\bm u$ \cite{miura.95}, returns spectra with a large artificial bottleneck at high wavenumbers, originating due to strong autocorrelation of $\sqrt{\rho}$, which has a broad distribution with a width of $\sim3$ dex in these models (Fig.~\ref{f-dpdf}-left). Our choice of the cospectrum $P(\bm j,\bm u;k)$ is motivated by recently derived exact relations for compressible isothermal turbulence \cite{galtier.11} and numerical experiments \cite{kritsuk..13b}. 
}

Figure~\ref{f-cosp} shows the kinetic and magnetic energy spectra as well as their sum, which scale $\propto k^{-1.5}$ at this resolution, for cases A, B, and C.  If the inertial range is sufficiently resolved, detailed kinetic-to-magnetic energy equipartition $K(k)\approx M(k)$ should hold and both $K(k)$ and $M(k)$ are expected to have scaling exponents in a range from $-2$ to $-3/2$. All three cases indeed show that energy equipartition $K(k)\sim M(k)$ holds within a factor of two or less with magnetic and kinetic energy spectra running roughly parallel to each other over $>1$~dex in $k$ in case C and $>1.5$~dex in case A.

\begin{figure}[t]
\centering
           \includegraphics[scale=0.6]{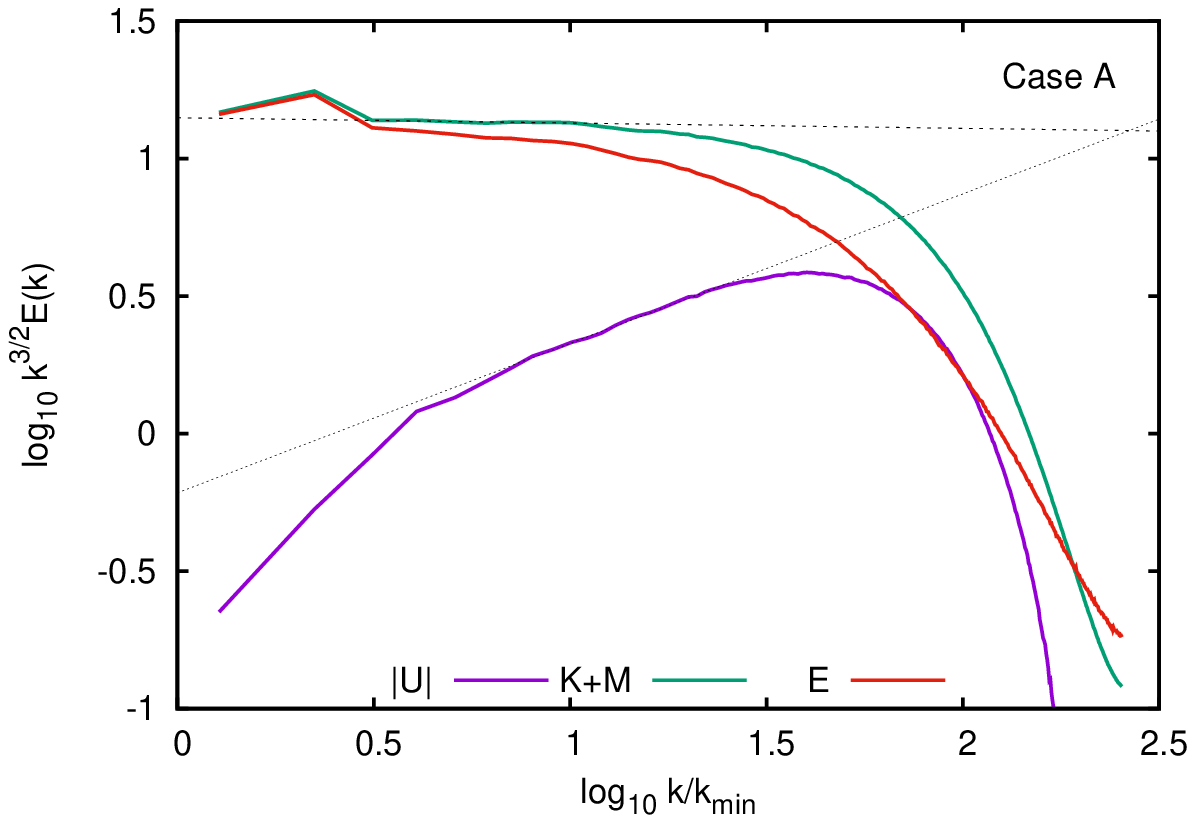}
           \includegraphics[scale=0.6]{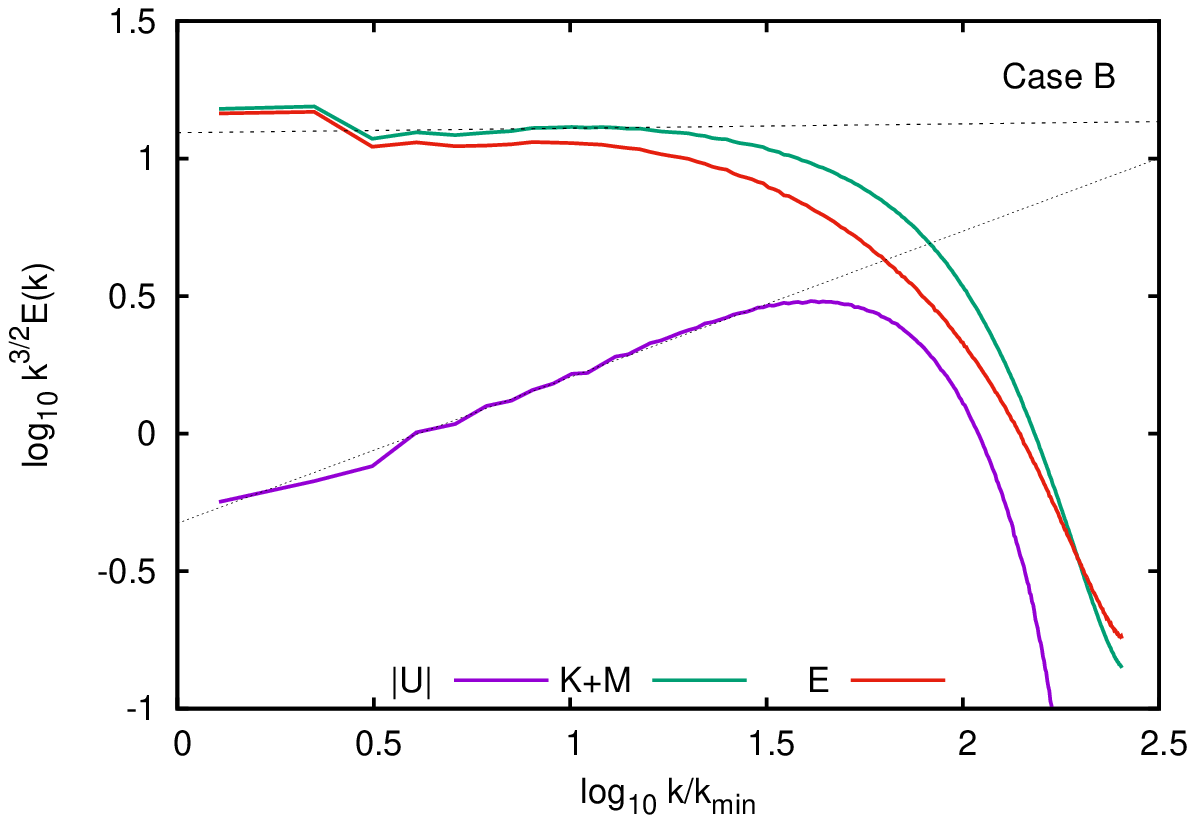}
           \includegraphics[scale=0.6]{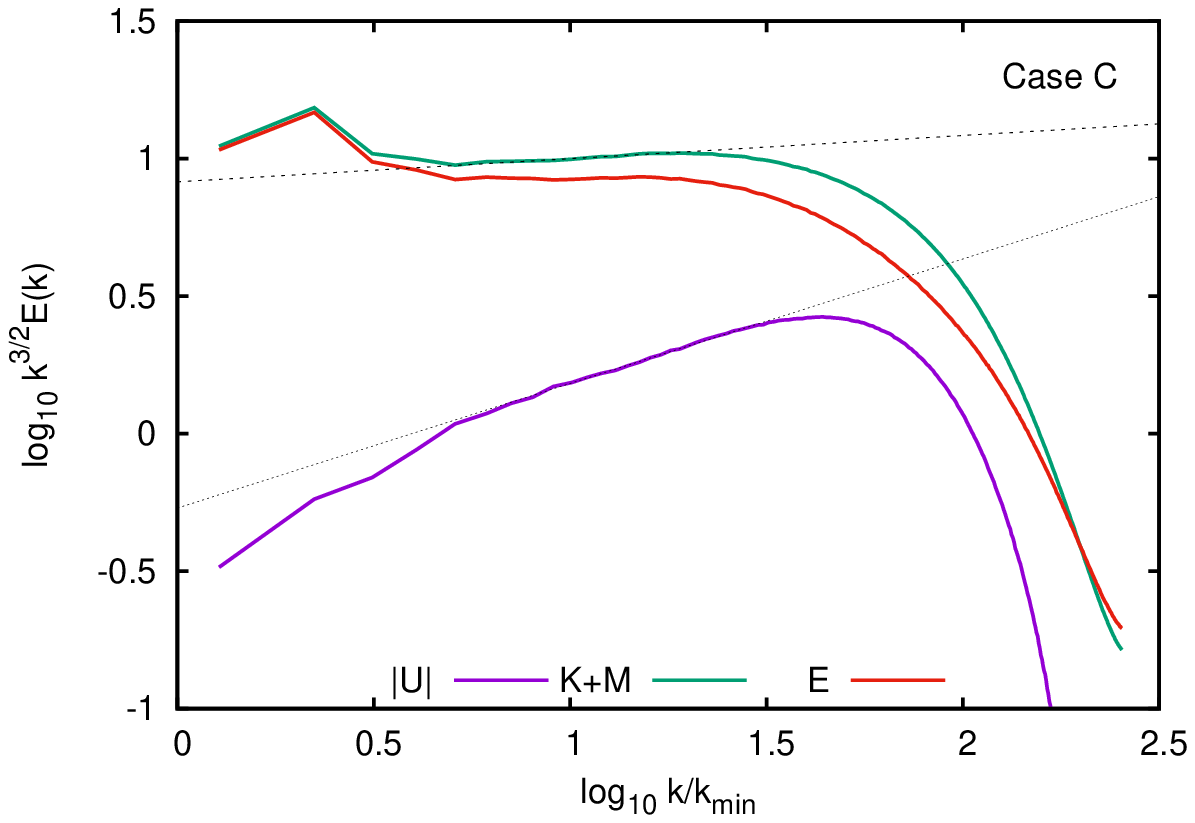}
           \caption{Compensated spectra of the total energy $E(k)$, $K(k)+M(k)$, and $|U(k)|$ for cases A, B, and C. Formal extrapolation of the spectra for $K(k)+M(k)$ and $U(k)$ defines (although with a high degree of uncertainty) a crossover  scale $\lambda_p\sim0.8$~pc, $0.4$~pc, $0.1$~pc for cases A, B, C, respectively.}
 \label{f-uco}
\end{figure}
While at length scales $\sim100$~pc the internal energy is clearly subdominant, it may still play a role on smaller scales, e.g., where turbulence in the bulk of the volume becomes subsonic. To probe the scale-dependent part of $U$, we define the spectral density of internal energy, following a template for isothermal supersonic turbulence developed in \cite{banerjee.17}, by $U(k)\equiv P(\rho,e;k)/2+\delta(k)U/2$.\footnote{Here, as in the case of kinetic energy spectral density, the cospectrum $P(\rho,e;k)$ is the Fourier transform of the symmetric part of the cross-covariance function ${\cal C}_{\rho e}(\bm r)=\langle\rho e'+\rho'e\rangle/2$ integrated over spherical shells in $k$-space. Also note the $k=0$ component, which is similar the mean-field energy in the expression for $M(k)$, and that Parseval's theorem $U=\int_0^{\infty}U(k)dk$ holds in this case as well.} 
Internal energy spectra computed for cases A, B, and C are very shallow, $U(k)\propto k^{-1}$ (Fig.~\ref{f-uco}), most likely because thermal processes are not properly resolved and the cold phase is highly fragmented. Unlike in the isothermal case, where $\rho$ and $e=c_{\rm s}^2\ln(\rho/\rho_0)$ are always positively correlated, the cospectra here are negative at large scales due to anti-correlation of density and temperature perturbations in the quasi-isobaric TI regime. The spectra then switch to positive on small scales (not shown) under quasi-adiabatic conditions. 
The shallow $U(k)\propto k^{-1}$ scaling in combination with the quasi-isobaric conditions in the multiphase gas create a depression in the total energy spectrum $E(k)$ centered at $k_p\sim 50k_{\rm min}$. The negative thermal pressure `support' at the associated `crossover' length scale $\lambda_p=2\pi/k_p$ would make objects of this size more prone to collapse, if their mean density is sufficiently high. Our models, however, do not resolve this scale properly. Thus we can only resort to extrapolation to get a rough estimate for $\lambda_p$, which returns a fraction of a parsec (Fig.~\ref{f-uco}). Higher resolution simulations would show whether or not this crossover scale defines the characteristic width of star forming filaments measured by the {\em Herschel} satellite \cite{arzoumanian+11}.

Finally, in order to make predictions for self-gravitating turbulence, we can readily estimate the gravitational potential energy spectral density, corresponding to the density spectrum in Fig.~\ref{f-dpow}, as $W(k)=-2\pi G k^{-2}P(\rho,k)\propto k^{-2.2}$ \cite{banerjee.17}. If we take the stationary state of developed turbulence as initial conditions for a simulation with self-gravity, the gravitational potential energy would initially have a slightly steeper scaling compared to the sum of kinetic and magnetic energy. Also, the (negative) spectral density of the potential energy is smaller in absolute value than $E(k)$, in our cases with $L=200$~pc, ensuring global gravitational stability of the ISM and hence purely local character of star formation.

\section{Conclusions\label{s-con}}
We explored the statistics of multiphase, magnetized, ISM turbulence and the formation mechanisms of molecular clouds with idealized numerical simulations, using the PPML solver \cite{ustyugov...09}. Our periodic box models employed a large-scale solenoidal force to stir the turbulence up to the rms velocities consistent with observations. After a few dynamical times of forcing, the fully developed turbulence reached a steady state subject to our statistical analysis. A fraction of the kinetic energy supplied by the external forcing is stored in the form of induced turbulent magnetic energy, which tends to saturate at equipartition with the kinetic energy on small scales. Due to the Alfv\'en effect, a strong dynamic field alignment develops in the quasi-stationary state in most of the computational domain, at characteristic ISM densities $\sim1$~cm$^{-3}$. The models indicate that this factor may be responsible for the super-Alfv\'enic nature of turbulence in the cold and dense parts of molecular clouds. Being a consequence of self-organization in highly compressible MHD turbulence, this result does not depend on the way the turbulence is initiated or fed in our models. The simulations capture basic physics and {help to constrain the range in parameter space, where the model} is overall successful in reproducing the following observables of the local ISM, including molecular clouds:
\begin{itemize}
\item the ratio of the turbulent magnetic field component versus the regular field measured at length scales $\sim100$~pc;
\item the mass and volume fractions of thermally stable neutral H{\sc i} and the abundance of the unstable regimes;
\item the variety of sonic and Alfv\'enic Mach number regimes in different thermal phases;
\item the low rates of star formation per free-fall time controlled by the multiphase thermodynamics and large-scale turbulence;
\item the lognormal distribution of column densities and deviations from lognormality for some specific tracers;
\item the mass-weighted distribution of thermal pressure, including the mean value and the characteristic asymmetric shape;
\item the overall hierarchical filamentary morphology of the molecular gas and the alignment of filaments with respect to magnetic field lines; 
\item the linewidth-size relationship for molecular clouds, including the slope and the offset;
\item the ratio of solenoidal-to-compressive velocity power.
\end{itemize}
Our models also provide predictions for the shape of magnetic field PDFs, which are strongly non-Gaussian. We apply new probabilistic tools developed for self-gravitating compressible turbulence elsewhere \cite{banerjee.17} to assess the spectral energy distributions. These help to disentangle the effects of turbulence, thermodynamics, and self-gravity and single out characteristic length scales.
By design, our models are limited to the local ISM and molecular clouds. Exploring cases with more active star formation, starbursts, mergers, and other non-steady phenomena would require more sophisticated experiments with properly closed feedback loops.
Higher dynamic range modeling is required to further harden the conclusions.

\section*{Acknowledgments}
This research was supported in part by the National Science Foundation through grants AST-0607675, AST-0808184, AST-0908740, AST-1109570, and AST-1412271; used XSEDE allocation MCA07S014, and DOE Office of Science INCITE-2009 and DD-2010 awards allocated at NCCS (ast015/ast021). A.K. and M.N. acknowledge significant contribution to this work by the late Dr. Sergey D. Ustyugov, who developed PPML and ran the simulations. Christoph T. Lee carried out statistical data analysis to produce Figure~\ref{f-sf1}. The authors would like to thank both anonymous referees for constructive reports, which helped to improve the quality of presentation.

\section*{References}

\bibliographystyle{unsrt}

\end{document}